\documentclass[conference, letterpaper, 10pt]{IEEEtran}
\AtBeginDocument{%
  }
    
\usepackage[utf8]{inputenc}
\usepackage{algorithm}
\usepackage{algpseudocode}
\usepackage{graphicx}
\usepackage{textcomp}
\usepackage{xcolor}
\usepackage{booktabs}
\usepackage{amssymb}
\usepackage{stmaryrd}
\usepackage{amsmath,amsthm}
\newcommand{\Mod}[1]{\ \mathrm{mod}\ #1}
\usepackage{paralist}

\let\labelindent\relax
\usepackage{enumitem}

\usepackage{subcaption}
\usepackage{mathtools}

\usepackage{hyperref}

\usepackage[acronym, shortcuts]{glossaries}
\glsdisablehyper

\usepackage[utf8]{inputenc}
\usepackage[english]{babel}
\addto\extrasenglish{
    
}

\theoremstyle{remark}
\newtheorem*{finding}{Finding}

\theoremstyle{remark}

\DeclareMathOperator*{\argmin}{argmin}
\DeclareMathOperator*{\argmax}{argmax}

\begin{document}

\title{Sneaky Spikes: Uncovering Stealthy\\ Backdoor Attacks in Spiking Neural Networks\\ with Neuromorphic Data}

\author{\IEEEauthorblockN{Gorka Abad}
\IEEEauthorblockA{Radboud University, The Netherlands}
\IEEEauthorblockA{ \& Ikerlan Research Centre, Spain\\
abad.gorka@ru.nl}
\and
\IEEEauthorblockN{O\u{g}uzhan Ersoy}
\IEEEauthorblockA{Radboud University, \\The Netherlands\\
oguzhan.ersoy@ru.nl}
\and
\IEEEauthorblockN{Stjepan Picek}
\IEEEauthorblockA{Radboud University,\\}
\IEEEauthorblockA{The Netherlands\\
stjepan.picek@ru.nl}
\and
\IEEEauthorblockN{Aitor Urbieta}
\IEEEauthorblockA{Ikerlan Research Centre,\\
Spain\\
aurbieta@ikerlan.es}
}



\newacronym{DNN}{DNN}{deep neural network}
\newacronym{MLP}{MLP}{multilayer perceptron}
\newacronym{ReLU}{ReLU}{rectified linear unit}
\newacronym{AE}{AE}{autoencoder}
\newacronym{DL}{DL}{deep learning}
\newacronym{AI}{AI}{artificial intelligence}
\newacronym{ML}{ML}{machine learning}
\newacronym{NN}{NN}{neural network}


\newacronym{STDP}{STDP}{spike timing-dependent plasticity}
\newacronym{SNN}{SNN}{spiking neural network}
\newacronym{DVS}{DVS}{dynamic vision sensor}
\newacronym{LIF}{LIF}{leaky integrate-and-fire}
\newacronym{ATIS}{ATIS}{asynchronous time-based image sensor}


\newacronym{LR}{LR}{learning rate}
\newacronym{MSE}{MSE}{mean squared error}
\newacronym{ASR}{ASR}{attack success rate}
\newacronym{SSIM}{SSIM}{structural similarity index}
\newacronym{SVD}{SVD}{singular value decomposition}



\newacronym{ABS}{ABS}{artificial brain stimulation}
\newacronym{STRIP}{STRIP}{strong intentional perturbation}
\newacronym{NC}{NC}{neural cleanses}


\maketitle

\begin{abstract}
Deep neural networks (DNNs) have demonstrated remarkable performance across various tasks, including image and speech recognition. However, maximizing the effectiveness of DNNs requires meticulous optimization of numerous hyperparameters and network parameters through training. Moreover, high-performance DNNs entail many parameters, which consume significant energy during training. To overcome these challenges, researchers have turned to spiking neural networks (SNNs), which offer enhanced energy efficiency and biologically plausible data processing capabilities, rendering them highly suitable for sensory data tasks, particularly in neuromorphic data. Despite their advantages, SNNs, like DNNs, are susceptible to various threats, including adversarial examples and backdoor attacks. Yet, the field of SNNs still needs to be explored in terms of understanding and countering these attacks.
This paper delves into backdoor attacks in SNNs using neuromorphic datasets and diverse triggers. Specifically, we explore backdoor triggers within neuromorphic data that can manipulate their position and color, providing a broader scope of possibilities than conventional triggers in domains like images. We present various attack strategies, achieving an attack success rate of up to 100\% while maintaining a negligible impact on clean accuracy.
Furthermore, we assess these attacks' stealthiness, revealing that our most potent attacks possess significant stealth capabilities.
Lastly, we adapt several state-of-the-art defenses from the image domain, evaluating their efficacy on neuromorphic data and uncovering instances where they fall short, leading to compromised performance.
\end{abstract}




\section{Introduction}
\label{sec:introduction}

\Acp{DNN} have achieved excellent performance in \ac{ML} tasks in different domains, like computer vision~\cite{krizhevsky2012imagenet}, speech recognition~\cite{graves2013speech}, and text generation~\cite{brown2020language}. One key aspect of DNNs that has contributed to their success is their ability to learn from large amounts of data and discover complex patterns. This is achieved through multiple layers of interconnected neurons. The connections between these nodes are weighted and adjusted during training to minimize error and improve the model's accuracy. \Acp{DNN} have many hyperparameters that can be tuned to achieve top performance on a given task, but careful optimization of these hyperparameters is crucial.
Training a well-performing \ac{DNN} can be time and energy expensive, requiring tuning many parameters with large training data. For example, training the GPT-3 model consumed about 190\,000 kWh of electricity~\cite{dhar2020carbon}.
These models' increasing complexity and computational requirements have led researchers to explore alternative approaches, such as \acp{SNN}~\cite{ghosh2009spiking, tavanaei2019deep, eshraghian2021training, fang2021incorporating}. 

\Acp{SNN} can significantly reduce the energy consumption of \acp{DNN}. For instance, Kundu et al.~\cite{kundu2021spike} achieved better compute energy efficiency (up to $12.2 \times$) compared to \acp{DNN} with a similar number of parameters. In addition to their energy efficiency, \acp{SNN} have several other benefits. 
\Acp{SNN} can be more robust to noise and perturbations, making them more reliable in real-world situations~\cite{lee2016training}. More precisely, data obtained by a \ac{DVS}~\cite{serrano2013128}---which \acp{SNN} can process---captures per pixel brightness changes asynchronously, instead of the absolute brightness in a constant rate---as in images.
Compared to standard cameras, \ac{DVS} cameras have low power consumption and capture low latency data, i.e., neuromorphic data, which also has high temporal resolution~\cite{chen2020event, liu2019event}. Thus, \acp{SNN} can process data in a more biologically plausible manner, for example, by processing neuromorphic data, making them well-suited for tasks involving sensory data processing. In computer vision, significant advancements have been made in the context of autonomous driving, as evidenced by the exceptional performance attained~\cite{yurtsever2020survey}. The surrounding environment is captured by employing one or more vehicular cameras, with the following data being processed via a \ac{DNN}. The decisions made by this \ac{DNN} facilitate the autonomous operation of the vehicle. In recent works, alternative approaches have been proposed, wherein event-based neuromorphic vision is advocated to accomplish the same objective~\cite{chen2020event, viale2021carsnn}.
Event-based data allows solving challenging scenarios where regular data (captured by standard cameras) cannot perform well~\cite{maqueda2018event, zhu2018multivehicle}, such as high-speed scenes, low latency, and low power consumption scenarios. 
Moreover, \acp{SNN} are widely applicable, being used in domains like medical diagnosis~\cite{ghoshdastidar2007improved} and computer vision~\cite{WYSOSKI2010819, 4370930}. Finally, while \acp{DNN} are often considered to perform better than \acp{SNN} in terms of accuracy, recent results show this performance gap is reducing or even disappearing~\cite{tavanaei2019deep}.

\Acp{DNN} are vulnerable to various privacy and security threats, including adversarial examples~\cite{szegedy2013intriguing}, inference attacks~\cite{carlini2021extracting}, model stealing~\cite{jagielski2020high}, and backdoor attacks~\cite{gu2019badnets}. However, despite the widespread application domain of \acp{SNN}, their security aspects have yet to receive a comprehensive evaluation. Recent investigations~\cite{marchisio2020is} have conducted a comparative analysis of the security vulnerabilities inherent in \acp{SNN} to date, revealing their susceptibility to adversarial examples. Furthermore, subsequent research~\cite{venceslai2020neuroattack} has demonstrated that \acp{SNN} are also susceptible to hardware attacks, where the deliberate induction of bit-flips can lead to misclassification.

Backdoor attacks are a threat where malicious samples containing a trigger are included in the dataset at training time. After training, a backdoor model correctly performs the main task at test time while achieving misclassification when the input contains the trigger.
Backdoor attacks on \acp{DNN} are well studied with several improvements such as stealthy triggers~\cite{liu2020reflection,zhang2022poison} and dynamic triggers unique per data sample~\cite{nguyen2020input,salem2022dynamic}. Moreover, multiple works consider the backdoor defenses trying to detect and prevent backdoor attacks by inspecting \ac{DNN} models~\cite{wang2019neural,liu2019abs, chen2000linkbreaker, li2021neural}.

However, the existing backdoor attacks and defenses on \acp{DNN} do not directly apply to \acp{SNN} because of the different structures of SNNs and their usage of neuromorphic data. Unlike \acp{DNN}, \acp{SNN} do not have activation functions but spiking neurons, which could reduce or even disable the usage of existing attacks and defenses in \acp{DNN} that rely on them, as discussed in~\autoref{sec:evaluation}.
Additionally, the time-encoded behavior of neuromorphic data allows a broader range of possibilities when generating input perturbations. At the same time, the captured data is encoded in much smaller pixel space (2-bit pixel space) than in regular images, which can handle up to 255 pixel possibilities per channel. 
The challenges regarding the application of backdoor attacks in \acp{SNN} are detailed in~\autoref{sec:attack}.

To our knowledge, only one work explores backdoor attacks on \acp{SNN}~\cite{abad2022poster}.
The triggers used in the attack are static and moving square placed in the image during training time, which is not stealthy and is easily visible by human inspection, as investigated in~\autoref{sec:stealthiness}. Furthermore, their attack setup is limited, only considering three different poisoning rates and a single trigger size. Finally, the authors did not consider any backdoor defense.
 
This paper thoroughly investigates the viability of backdoor attacks in \acp{SNN}, the stealthiness of the trigger, and the robustness against the defenses.
First, we improve the performance of the static and moving triggers proposed in~\cite{abad2022poster}.
Next, we propose two new trigger methods: smart and dynamic triggers.
Our novel methods significantly outperform the existing backdoor attacks in \acp{SNN}.
Our main contributions are:
\begin{compactitem} 
	\item We explore different backdoor injecting methods on \acp{SNN}, achieving at least 99\% accuracy in both main and backdoor tasks. We first explore static and moving triggers, which led to developing a smart attack that selects the optimal trigger location and color.
	\item We introduce the first dynamic backdoor attack on the neuromorphic dataset, which is highly stealthy.
	\item We analyze the stealthiness of backdoor attacks using the \ac{SSIM} metric, showing that our dynamic trigger achieves up to 99.9\% \ac{SSIM}, outperforming static and moving triggers at 98.5\% \ac{SSIM}. Additionally, we conduct a user study to measure the stealthiness of our attacks and compare their effectiveness with \ac{SSIM}.
	\item We adapt image domain defenses for \acp{SNN} and neuromorphic data, observing their ineffectiveness against backdoor attacks.
 \end{compactitem}
 
We share our code to allow the reproducibility of our results.\footnote{\url{https://github.com/GorkaAbad/Sneaky-Spikes}}
Moreover, we show our triggers' dynamic motion and stealthiness as a live demo in the repository.

\section{Background}
\label{sec:background}



\subsection{Backdoor Attacks}

Backdoor attacks modify the behavior of a model during training, so at test time, it behaves abnormally~\cite{gu2019badnets}. A backdoored model misclassifies the inputs with a trigger while behaving normally on clean inputs. In a data poisoning backdoor, the training set is modified to include malicious data samples with the trigger. In the image domain, the trigger can be a pixel pattern in a specific image part. When the algorithm is trained on a mixture of clean and backdoor data, the model learns only to misclassify the inputs containing the pixel pattern, i.e., the trigger, to a particular target label.

Formally, an algorithm $f_\theta(\cdot)$ is trained on a mixture dataset containing clean and backdoor data, which rate is controlled by $\epsilon = \frac{m}{n}$ where $n$ is the size of the clean dataset, $m$ is the size of the backdoor dataset, and $m \ll n$. The backdoor dataset $\mathcal{D}_{bk}$ is composed of backdoor samples $\{(\hat{\mathbf{x}}, \hat{y})\}^m \in \mathcal{D}_{bk}$, where $\hat{\mathbf{x}}$ is the sample containing the trigger and $\hat{y}$ is the target label. 
For a clean dataset of size $n$, the training procedure aims to find $\theta$ by minimizing the loss function $\mathcal{L}$:
\begin{align}
\footnotesize
    \label{eq:train}
    \theta' = \argmin_\theta \sum_{i=0}^n \mathcal{L}(f_\theta(\mathbf{x}_i),y_i),
\end{align}
where $\mathbf{x}$ is input and $y$ is label.
During the training with backdoor data,~\autoref{eq:train} is modified to include the backdoor behavior expressed as:
\begin{align*}
\footnotesize
    \theta' = \argmin_\theta \sum_{i=0}^n \mathcal{L}(f_\theta(\mathbf{x}_i),y_i) + \sum_{j=0}^m \mathcal{L}(f_\theta(\hat{\mathbf{x}}_j),\hat{y}_j).
\end{align*}

\subsection{Spiking Neural Networks \& Neuromorphic Data}

\Acp{SNN} are a class of neural networks that model the activity of biological neurons by using discrete events, or spikes, to represent the output of a neuron when it reaches a certain threshold. In contrast to traditional artificial neural networks, which operate on continuous-valued signals, \acp{SNN} are event-driven and incorporate the temporal dynamics of neural activity~\cite{tavanaei2019deep}.
\Acp{SNN} comprise layers of interconnected spiking neurons, including input, hidden, and output layers---similar to \acp{DNN}. 

Mathematically, the behavior of a spiking neuron can be modeled using the \ac{LIF} model, which describes the neuron's membrane potential as a function of time~\cite{hunsberger2015spiking}. The membrane potential of a neuron is a result of the sum of its inputs. When the membrane potential reaches a certain threshold $\Theta$, the neuron emits a spike, and its membrane potential is reset to a resting potential:
\begin{align}
\footnotesize
\label{eq:lif}
    h(x)= 
\begin{cases}
    1,& \text{if } x\geq \Theta\\
    0,              & \text{otherwise,}
\end{cases}
\end{align}
similar to the commonly used in \ac{DL} \ac{ReLU} activation function~\cite{fukushima1975cognitron}. However, note that~\autoref{eq:lif} is non-differentiable, which would lead to the impossibility of the derivative calculation during backpropagation, the \textit{de facto} training algorithm in \acp{DNN}. Thus, there are two main options to train \acp{SNN}: \ac{STDP}~\cite{abbott2000synaptic} or surrogate gradients~\cite{neftci2019surrogate}. \Ac{STDP} modifies the strength of connections between neurons based on the relative timing of their spikes. Specifically, if neuron \emph{A} fires before neuron \emph{B}, the connection between \emph{A} and \emph{B} is strengthened, whereas if neuron \emph{B} fires before neuron \emph{A}, the connection is weakened. This allows the network to adapt to changing input patterns and improve performance. Surrogate training, conversely, allows the approximation of the derivatives to perform backpropagation, enabling compatibility with Adam~\cite{kingma2014adam} or stochastic gradient descent~\cite{amari1993backpropagation}. The latter achieves better performance~\cite{lee2016training} and is the method we follow in our experimentation.

\Acp{SNN} commonly operate on neuromorphic data, a time-encoded representation of the illumination changes of an object/subject captured by a \ac{DVS} camera. The \ac{DVS} camera captures a flow of spiking events that dynamically represents the changing visual scene. The advantage of using \ac{DVS} cameras is that they provide a compressed representation of the visual scene that can be processed almost instantaneously, as reported in~\cite{serrano2013128}. More precisely, neuromorphic data is encoded in $T$ frames and $p$ polarities. In neuromorphic data processing, polarity refers to the direction of the electrical signals representing data. These signals can be positive or negative, denoted as the \emph{ON polarity} and \emph{OFF polarity}, respectively; each polarity generates a different color. Using polarity in neuromorphic data processing can help reduce the energy required for computation. Using only positive or negative spikes reduces the number of bits required to represent data and thus reduces the system's power consumption~\cite{tavanaei2019deep}. \Acp{SNN} have shown promising results in various tasks~\cite{ponulak2011introduction}, including speech recognition, image classification, and robotics.

\section{Backdoor Attacks to SNNs}
\label{sec:attack}

\subsection{Threat Model}
\label{sec:threat}



We consider the same threat model as in prior studies~\cite{gu2019badnets, doan2021lira, salem2022dynamic, abad2022poster, liu2020reflection}, which assumes that the attacker can have full access to the model. Additionally, the attacker has access to the training dataset provided by the client. More precisely, we consider data poisoning-based, dirty label methodology for injecting the backdoor. We also limit our research to the digital image domain, where triggers are intended to be injected in digital samples rather than applied physically in the wild~\cite{bagdasaryan2021blind}.

As a use case, we assume that a client wants to train an \ac{SNN} on an owned dataset but does not have the resources, e.g., GPU cards, to train it. Therefore, the client outsources the training to a third party that provides on-cloud training services, such as Google Cloud\footnote{\url{https://cloud.google.com}} or Amazon Web Services\footnote{\url{https://aws.amazon.com}}, by sharing the model architecture and the training dataset. We assume that the attacker is the third-party provider, thus having access to the training procedure, the model, and the dataset. The attacker then injects the backdoor during training and shares the model with the client. The client can check the model's performance using a holdout dataset.



\subsection{Challenges in SNNs}


\Acp{SNN} have shown promising results in various domains, including image recognition~\cite{kheradpisheh2018stdp} or image segmentation~\cite{patel2021spiking} with several applications ranging from autonomous driving to medical diagnosis, to name a few. However, as we demonstrate, \acp{SNN} are vulnerable to security threats, which can have serious consequences. \acp{SNN} have a unique network structure and utilize neuromorphic data. The information propagation is the key difference between ``classical'' \acp{DNN} and \acp{SNN}. \Acp{SNN} do not work with continuously changing time values (like \acp{DNN}) but operate with discrete events that occur at certain points in time. The training is different, making the attacks happening in the training phase (potentially) challenging to deploy.

For instance, the triggers used in the image domain are encoded in 255 possibilities per channel, which gives many combinations of color. In neuromorphic data, however, the trigger space is reduced to 4 possibilities encoded by the two different polarities. 
Furthermore, in the image domain, the trigger is commonly static, i.e., no time-encoded data is used. In neuromorphic data, we encode the trigger using the time window, allowing us to create triggers that change the location through time. Next, we list the main challenges:

\begin{enumerate}[itemsep=0mm, label=\textit{C.\arabic*:}, ref={C.\arabic*}, leftmargin=0ex, labelsep=\widthof{~}, itemindent=\widthof{C.2:~\hspace{.125ex}}]

\item \label{itm:trigger} \textit{Designing and optimizing the trigger.}
Selecting and optimizing the trigger in the context of neuromorphic systems poses significant challenges due to the temporal nature and multiple frames per data point. The diverse range of options raises important questions: \emph{How can we efficiently identify the trigger that maximizes the efficacy of the attack while minimizing its impact on clean accuracy? When designing a backdoor trigger specific to \acp{SNN}, what are the crucial parameters?}

\item \label{itm:invisible} \textit{Generating stealthy triggers.}
Generating stealthy triggers for neuromorphic systems presents difficulties since each trigger pixel can only assume four distinct values, making smoothing the trigger over clean data more complex than images with higher value ranges, such as $3\times 256$ in regular data. This limitation raises the questions: \emph{How can we design a backdoor trigger that exploits the time-encoded data to create a unique, imperceptible trigger for each input sample and frame? What influence do the selected parameters exercise on the stealthiness of the trigger?}

\item \label{itm:defenses} \textit{Backdoor defenses.}
Given that \acp{SNN} do not incorporate activation functions commonly employed in backdoor defense mechanisms, existing state-of-the-art defenses may not directly apply. These defenses are typically designed for image data, while neuromorphic data comprises multiple frames per data point and distinct color encoding. These dissimilarities lead to the following questions: \emph{How effective are the current defense strategies when applied to \acp{SNN}? How can we adapt these defenses to datasets encompassing multiple frames?}

\item \label{itm:stealth} \textit{Assessing stealthiness.}
It is non-trivial to assess the stealthiness of a backdoor trigger via a subjective human perspective. \emph{Can we objectively assess the stealthiness of a trigger for neuromorphic data? If yes, how?}

\end{enumerate}


Knowing the limitations in color and the flexibility in changes through time, we propose different techniques for injecting a backdoor in \acp{SNN}. With this information, we improve two existing attacks (static and moving backdoors), and we propose two novel attacks: smart and dynamic backdoors.



\subsection{Static Backdoor}


Backdoor attacks in \acp{SNN} or neuromorphic data were not explored before the work of Abad et al.~\cite{abad2022poster}. Inspired by backdoor attacks in the image domain~\cite{gu2019badnets}, the authors replicated the square trigger used in neuromorphic data. By completely discarding the time-encoded advantages neuromorphic data have, the authors included the same trigger (same position and polarity) in all the frames, thus making a static trigger. We will investigate this trigger type as a baseline for subsequent ones, thoroughly investigating the trigger position, polarity, and size in a wide range of cases.

\begin{figure}[!htb]
     \centering
     \begin{subfigure}[b]{0.22\linewidth}
         \centering
         \includegraphics[width=\textwidth]{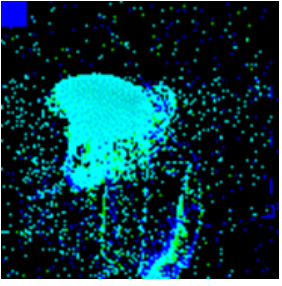}
         \caption{Top}
         \label{fig:s_trigger_1}
     \end{subfigure}
     \begin{subfigure}[b]{0.22\linewidth}
         \centering
         \includegraphics[width=\textwidth]{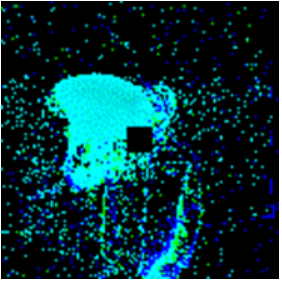}
         \caption{Middle}
         \label{fig:s_trigger_3}
     \end{subfigure}
     \begin{subfigure}[b]{0.22\linewidth}
         \centering
         \includegraphics[width=\textwidth]{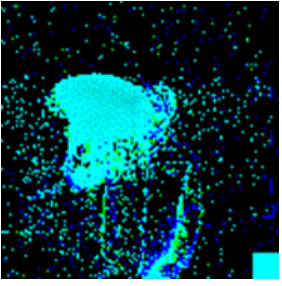}
         \caption{Bottom}
         \label{fig:s_trigger_5}
     \end{subfigure}
     \begin{subfigure}[b]{0.28\linewidth}
         \centering
         \includegraphics[width=\textwidth]{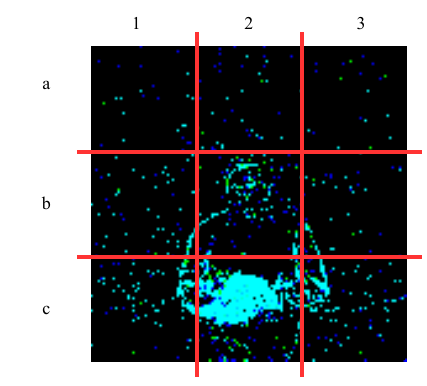}
         \caption{Smart}
         \label{fig:smart}
     \end{subfigure}
        \caption{Input samples containing a static trigger (\ref{fig:s_trigger_1}, \ref{fig:s_trigger_3}, and \ref{fig:s_trigger_5}) and a smart backdoor mask for $c=2$ (\ref{fig:smart}).}
        \label{fig:static_backdoor}
\end{figure}

We follow the same intuition of a pixel-squared trigger of a given color, which is now set by the polarity for neuromorphic data. The data samples contain two polarity values, either \textit{ON polarity} or \textit{OFF polarity} corresponding to the black and light blue. However, when pixels from different polarities are overlapped, it generates another two color polarities, i.e., dark blue and green. The polarity $p$ in neuromorphic datasets is a two-bit discrete value, creating up to four different combinations; we rename the polarities for simplicity to $p_0$, $p_1$, $p_2$, and $p_3$. Thus, the trigger gets a different color for different polarity combinations, i.e., black, dark blue, green, or light blue. Additionally, it can be placed in arbitrary locations of the input, e.g., top-right, middle, bottom-left, random, or any other desired location $l$, see~\autoref{fig:static_backdoor} as an example.
For our attacks, we also consider the trigger size, $s$, as the percentage of the input size for constructing the trigger. Still, the input samples are divided into $T$ frames, so the trigger $k$ is replicated for each frame and sample, i.e., the trigger does not change the location. Consequently, it is static.

We consider all discussed parameters in the backdoor creation function $\mathcal{A}(\mathbf{x}, p, s, l)$ for creating a set of backdoor samples $\mathcal{D}_{bk}: \hat{\mathbf{x}} \in \mathcal{D}_{bk}$ containing the trigger $k$. By controlling the $\epsilon$ value $\epsilon = \frac{m}{n}; m\ll n$ with $n$ the size of $\mathcal{D}_{clean}$ and $m$ the size of $\mathcal{D}_{bk}$, the attacker controls the amount of backdoored data during training. 

\subsection{Moving Backdoor}


As previously seen, static backdoors replicate the trigger from backdoor triggers in the image domain. However, a unique characteristic of neuromorphic data allows the attacker to develop a better version of the trigger. To this end, moving triggers inject a trigger per frame in different locations, exploiting the time-encoded nature of neuromorphic data.
The nature of neuromorphic data is driven by polarity, i.e., movement, which contradicts the static behavior of the naïve static attack. Driven by this discrepancy and the aim of creating a more stealthy attack that cannot be detected easily by human inspection (more about attack stealthiness is in~\autoref{sec:stealthiness}).

The moving backdoor primary leverages the ``motion'' nature of neuromorphic datasets to create moving triggers along the input. Precisely, for a given polarity $p$, a location $l$, and size $s$, the trigger $k$ smoothly changes from frame to frame, creating a moving effect. Formally, the backdoor creation function takes the parameters mentioned above $\mathcal{A}(\mathbf{x}, p, s, l, T)$ for creating a backdoor set of inputs $\mathcal{D}_{bk}$, where $T$ is the total number of frames that the input is divided. 
The backdoor creation function also considers the number of frames $T$, such that for each frame, $\mathcal{A}(\cdot)$ calculates a location at $t+1 \in T$ close to the previous frame $t$. This allows the trigger to simulate a smooth movement in the input space. Unlike the static trigger, the moving trigger can be placed atop the input ``activity area'' for better stealthiness. 

To create a moving trigger, we use a \emph{binary mask}\footnote{A binary mask is a data structure, an image in this case, that consists of binary values (0 or 1) per pixel. A value of one means that a pixel from the trigger will be used, and when the value is zero, the input's pixel is used. It is widely used in the literature to decide the position of the trigger~\cite{yao2019latent, dong2021black}.}. The value $m$ has the same dimensions as the input $\mathbf{x}$, and it defines the trigger location and size. For example, given that at frame $t=0$, the upper-left corner of the trigger is located in pixel $(\alpha, \beta)$, the size of the trigger is $s$, the width of the image is $w$, and we shift the trigger 2 pixels to the right, then the mask is:
\begin{align*}
    \footnotesize
        m(t)_{i,j}= 
    \begin{cases}
        1,& \text{if } i\in [\alpha \Mod{w}, (\alpha + s) \Mod w] \text{ and } \\
        & j \in [(\beta + 2\times t) \Mod{w}, (\beta + s + (2\times t) ) \Mod w]\\
        0,& \text{otherwise}.
    \end{cases}
\end{align*}

We chose a shift of two pixels as it is used in the literature~\cite{abad2022poster}, and we verified through visual inspection that the trigger moved naturally in this way. If we used one pixel, the trigger moved very slowly; sometimes, it seemed static. Each pixel of the poisoned image ($\mathbf{\hat{x}}_{i,j}$) for every frame $t$ is given by:
\begin{align*}
\footnotesize
    \mathbf{\hat{x}}_{t,i,j}= 
\begin{cases}
    \mathbf{x}_{t,i,j},& \text{if } m_{t,i,j} = 0\\
    k_{i,j},& \text{if }  m_{t,i,j} = 1.
\end{cases}
\end{align*}


To improve previous work~\cite{abad2022poster}, we conducted a complete experimental setup to find the best moving trigger. Additionally, we analytically measure and quantify the stealthiness of the triggers. Finally, we correlate the stealthiness to the triggers' ability to evade state-of-the-art defenses adapted from the image domain. Additional information about stealthiness can be found in~\autoref{sec:stealthiness}.


\subsection{Smart Backdoor}

So far, the proposed techniques, i.e., static and moving backdoors, inject the backdoor in the model correctly.
With the \textit{smart backdoor} approach, we aim to optimize the backdoor performance, simplicity, stealthiness, and understanding by removing the two hyperparameters: polarity $p$ and trigger location $l$. For a better understanding of the effects of the trigger location, we split the image by drawing $c$ vertical and horizontal lines (see~\autoref{fig:smart}). These will divide the image into $(c+1)^2$ chunks, which we call \textit{masks}; the collection of masks is denoted as $\mathcal{S}_{mask}$. Note that a larger $c$ value would create more masks, allowing the attacker more control over the ``optimal'' spot of trigger placement. The smart backdoor attacks leverage the inputs' polarity changes to find the most \textit{active mask}. We define the \textit{mask activation} as the sum of all the polarity changes happening in a mask for all the frames, excluding the polarity $p_0$, which represents no movement, i.e., the black color or the background. This way, the smart backdoor finds the most active mask in the input sample. For example, the change from $p_2$ to $p_1$ or $p_0$ to $p_1$ is counted as $1$. In contrast, the change from $p_2$ to $p_0$ does not count due to the representation of a ``movement'' state to a ``no movement'' state. Instead of calculating the activity per sample, we sum the activity over various samples. Note that per sample calculation of the activity would result in a sample-specific trigger, which is not the goal of the smart trigger. Sample-specific triggers are studied in~\autoref{sec:dynamic}. We denote each polarity switch from \(p_n\) to \(p_m\) as \(p_n \to p_m\). Formally, given a collection of masks, $\mathcal{S}_{mask}$, for each mask $v_z$ where $H$ is the height and $W$ is the width of the mask:
\begin{align*}
\footnotesize
f_{v_z} = \sum_{i=1}^{W} \sum_{j=1}^{H} \sum_{t=1}^{T} f_{i,j}
\end{align*}
\begin{align*}
\footnotesize
\text{where } f_{i,j} = 
\begin{cases} 
0 & \text{if } p_n \to p_0 \\
1 & \text{otherwise,}
\end{cases}
\end{align*}
where \(f_{v_z}\) demonstrates the calculated changes for the regions within that mask \(v_z\) during all frame steps \(t\), and \(f_{i,j}\) demonstrates the change just for pixel \( (i, j)\) in the image.

The most active mask $v'$ is found by summing the activity of all the masks in the malicious set $m$:
\begin{align*}
\footnotesize
v' = \argmax_{v_z} \sum_{j=1}^{m} f_{v_z}.
\end{align*}




For example, in~\autoref{fig:smart}, $c=2$ lines horizontally and vertically split the image into nine masks, and the smart attack will decide which mask is the most active. Let us take~\autoref{fig:smart} as an example, where the bottom middle mask, i.e., ``2,c'' is the most active one.\footnote{We also investigate the effect of trigger injection in the least active area in~\autoref{sec:evaluation}.}
Once the location is chosen, the smart attack also selects the best---performance-wise---polarity for the trigger, i.e., the least used polarity in the mask, denoted as $p'$.
Therefore, $p'$ is used for the trigger's polarity $k$ and is injected in $v'$, randomly and smoothly moving around the mask for all the frames.

Note that the trigger polarity $p'$ and the mask $v'$ are calculated for the poisoned dataset $m$, i.e., all the poisoned samples have the same trigger location and polarity. Formally, the smart backdoor creation function is defined as $\mathcal{A}(\textbf{x}, p', v', T)$, generating a set of moving triggers that are combined with the clean samples to create a set of poisoned samples $\mathcal{D}_{bk}$.

Additionally, we investigate the effect of injecting the trigger in the least active masks, and we consider the usage of the most used polarity in the mask.

\subsection{Dynamic Backdoor}
\label{sec:dynamic}

\begin{figure}[!htb]
    \centering
    \includegraphics[width=\linewidth]{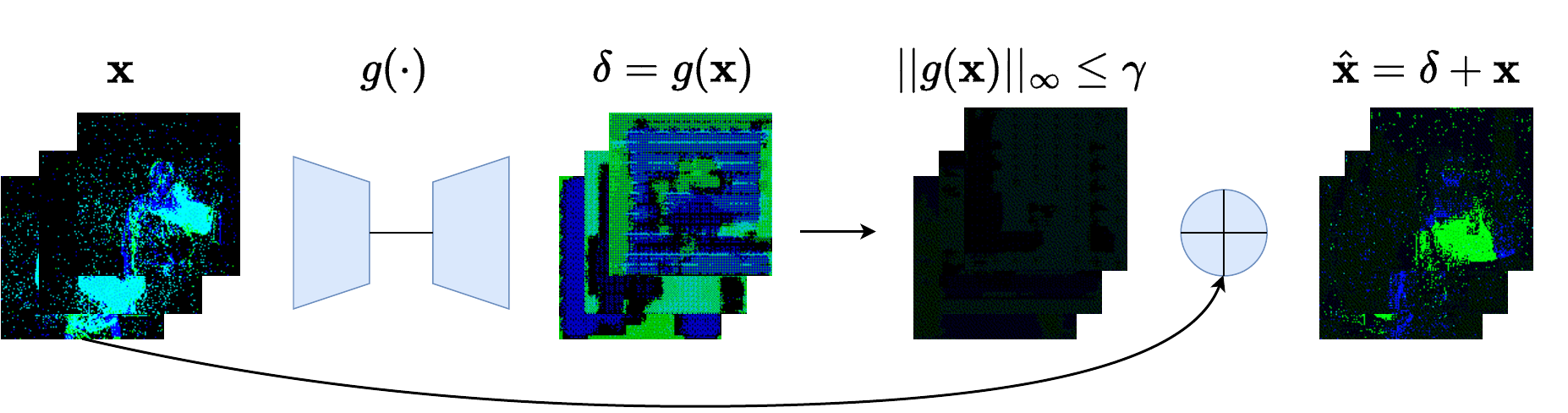}
    \caption{Overview of the dynamic moving attack.}
    \label{fig:dynamic}
\end{figure}


Having explored how to inject a backdoor in \acp{SNN} using static triggers, exploiting the time-encoded nature of neuromorphic data with moving backdoors, and optimizing the trigger polarity and location with the smart trigger, we propose a stealthy, invisible, and dynamic trigger. More precisely, motivated by dynamic backdoors in the image domain~\cite{nguyen2020input, doan2021lira}, we investigate \textit{dynamic moving backdoors} where the triggers are invisible, unique for each image and frame, i.e., the trigger changes from frame to frame. Note that generating a trigger that alternates in shape and color per sample and frame has not been previously investigated in the literature. Dynamic triggers are different from backdoors in the video domain as in~\cite{sato2021dirty}, where the trigger is a patch, is not neuromorphic, does not change its location in time, and is not pursuing stealthiness. Neuromorphic data allows us to generate a dynamic trigger specific to a sample that is also unique per frame. To achieve this, we use a spiking \ac{AE}---which shares the common structure and usage as a standard autoencoder but with spiking neurons---to generate the optimal noise, as large as the image, which maximizes the backdoor performance, maintains a clean accuracy, and is invisible. More precisely, one of the weakest points of the previous backdoor triggers is that they are detectable under human inspection subject to the trigger location and polarity, see~\autoref{sec:stealthiness}. Therefore, we aim to create an invisible trigger that is not constrained by the polarity or the location. We leverage \ac{AE}, which, for example, are used for denoising tasks, where we would usually require clean (denoised) and noisy versions of the image to train the \ac{AE}, i.e., the \ac{AE} is trained on image pairs~\cite{Goodfellow-et-al-2016}. However, we do not have the clean image and trigger pair to train the \ac{AE} for our attack. If we had the trigger, there would be no need for the \ac{AE}. Therefore, to fulfill these requirements, we must train the model and the \ac{AE} simultaneously to make the \ac{AE} generate a trigger unique for each sample and frame. Contrary to previous work~\cite{doan2021lira}, we do not need a fine-tuning phase to achieve a successful backdoor model, since \ac{ASR} and main task accuracy is already high.
During training, we maximize the main task accuracy as well as the \ac{ASR}, which is computationally more efficient.

Intuitively, the dynamic backdoor is designed as follows (see~\autoref{fig:dynamic}). At first, we generate the perturbation by passing a clean image to the spiking \ac{AE} $g(\cdot): \delta = g(\mathbf{x})$. The perturbation is then added to the clean image to construct a backdoor image $\hat{\mathbf{x}} = \mathbf{x} + \delta$. However, this naïve approach would saturate $\mathbf{x}$ with $\delta$, which makes the trigger visible. Thus, we project the perturbation to a $l_p$-ball\footnote{$l_p$-ball refers to a geometric shape that is defined as the set of all points in $n$-dimensional space that are within a certain distance of a given point, according to the $l_p$-norm.} of a given budget $\gamma: \Vert g(\mathbf{x})\Vert_\infty \leq \gamma$. Then, $g(\cdot)$ is updated aiming to maximize the backdoor accuracy of $f(\cdot)$, thus, during training, $g(\cdot)$ optimizes the parameters $\zeta$ that minimizes a loss function:
\begin{align*}
    \zeta' = \argmin_\zeta \sum_{i=0}^n \mathcal{L}(f_\theta(g_\zeta({\mathbf{x}}_i) + \mathbf{x}_i),\hat{y}_i), 
    \\ s.t. \ \Vert g_\zeta(\mathbf{x})\Vert_\infty \leq \gamma \quad \forall \; \mathbf{x},
\end{align*}
where $\hat{y}$ is the target label, $n$ is the length of the dataset, and $\mathcal{L}$ is a loss function (\ac{MSE} in our case). 

For training $f(\cdot)$, the parameters $\theta$ are updated by minimizing
\begin{align*}
    \theta' = \argmin_\theta \sum_{i=0}^n \alpha \mathcal{L}(f_\theta(\mathbf{x}_i)),y_i) +
    \\ (1 - \alpha) \mathcal{L}(f_\theta(g_\zeta({\mathbf{x}}_i)+ \mathbf{x}_i),\hat{y}_i), 
    \\ s.t. \ \Vert g_\zeta(\mathbf{x})\Vert_\infty \leq \gamma \quad \forall \; \mathbf{x},
\end{align*}
where $\alpha$ controls the trade-off between the clean and the backdoor performance; $\alpha=1$ to train $f(\cdot)$ only with clean data. $\gamma$ controls the visibility of the trigger. 
We discuss the influence of $\alpha$ and $\gamma$ in~\autoref{sec:evaluation}. 

\section{Evaluation}
\label{sec:evaluation}

\subsection{Experimental Results}

In this section, we provide the results for four different attacks, emphasizing their strong and weak points\footnote{Due to space limitations, for more detailed results on the N-Caltech101 dataset, refer to Appendix~\ref{sec:caltech}.}. For details on the datasets, models, and training settings, we refer readers to Appendix,~\ref{app:datasets},~\ref{app:network}, and~\ref{sec:train}.

We evaluate the attacks with the commonly used metrics:
\begin{compactitem}
    \item \textbf{\Ac{ASR}} measures the backdoor performance of the model based on a holdout fully backdoored dataset.
    \item Model utility or \textbf{clean accuracy} is the performance of the model test on a holdout clean dataset. 
    \item \textbf{Clean accuracy degradation} is the accuracy drop (in percentage) from the clean and backdoor models. It is calculated as $\frac{V_2 - V_1}{V1} \times 100$, where $V_1$ is the clean baseline accuracy, and $V_2$ is the clean accuracy after the attack.
\end{compactitem}

\subsubsection{Static Backdoor}

To first evaluate the viability of backdoor attacks in \acp{SNN}, we explore the basic BadNets~\cite{gu2019badnets} approach by placing a static trigger in different locations of the input space, 
using different poisoning rates, triggers sizes, and polarities.
We test the static attack with $\epsilon$ values of 0.001, 0.005, 0.01, 0.05, and 0.1. We set the trigger sizes to 1\% and 10\% of the input image size. Lastly, we experiment with three trigger locations: bottom-right, middle, and top-left, and four polarities. 

Our results show that static backdoors require a trigger size as big as 10\% of the input size to inject the backdoor behavior in complex datasets like CIFAR10-DVS or DVS128-Gesture. When the trigger is 1\% of the input size, the backdoor is only successfully injected in N-MNIST when the polarity is different from 0. However, when the trigger is in the middle, we observe that $p=0$ achieves up to 100\% \ac{ASR}. This is caused because the data is centered; thus, the black trigger is on top of the image, contrasting and allowing the model to distinguish the trigger. In subsequent sections, we further investigate the importance of injecting the trigger in the most important or least important location.
Increasing the trigger size makes the backdoor achieve an excellent \ac{ASR} (up to 100\%) when the trigger is placed in the corners. See~\autoref{fig:static_bottom_right},~\autoref{fig:static_middle}, and~\autoref{fig:static_top_left}, in Appendix~\ref{sec:static_moving_results} for the results of top-left, middle, and bottom-right placed triggers.

Since the DVS128-Gesture dataset is small, the $\epsilon$ value drastically affects \ac{ASR}. When $\epsilon = 0.01$, only a single sample will contain the trigger, which is insufficient to inject the backdoor when the trigger size is small. We further experiment with a larger trigger size, i.e., 0.3, achieving 99\% \acp{ASR} with $\epsilon = 0.01$, in the top-left corner and using $p = 0$.
CIFAR10-DVS achieves 100\% \ac{ASR} in all the settings when the trigger size and the poisoning rate are 0.1. CIFAR10-DVS is the only dataset that achieves 100\% \ac{ASR} in the bottom-right, with the polarity 0. This is caused by the dataset itself, which is noisy; thus, the black trigger can contrast with the background.



Regarding the clean accuracy degradation, we notice a slight degradation in most cases concerning the clean accuracy baseline. See~\autoref{fig:static_bottom_right_degradation},~\autoref{fig:static_middle_degradation}, and~\autoref{fig:static_top_left_degradation}, for the results of the clean accuracy degradation of top-left, middle, and bottom-right placed triggers.
N-MNIST does not show any degradation, i.e., 0\% and in some cases even improving the accuracy by 0.1\%, while DVS128-Gesture and CIFAR10-DVS are more prone to degrade the main task up to 5\%.
Overall, our static backdoors implementation in \acp{SNN} shows excellent performance, improving previous work~\cite{abad2022poster} due to more precise tuning.
Specifically, placing a static trigger in a moving input is unnatural, and it could be detected by checking if a part of the image is not moving or by inspecting changes in polarity between pixels. We address this limitation in the following sections.

\subsubsection{Moving Backdoor}

We investigate the effect of moving triggers to overcome the stationary behavior of static backdoor attacks. The moving backdoor changes the trigger position per frame horizontally, moving in a constant loop. We experiment with the same setting as static backdoors. However, the trigger location varies in time by horizontally moving using top-left, middle, and bottom-right as initial locations. The trigger changes location in two pixels every frame. Thus, the triggers change according to T---16 times in our experiments.

We observe that moving the backdoor overcomes the limitation of static triggers when placed on top of the image action. Since the trigger is moving, it is not always on top of the active area, thus allowing the model to capture both clean and backdoor features. 
Interestingly, as seen in~\autoref{fig:moving_bottom_right} and contrary to the static backdoor (see~\autoref{fig:static_bottom_right}), triggers in the bottom-right corner with background polarity do not work with complex datasets because they merge better with the image, making it impossible for the model to recognize them. However, in N-MNIST, we can achieve 100\% \acp{ASR} with $p\neq 0$ and large $\epsilon$. Moreover, triggers with background polarity in the bottom-right position do not inject the backdoor successfully for the DVS128-Gesture dataset, contrary to static backdoors. That effect is intuitively explained as moving backdoors are more difficult to inject than static ones. The model has to find a more complex relation between the trigger, samples, and label, thus requiring large datasets.

We observe the opposite behavior with triggers in the top-left corner (see~\autoref{fig:moving_top_left}). We investigate the data samples and conclude that images are usually centered; thus, the main action of the image is also in the middle of the image. For DVS128-Gesture or CIFAR10-DVS, the action is also contained in the corners of the image. From here, we can intuitively explain that injecting the trigger in an active or inactive area of the image could enable or disable the backdoor effect. We investigate the backdoor effect when placing the trigger in the most and least active areas in~\autoref{sec:smart}.
Lastly, by placing the triggers in the middle (see~\autoref{fig:moving_middle}), we observe that the DVS128-Gesture dataset achieves 100\% \acp{ASR} when polarity is 1 or 2. These results also suggest that the trigger's polarity strongly affects the backdoor's performance. Furthermore, depending on where the trigger is placed, a given polarity could have a different effect, as observed with polarity 2 in~\autoref{fig:moving_middle} and~\autoref{fig:moving_bottom_right} for the DVS128-Gesture. \autoref{sec:smart} investigates this effect in more detail.




Regarding the clean accuracy degradation, we notice a slight degradation in most cases concerning the clean accuracy baseline and even improving the accuracy of the main task in some settings. 
See~\autoref{fig:moving_bottom_right_degradation},~\autoref{fig:moving_middle_degradation}, and~\autoref{fig:moving_top_left_degradation} for the results of the clean accuracy degradation of top-left, middle, and bottom-right placed triggers.
N-MNIST does not show any degradation, while DVS128-Gesture and CIFAR10-DVS are more prone to degrade the main task up to 7\%, contrary to static triggers, which may degrade the clean accuracy up to 10\%. 
We believe this happens as those datasets are more complex, and adding backdoors makes the main task more challenging.

\subsubsection{Smart Backdoor}
\label{sec:smart}

To explore the effects of the trigger location and the trigger polarity, we designed a novel attack that chooses the best combination of both. The smart attack removes the trigger location and the polarity selection by choosing either the most active or least active area of the image. Then, it chooses the least or most common polarity in that mask, where the least common polarity would contrast while the most common one would be more stealthy---enabling a more optimized attack.
We experiment with different settings, such as poisoning rates, trigger sizes, and most and least active masks. We also investigate the trigger polarity's effect using the most/least active polarity in the selected mask. We split the image using $c=2$, see~\autoref{fig:smart}. 

\begin{figure}
    \centering
    \begin{subfigure}[b]{0.23\textwidth}
        \includegraphics[width=\textwidth]{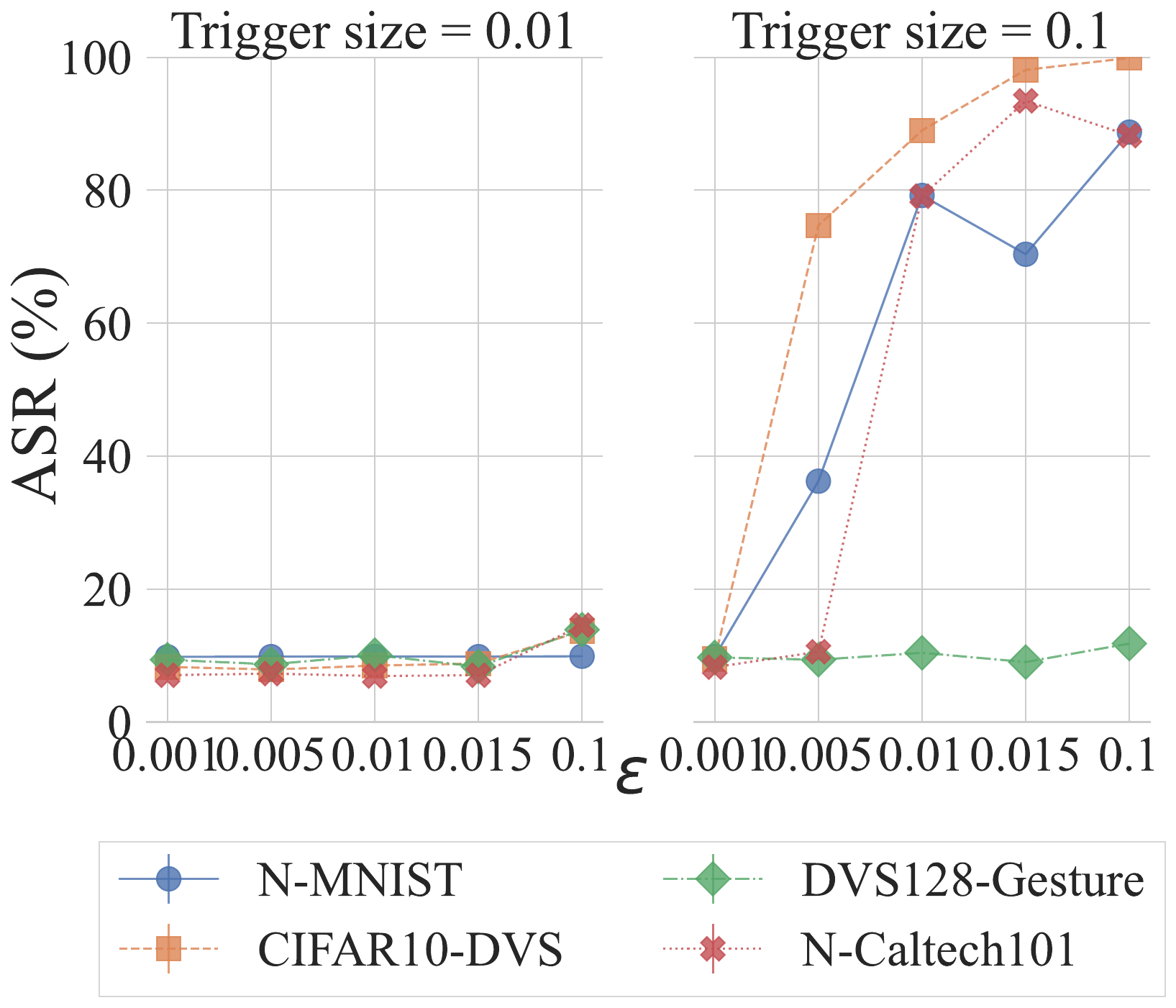}
        \caption{}
        \label{fig:asr_smart_false_false}
    \end{subfigure}
    \begin{subfigure}[b]{0.23\textwidth}
        \includegraphics[width=\textwidth]{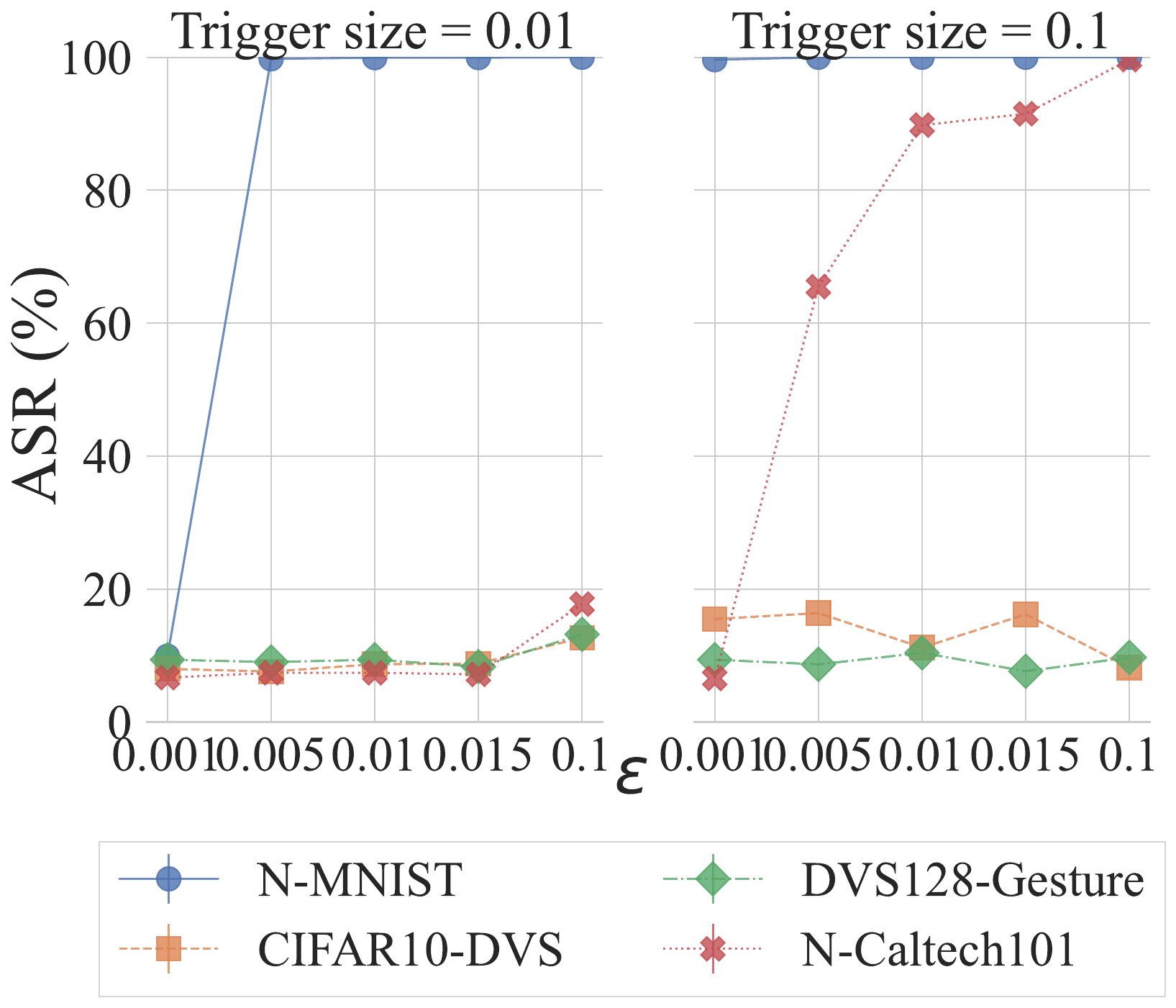}
        \caption{}
        \label{fig:asr_smart_true_false}
    \end{subfigure}
    \vfill
    \begin{subfigure}[b]{0.23\textwidth}
        \includegraphics[width=\textwidth]{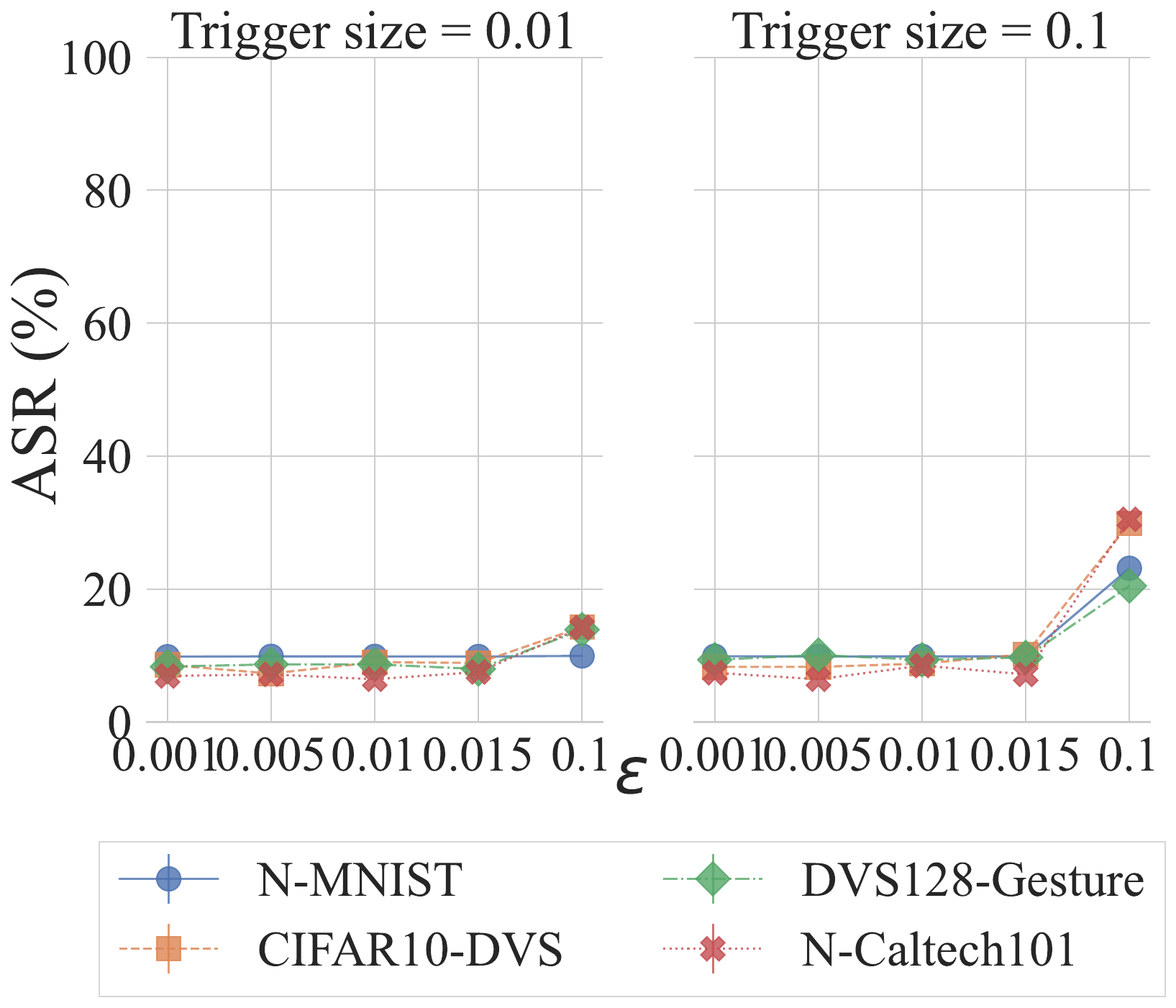}
        \caption{}
        \label{fig:asr_smart_false_true}
    \end{subfigure}
    \begin{subfigure}[b]{0.23\textwidth}
        \includegraphics[width=\textwidth]{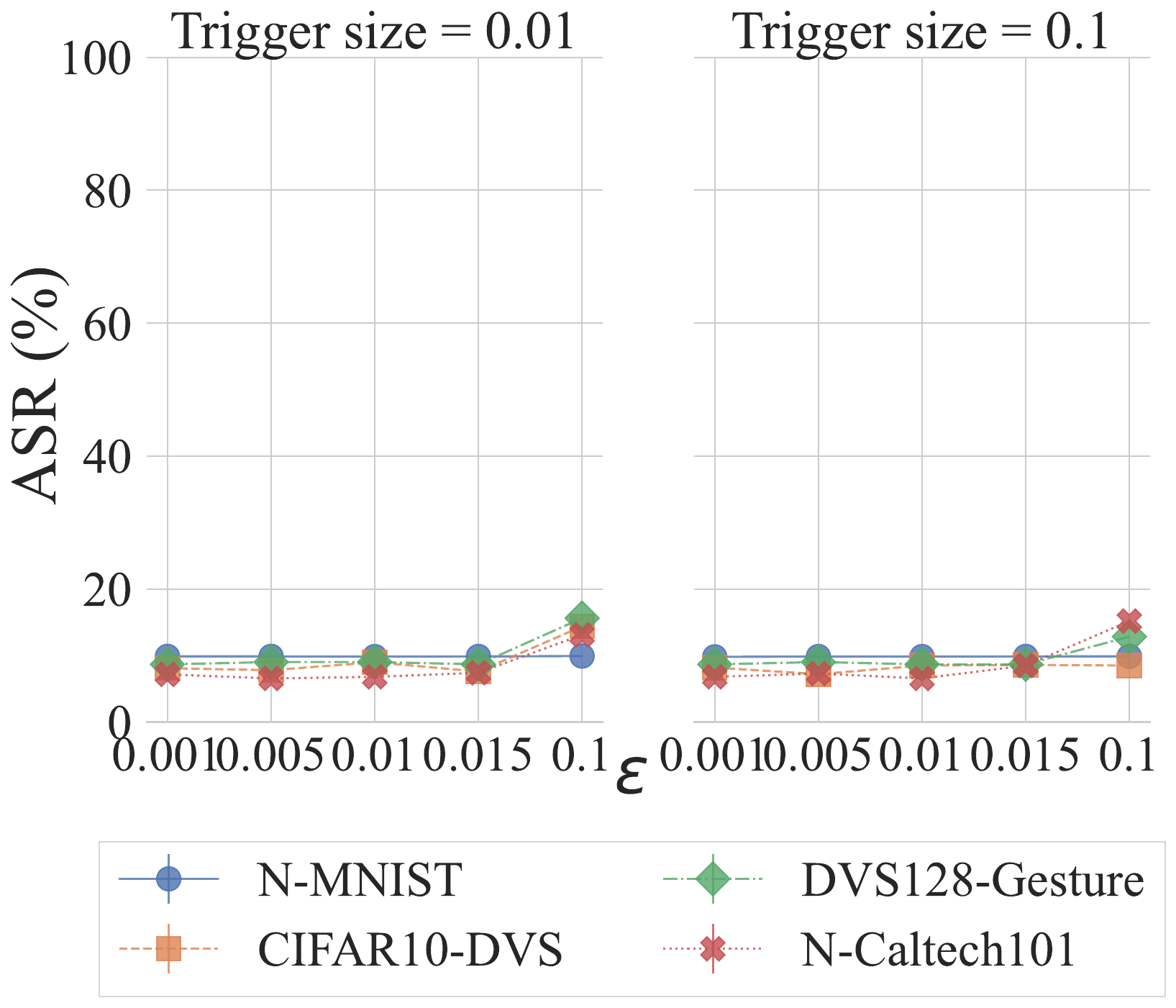}
        \caption{}
        \label{fig:asr_smart_true_true}
    \end{subfigure}    
    \caption{\ref{fig:asr_smart_false_false} and~\ref{fig:asr_smart_true_false} show the smart triggers in the most active area, using the least and most common polarity. In~\ref{fig:asr_smart_false_true} and~\ref{fig:asr_smart_true_true}, we show the smart triggers in the least active area, using the least and most common polarity.}
    \label{fig:smart_triggers}
\end{figure}

We observe that the backdoor is not successful with the DVS128-Gesture dataset. The few samples in the dataset make the choice of the most/least active mask of the image not precise. Note that the activity sum is done over all the images in the training set. The more samples, the more precise the selected mask is.
Experimentation using the most active area shows excellent performance when the least common trigger polarity is used; see~\autoref{fig:asr_smart_false_false}. Intuitively, the least common trigger polarity is preferred to increase \ac{ASR} because of the trigger contrast compared to the background image. 

Using 1\% of the image size for the trigger only shows promising results with N-MNIST with $\epsilon = 0.1$. A larger trigger size improves the backdoor success using the least poisoned samples. Interestingly, injecting the trigger in the least active area with the least active trigger polarity, see~\autoref{fig:asr_smart_true_false}, shows excellent backdoor performance even with a small trigger size.
Finally, experimenting with the most active trigger polarities shows that the trigger merges with the actual image, not allowing the model to capture both the clean image and the trigger, see~\autoref{fig:asr_smart_false_true} and~\autoref{fig:asr_smart_true_true}.

\begin{finding}
    The triggers are best injected in the most active area with the least common polarity.
\end{finding}

We also experimented with the clean accuracy degradation with the smart trigger. As shown in~\autoref{fig:smart_False_False_degradation},~\autoref{fig:smart_True_False_degradation},~\autoref{fig:smart_False_True_degradation}, and~\autoref{fig:smart_True_True_degradation}, we observe a maximum of 4\% degradation when the trigger size is 0.01 and the poisoning rate is 0.1 for the most active area. Results also show that when the trigger size is larger, the clean accuracy drop is negligible in all the cases, even improving it slightly. Injecting the trigger in the least active area shows similar performance; however, the accuracy degradation on the main task is smaller. This could be caused by the trigger not overlapping the active area, allowing the model to capture both tasks. This addresses Challenge~\ref{itm:trigger}.

\subsubsection{Dynamic Backdoor}

In this study, we perform experiments using the dynamic backdoor technique to investigate the impact of different values of $\alpha$ and $\gamma$ on the clean/backdoor trade-off and trigger intensity, respectively. We vary $\alpha$ in the range of 0.5 to 0.9 and $\gamma$ in the range of 0.01 to 0.1, excluding $\alpha < 0.5$ due to its impracticality in real-world scenarios where assigning more weight to the backdoor than the clean task is unrealistic. We aim to maximize clean accuracy and achieve a high \ac{ASR} by selecting epochs with higher values. It should be noted that the measurement accuracy across different epochs may introduce a larger standard error in the results.

\begin{figure}
    \centering
    \includegraphics[width=0.8\linewidth]{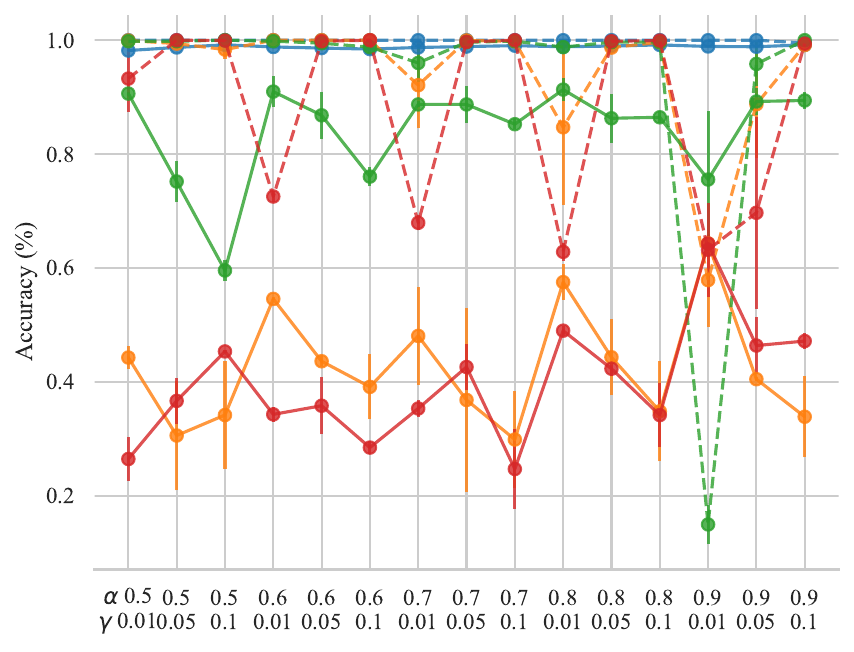}
    \caption{ASR and clean accuracy degradation of dynamic triggers. Dashed lines represent the ASR, and solid lines represent the clean accuracy. Blue corresponds to N-MNIST, orange to CIFAR10-DVS, green to  DVS128-Gesture, and red to N-Caltech101.}
    \label{fig:dynamic_combined} 
\end{figure}

The experimental results presented in~\autoref{fig:dynamic_combined} demonstrate the effectiveness of our dynamic attack strategy. We achieve 100\% \ac{ASR} for all tested settings for the N-MNIST dataset while maintaining a clean, high accuracy. Particularly, when $\gamma$ is small, no degradation in clean accuracy is observed. However, as $\gamma$ increases, there is a noticeable decline in clean accuracy, as shown in~\autoref{fig:dynamic_combined} and other datasets.

For the more complex DVS128-Gesture dataset, depicted in~\autoref{fig:dynamic_combined}, we achieve a minimum of 95\% \ac{ASR} (excluding $\alpha = 0.9$). However, we observe a slight degradation in clean accuracy when $\alpha$ is set to 0.5 or when $\gamma$ is large. The reduction in ASR is most significant with $\alpha = 0.9$ and $\gamma = 0.01$ since it prioritizes the main task and triggers stealthiness. Moreover, as $\alpha$ increases, the drop in accuracy becomes less pronounced, reducing the impact of $\gamma$.

We observe lower performance in the case of CIFAR10-DVS (\autoref{fig:dynamic_combined}). Here, $\gamma$ plays a critical role in \ac{ASR} and clean accuracy. The trigger became more visible with increasing $\gamma$, resulting in higher \ac{ASR} and reduced clean accuracy. Conversely, we observe the opposite behavior when the trigger is nearly invisible ($\gamma = 0.01$). Similar to the DVS128-Gesture dataset, setting $\alpha = 0.9$ and $\gamma = 0.01$ reduces \ac{ASR}. A clear trend is evident in this plot, where increasing $\alpha$ leads to a reduction in \ac{ASR} while achieving higher clean accuracy. It is important to note that the baseline accuracy for CIFAR10-DVS is not as high as in the other datasets, which leads to more misclassification. This also incurs a more severe degradation of the clean accuracy after the attack.

Our findings indicate that \ac{ASR} remains consistently close to 100\% in most settings. However, the attacker must carefully select appropriate values for $\alpha$ and $\gamma$ to balance clean accuracy degradation and trigger invisibility. Notably, $\gamma$ controls the visibility of the backdoor image, which becomes indistinguishable from the clean image when $\gamma = 0.01$. Our experiments suggest that using a small $\gamma$ and a large $\alpha$ yield optimal results regarding clean accuracy and \ac{ASR} while ensuring that the trigger remains unnoticed, thereby addressing Challenge~\ref{itm:invisible}. Further details on the stealthiness aspect of our approach are discussed in Section~\ref{sec:stealthiness}.

\subsection{Evaluating State-of-the-Art Defenses}

In this section, and due to the lack of specially crafted defenses for \acp{SNN}, we discuss state-of-the-art backdoor defenses in \acp{DNN} and how they could be adapted to \acp{SNN} and neuromorphic datasets. As discussed in the following sections, defenses for \acp{DNN} have core problems since they are based on \ac{DL} assumptions or consider regular static data. We select four representative defenses based on model inspection: \ac{ABS}~\cite{liu2019abs}, STRIP~\cite{gao2019strip}, spectral signatures~\cite{tran2018spectral}, and fine-pruning~\cite{liu2018fine}.

\subsubsection{ABS}

The \ac{ABS} method, as introduced in Liu et al.~\cite{liu2019abs}, is a method for identifying backdoors in neural networks using a model-based approach. It works by stimulating neurons in a specific layer and examining the resulting outputs for deviations from expected behavior. \ac{ABS} is based on the idea that a class can be represented as a subspace within a feature space and that a backdoored class will create a distinct subspace throughout the feature space. Therefore, \ac{ABS} hypothesizes that a poisoned neuron activated with a target input will tend to produce a larger output than a non-poisoned neuron.

We adapt \ac{ABS} to handle neuromorphic data and \acp{SNN}. Specifically, we modify the code to process all frames of an image together rather than treating each frame individually since neuromorphic data contains time-encoded information. However, \ac{ABS} also does not support dynamic, moving, or smart triggers, which are types of backdoors that can change position or be unique to each image. 
Since the trigger position changes could be interpreted as multi-trigger backdoors---attacks that contain more than one trigger within a single input---\ac{ABS} cannot handle them. 
Additionally, dynamic backdoors present a twofold problem for \ac{ABS}. First, dynamic triggers can also be interpreted as multi-trigger. Second, \ac{ABS} requires the trigger to be the same every time, i.e., the trigger is not unique per sample. However, the dynamic attack creates input-specific triggers, which can surpass \ac{ABS} from its original design.

We observe several false positives when testing \ac{ABS} against static backdoors. When applied to a clean model, \ac{ABS} marked it as compromised, and when applied to a poisoned model, \ac{ABS} identified it as compromised but with the wrong target class. This behavior was consistent across all datasets. One possible explanation for this issue is the core assumption of \ac{ABS}. \ac{ABS} relies on ``turn-points'' created by the activation functions in the model, such as \ac{ReLU}. However, the lack of \ac{ReLU} activation functions in \acp{SNN} makes \ac{ABS} malfunction, providing inaccurate results.

\subsubsection{STRIP}
\label{sec:strip}

Unlike repairing or flagging a model as compromised, Gao et al.~\cite{gao2019strip} investigated the detection of backdoor inputs during test time. The authors proposed a method called \ac{STRIP} to identify backdoor inputs at runtime by intentionally perturbing incoming inputs and observing the randomness of predicted classes. Low entropy in the predicted classes indicates the presence of malicious input. The experiments conducted by the authors demonstrated that a decision boundary could effectively separate benign samples. The authors assumed access to a set of clean data, typically the test set, and created a set of backdoor data by interpolating different samples from the dataset. The entropy of clean and backdoor data was calculated, revealing their distinct separability. We adapted this mechanism to neuromorphic data. For constructing the poisoned test set, we performed frame-by-frame interpolation between samples, i.e., pairing the first frame of one sample with the first frame of another.

\begin{figure}
    \centering
    \begin{subfigure}[b]{0.41\linewidth}
        \includegraphics[width=\linewidth]{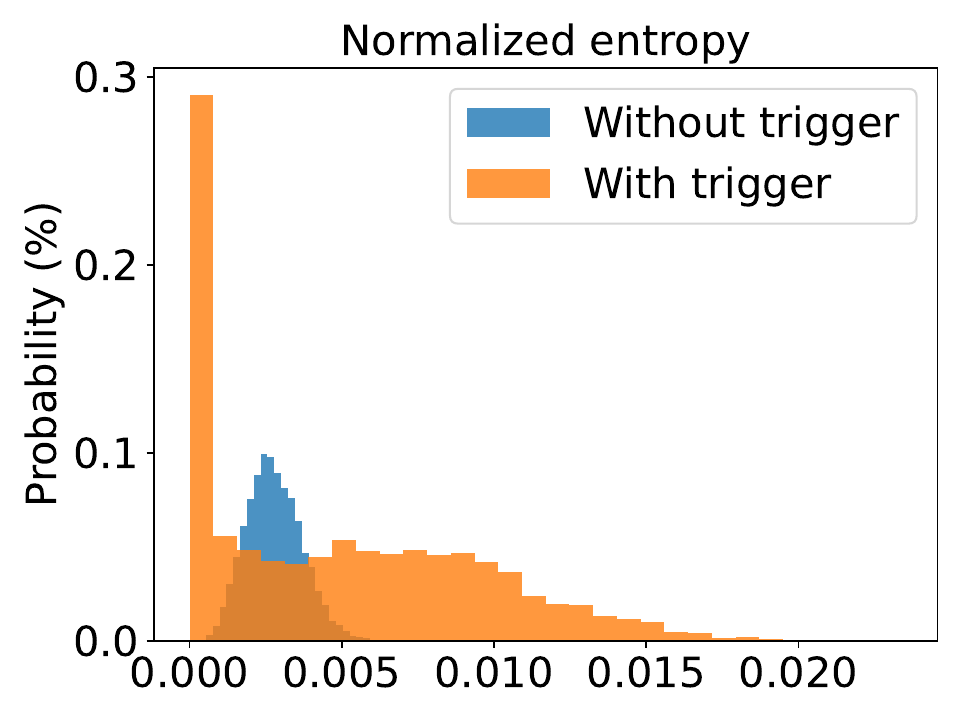}
        \caption{Static N-MNIST}
    \end{subfigure}
    \begin{subfigure}[b]{0.41\linewidth}
        \includegraphics[width=\linewidth]{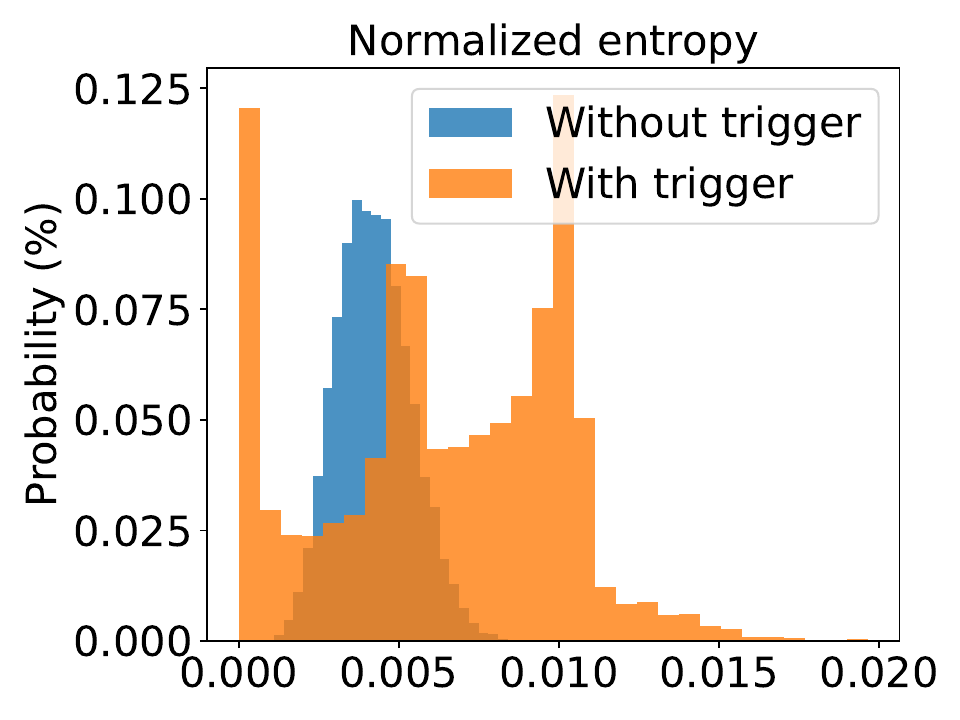}
        \caption{Dynamic N-MNIST}
    \end{subfigure}

    \vfill

    \begin{subfigure}[b]{0.41\linewidth}
        \includegraphics[width=\linewidth]{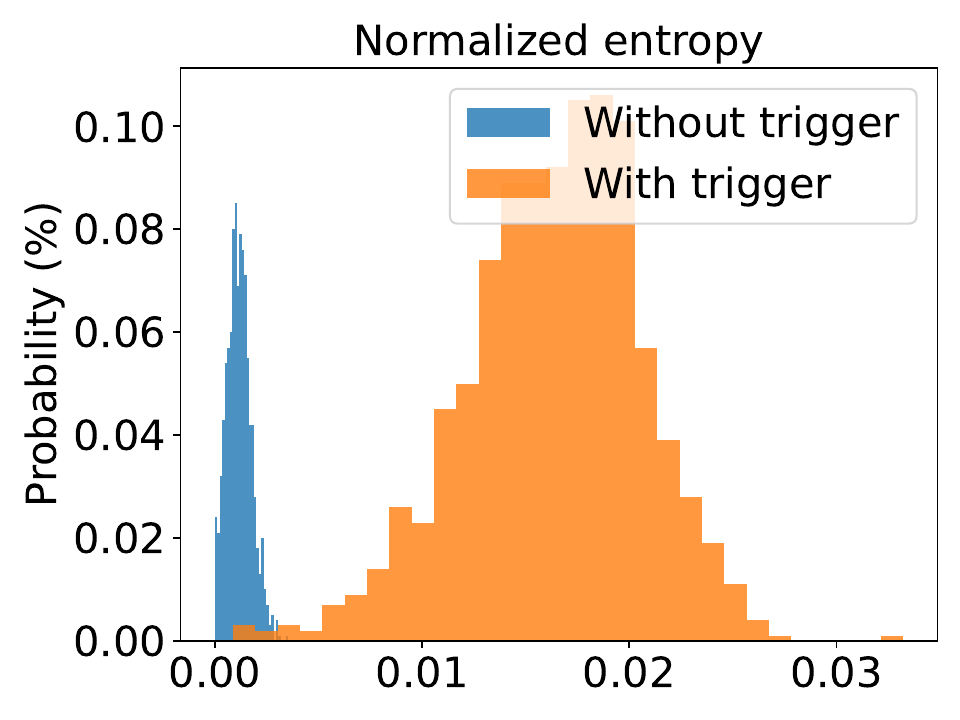}
        \caption{Static CIFAR10-DVS}
    \end{subfigure}
    \begin{subfigure}[b]{0.41\linewidth}
        \includegraphics[width=\linewidth]{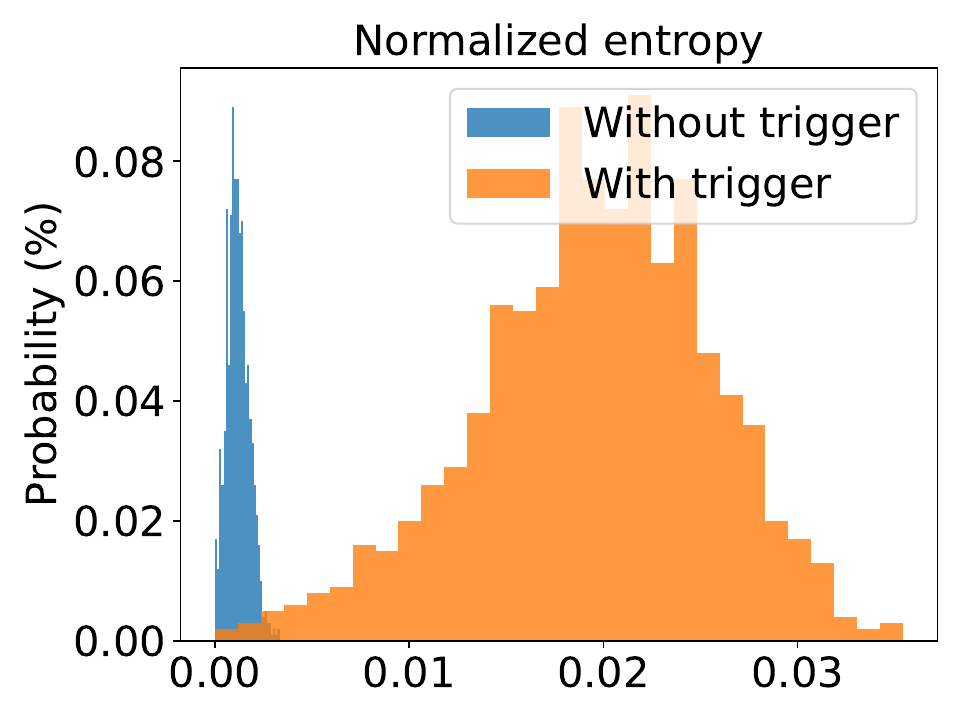}
        \caption{Dynamic CIFAR10-DVS}
    \end{subfigure}

    \vfill
    
    \begin{subfigure}[b]{0.41\linewidth}
        \includegraphics[width=\linewidth]{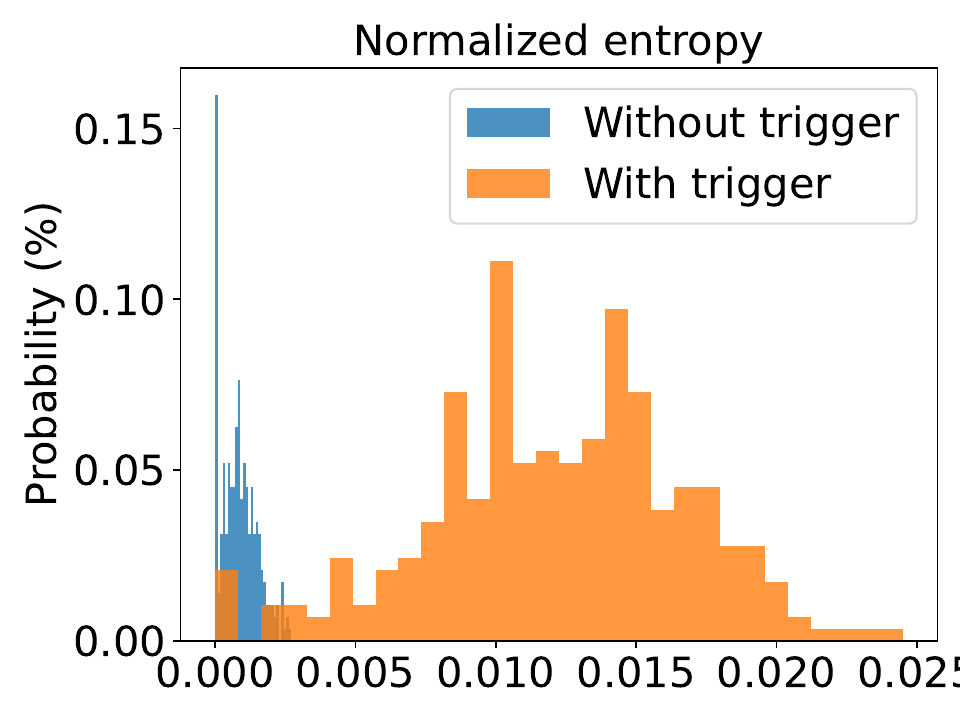}
        \caption{Static Gesture}
    \end{subfigure}
    \begin{subfigure}[b]{0.41\linewidth}
        \includegraphics[width=\linewidth]{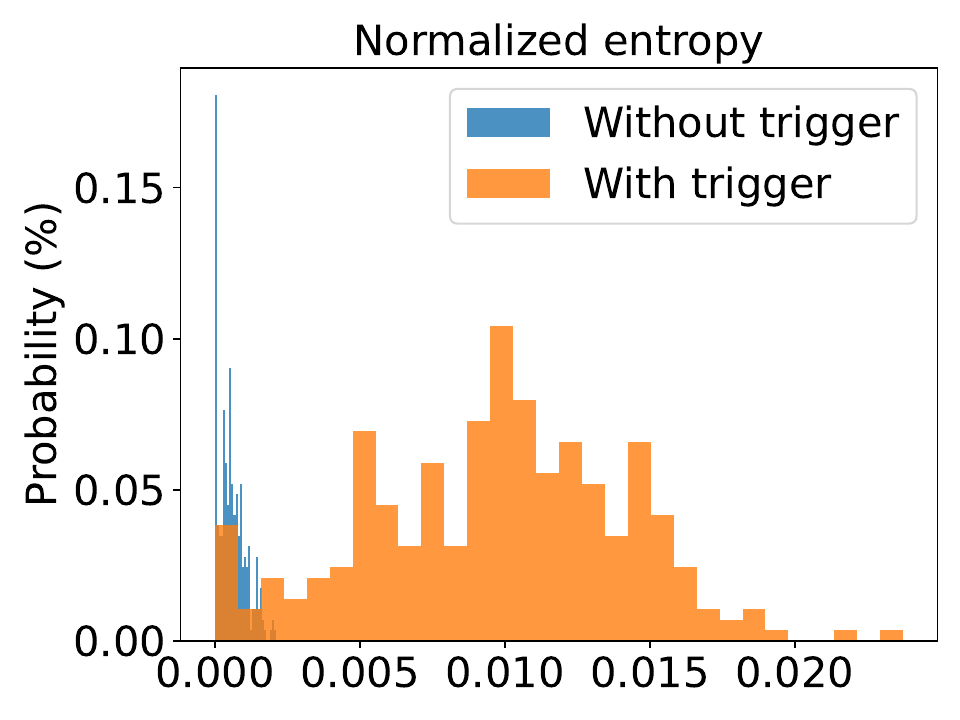}
        \caption{Dynamic Gesture}
    \end{subfigure}

    \vfill
    
    \begin{subfigure}[b]{0.41\linewidth}
        \includegraphics[width=\linewidth]{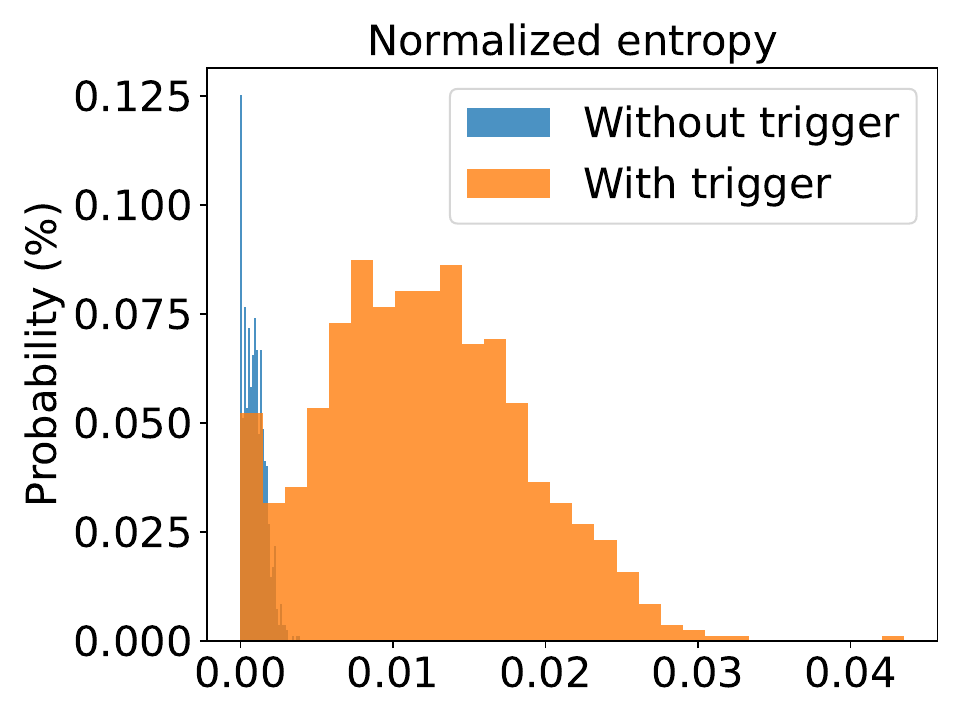}
        \caption{Static Caltech}
    \end{subfigure}
    \begin{subfigure}[b]{0.41\linewidth}
        \includegraphics[width=\linewidth]{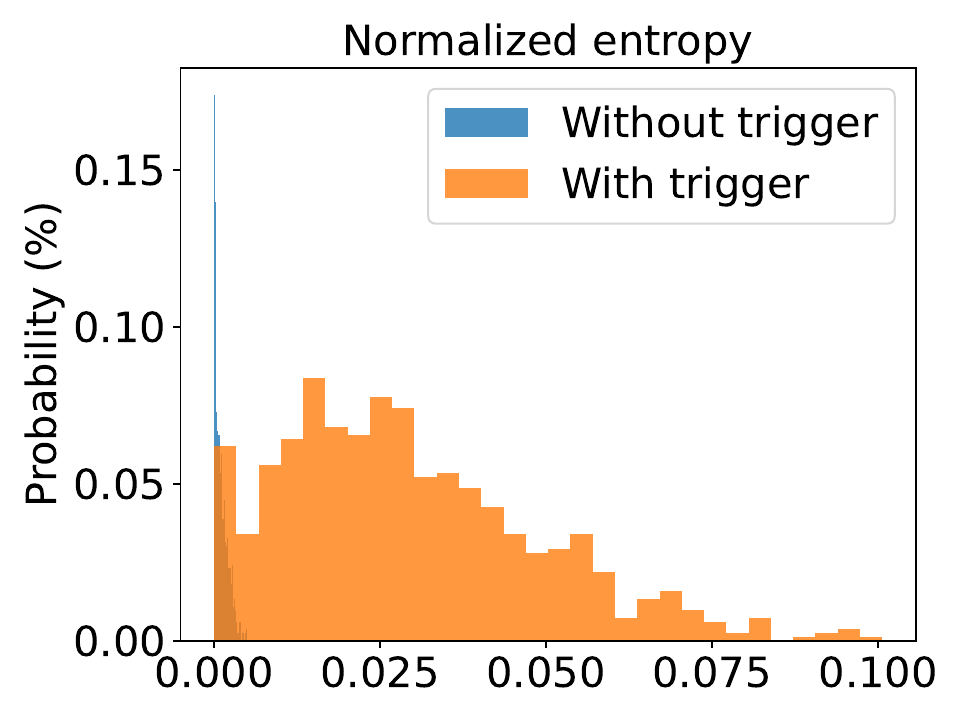}
        \caption{Dynamic Caltech}
    \end{subfigure}

    \caption{Normalized entropy of different triggers and datasets.}
    \label{fig:strip}
\end{figure}

Our experiments demonstrated that the entropy levels of neuromorphic data are significantly lower than those of \emph{regular} data, enabling a reasonably confident differentiation between clean and malicious samples. \autoref{fig:strip} illustrates the entropy levels for various attacks and datasets, with entropy measured on each test set sample. For results on smart and moving triggers, refer to~\autoref{fig:strip_appendix} in the Appendix~\ref{sec:appdx_strip}. 
We derived three main observations from our experiments. First, the claim made by Gao et al. that poisoned data exhibits lower entropy than clean data does not always hold. In certain cases, clean data demonstrates lower entropy than poisoned data. Second, in the remaining cases, the entropy of clean and backdoor data overlaps, rendering them indistinguishable and inseparable. Third, overall entropy levels are much lower in neuromorphic data than in the regular data tested in~\cite{gao2019strip}. For example, in the CIFAR-10 dataset, the mean entropy is approximately one, while in neuromorphic data, it is around 0.01.

\subsubsection{Spectral Signatures}

Recent research conducted by Tran et al.~\cite{tran2018spectral} has focused on mitigating dataset poisoning by identifying and eliminating compromised sub-populations within the dataset. The authors utilized statistical techniques such as \ac{SVD} to identify crucial input features and magnify the distribution difference in the latent space of the last convolutional layer. This approach facilitated the removal of the backdoor effect by retraining the network using clean data. It is important to note that this defense mechanism relies on having access to the compromised dataset, which may not be feasible in many scenarios.

We assessed the effectiveness of this mechanism against static and moving attacks.\footnote{Spectral signatures do not apply to dynamic backdoor attacks because the trigger is generated on the fly, and thus, the poisoned samples are not available for inspection.} 
We selected attack parameters that achieved high accuracy in clean and backdoor tasks. Following the authors' suggestion, we set the target label to 0 and the percentile to 85\%. As shown in~\autoref{tab:spectral}, we observed no significant degradation or improvement in clean accuracy or \ac{ASR}. However, we discovered that spectral signatures incorrectly flagged some legitimate samples as backdoors, resulting in compromised data remaining in the dataset used for model retraining. This aligns with our previous experience with \ac{STRIP}, see~\autoref{sec:strip}, where the entropy of clean and backdoor samples exhibited similarities. Further investigation is necessary to establish reliable mechanisms to effectively remove backdoor samples from compromised datasets.

\begin{table}[htb]
\centering
\caption{Comparison of the clean accuracy and the ASR between the baseline attack and after applying spectral signatures.}
\label{tab:spectral}
\resizebox{\columnwidth}{!}{%
\begin{tabular}{ccccccccc}
\hline
               & \multicolumn{4}{c}{Static}                                   & \multicolumn{4}{c}{Moving}                                   \\ \cline{2-9} 
               & \multicolumn{2}{c}{Baseline} & \multicolumn{2}{c}{Spectral}  & \multicolumn{2}{c}{Baseline} & \multicolumn{2}{c}{Spectral}  \\ \cline{2-9} 
               & Clean acc.        & ASR      & Clean acc.    & ASR           & Clean acc.        & ASR      & Clean acc.    & ASR           \\ \hline
               
N-MNIST        & 99.4              & 100      & \textbf{99.4} & \textbf{100}  & 99.3              & 100      & \textbf{99.3} & \textbf{100}  \\

CIFAR10-DVS    & 67.7              & 100      & \textbf{68.1} & \textbf{100}  & \textbf{68.2}     & 100      & 68.1          & \textbf{100}  \\

DVS128-Gesture & \textbf{92.0}     & 99.3     & 91.6          & \textbf{99.3} & \textbf{92.0}     & 95.8     & 91.6          & \textbf{96.2} \\ 

N-Caltech101 &  76.5    &   99.4   &    \textbf{76.9 }      &  \textbf{99.4} &  76.2   &  \textbf{98.5}   &   \textbf{76.4 }  &   98.3 \\ 

\hline

\end{tabular}%
}
\end{table}

\subsubsection{Fine-pruning}

Fine-pruning~\cite{liu2018fine} is a defense mechanism against backdoor attacks composed of two parts: pruning and fine-tuning. Existing works show that removing (pruning) some neurons of a \ac{DNN} 
makes \acp{DNN} simpler, thus, easing the training while the prediction capacity remains equal~\cite{han2016deep, yu2017scalpel}. The authors suggested that some neurons may contain the primary task information, others the backdoor behavior, and the rest a combination of main and backdoor behavior. Thus, the backdoor could be completely removed by removing the neurons containing the malicious information. The authors proposed ranking the neurons in the last convolutional layer based on their activation values by querying some data. A pruning rate $\tau$ controls the number of neurons to prune.
The second part of the defense is fine-tuning. Fine-tuning consists of retraining the (pruned) model for some (small) number of epochs on clean data. By doing this, the model could (i) recover its dropped accuracy during pruning and (ii) altogether remove the backdoor effect. The authors showed that by combining these two, \ac{ASR} of a poisoned model could drop from 99\% to 0\%.

\begin{figure}
    \centering

    \begin{subfigure}[b]{0.26\linewidth}
        \includegraphics[width=\linewidth]{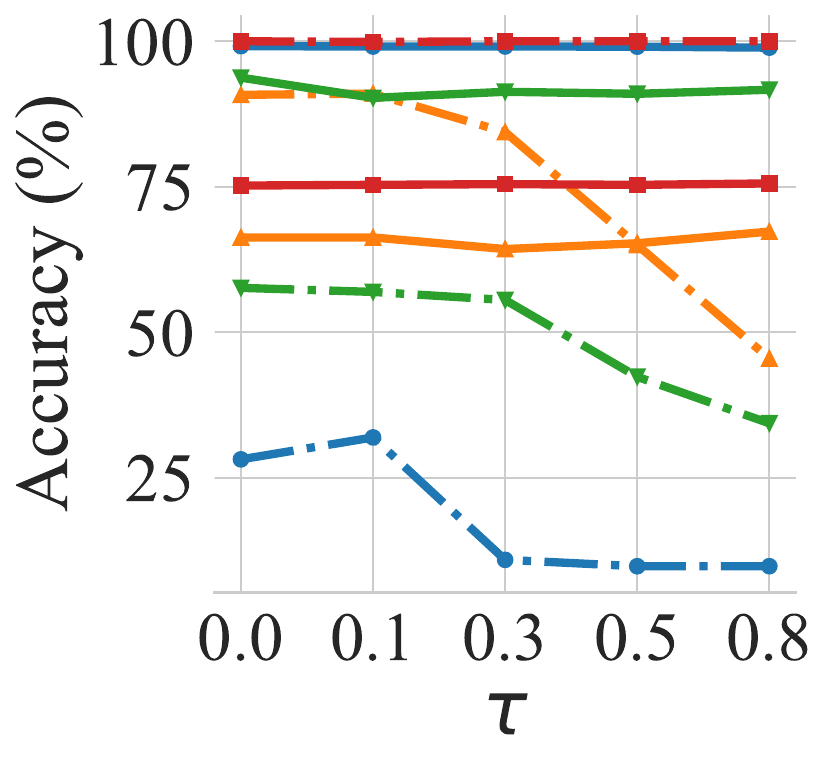}
        \caption{Static}
        \label{fig:fine_pruning_static}
    \end{subfigure}
    \begin{subfigure}[b]{0.23\linewidth}
        \includegraphics[width=\linewidth]{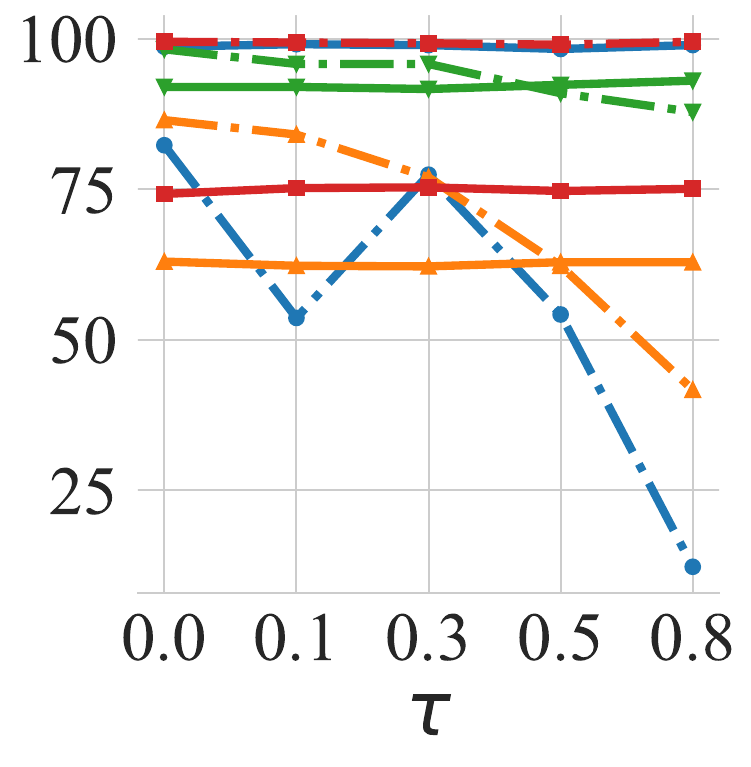}
        \caption{Moving}
        \label{fig:fine_pruning_moving}
    \end{subfigure}
    \begin{subfigure}[b]{0.23\linewidth}
        \includegraphics[width=\linewidth]{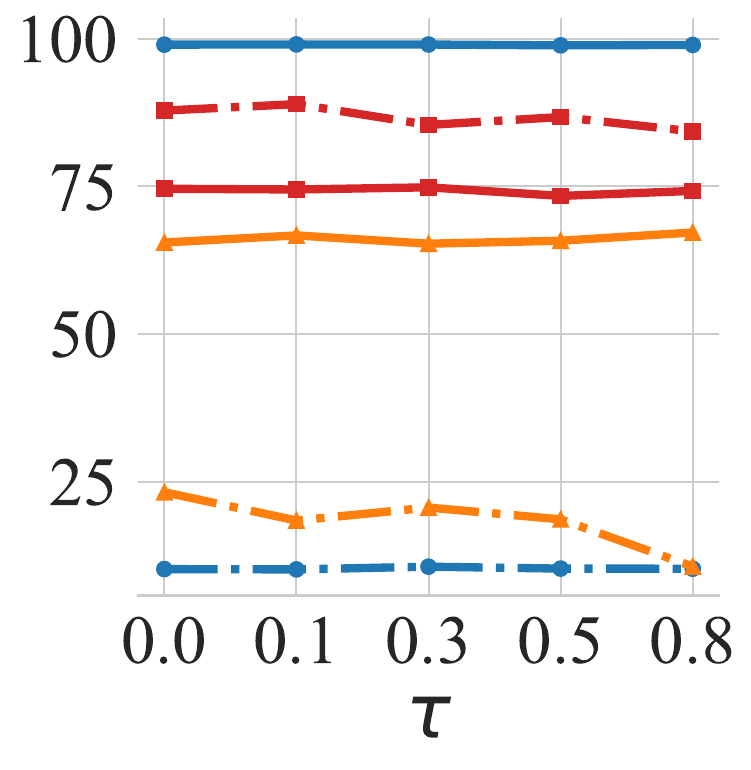}
        \caption{Smart}
        \label{fig:fine_pruning_smart}
    \end{subfigure}
    \begin{subfigure}[b]{0.23\linewidth}
        \includegraphics[width=\linewidth]{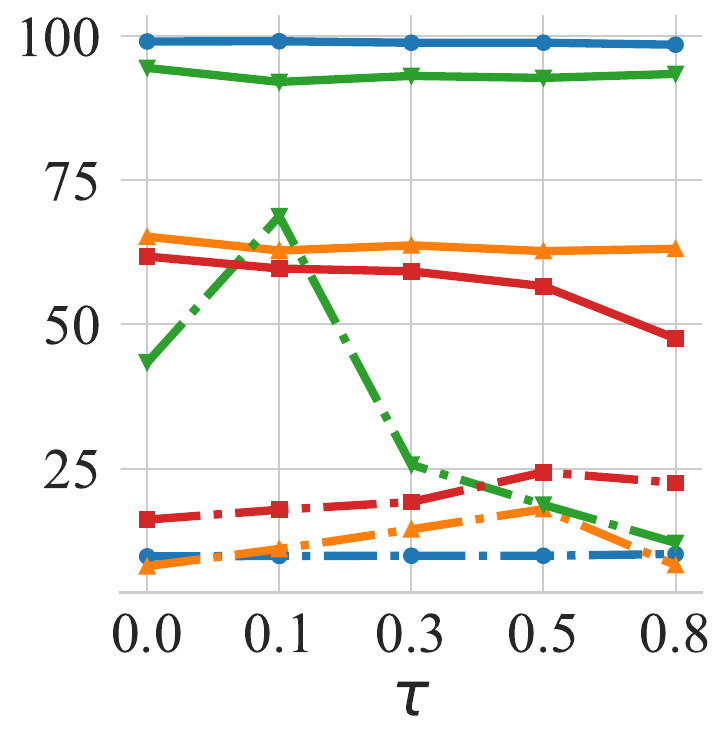}
        \caption{Dynamic}
        \label{fig:fine_pruning_dynamic}
    \end{subfigure}
    \caption{Effect of fine-pruning on the \ac{ASR} (dashed lines) and clean accuracy (full line) for different types of attacks, i.e., static, moving, smart, and dynamic. Blue corresponds to N-MNIST, orange to CIFAR10-DVS, green to DVS128-Gesture, and red to N-Caltech101.}
    \label{fig:fine-pruning}
\end{figure}

We implemented this defense for \acp{SNN} and adapted it to work with neuromorphic data. We investigate the effect of pruning\footnote{For results on pruning, refer to~\autoref{fig:pruning_appendix} in the Appendix.}, fine-pruning (pruning + fine-tuning), and fine-tuning only (when the pruning rate is 0). We also investigate various pruning rates, i.e., $\tau = \{0.0, 0.1, 0.3, 0.5, 0.8\}$ and analyze their impact, see~\autoref{fig:fine-pruning}.
Analyzing the results, we observe that pruning alone does not work. We notice that the clean accuracy drops drastically while \ac{ASR} remains high, for example, as seen in~\autoref{fig:pruning_static}. Depending on the trigger type, the drop in the clean accuracy is not that severe, but \ac{ASR} remains high, as seen in~\autoref{fig:pruning_dynamic}.
When combining pruning with a fine-tuning phase, i.e., fine-pruning, we observe that \ac{ASR} can be drastically reduced while the clean accuracy remains high. The effect can be similar when focusing solely on the effect of fine-tuning, i.e., no pruned neurons ($\tau = 0$). Thus, pruning will not necessarily affect the model's backdoor performance. However, we find that solely retraining the model with clean data reduces the backdoor effect. We conclude that fine-tuning could effectively reduce the backdoor performance while keeping clean accuracy high. Still, the effect is more pronounced for backdoors that aim to be more stealthy, making it an interesting trade-off.
Since neuromorphic data consists of several frames (16), we can inject a moving backdoor, which is not possible in a single image.
According to our experimental results shown in~\autoref{fig:fine_pruning_moving}, we consider that fine-pruning may fail to reduce the effect of moving triggers.

\begin{finding}
    Fine-pruning can be effective against backdoor attacks in \acp{SNN} using neuromorphic data. Still, this depends on the dataset characteristics and the trigger type.
\end{finding}

The performance of the selected defenses in detecting backdoors in neuromorphic data and \acp{SNN} is limited by their inability to handle dynamic, moving, and smart triggers and their reliance on activation functions not present in \acp{SNN}. Further research and development are necessary to address these limitations and improve the robustness of the defenses in these contexts. Specific defenses considering the nature of neuromorphic data and \acp{SNN} are necessary to detect and defend against these backdoors effectively. This addresses Challenge~\ref{itm:defenses}.

\paragraph{Adaptive Attacker}

Many defenses are intended to detect either malicious model parameters or samples by observing statistical differences between malicious and clean samples on (potentially) compromised models, e.g., \ac{NC}~\cite{wang2019neural} or fine-pruning~\cite{liu2018fine}. By observing the effect of fine-pruning on our attacks, we investigated the ability of an adaptive attacker who knows the existence of defenses applied by the client in advance. 
Fine-pruning assumes that the backdoor effect is retained in some neurons while the clean behavior is retained in others. By stimulating the neurons in the last convolutional layer, the neurons with higher activation are thus compromised. The backdoor effect is removed by pruning the neurons with the highest activation (to some extent controlled by the pruning rate $\tau$).

\begin{figure}
    \centering
    \begin{subfigure}[b]{0.42\columnwidth}
        \includegraphics[width=\linewidth]{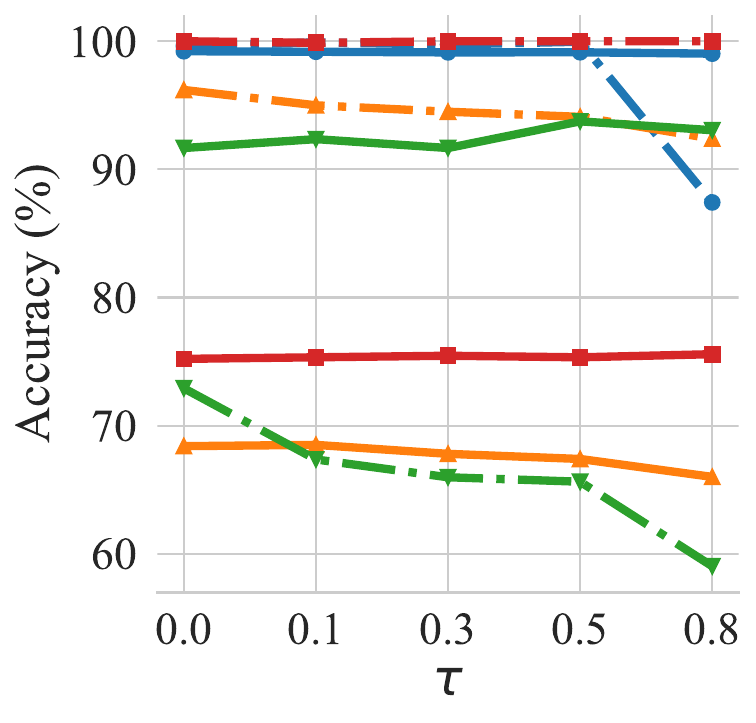}
        \caption{Static}
    \end{subfigure}
    \hfill
    \begin{subfigure}[b]{0.42\columnwidth}
        \includegraphics[width=\linewidth]{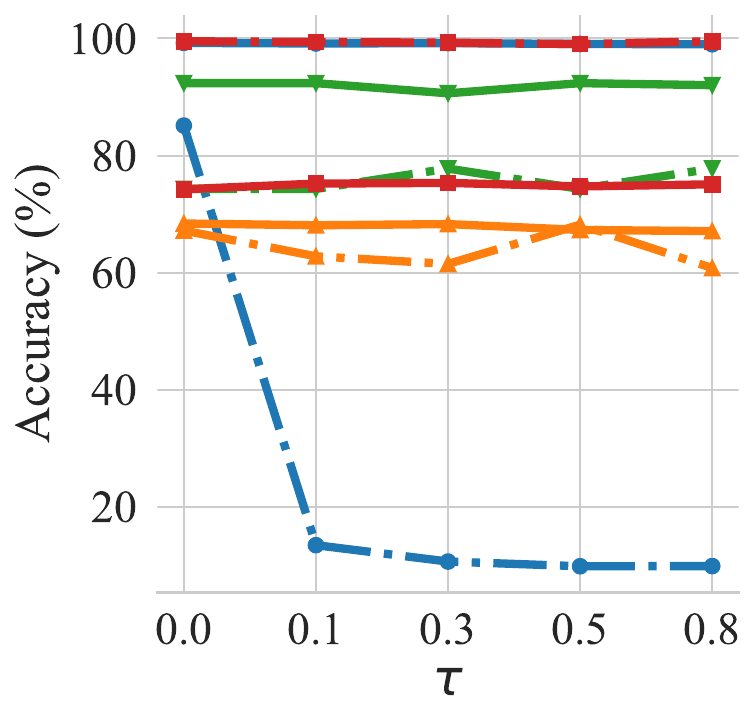}
        \caption{Moving}
    \end{subfigure}
    
    \caption{Results of different adaptive attacks after fine-pruning. \ac{ASR} (dashed lines) and clean accuracy (full line) for different types of attacks, i.e., static and moving. Blue corresponds to N-MNIST, orange to CIFAR10-DVS, green to DVS128-Gesture, and red to N-Caltech101.}
    \label{fig:adaptive}
\end{figure}

Recent work has made a substantial effort to develop techniques to bypass known defenses~\cite{tan2020bypassing, qi2023revisiting}. We observed that fine-pruning results are ineffective by using a low poisoning rate. We experimented with $\epsilon = \{0.001, 0.01\}$ for all the datasets in different attack settings. In most studied cases (see~\autoref{fig:adaptive}), the backdoor performance was kept high after fine-pruning, regardless of the pruning rate. Additionally, we investigate if the trigger size is relevant for fine-pruning. An adaptive attacker can bypass the defense even when increasing the trigger size to 30\% of the image size and lowering the poisoning rate to 0.001: we test this with the DVS128-Gesture dataset. However, fine-pruning prevents the backdoor effect in the simplest case of a static trigger with N-MNIST $\epsilon = 0.01$ and a trigger size of 10\%.

Following the same intuition of using a low poisoning rate in dynamic triggers, we can adjust the backdoor effect by tuning $\alpha$. To bypass fine-pruning, we experiment with $\alpha = 0.9$ with the DVS128-Gesture and N-MNIST datasets as use cases. The results show similar behavior, and the attack maintains both clean high accuracy and \ac{ASR}.
Additionally, fine-pruning is often performed by pruning only in the last convolutional layer. An attacker who knows this beforehand could exclude this layer during training with backdoor data, so the rest of the layers contain the logic of the backdoor. Therefore, after pruning, the backdoor behavior will not be affected. Our experiments verify this hypothesis. A clean model trained on clean data on the DVS128-Gesture dataset achieved 90\% accuracy. We retrain using backdoor data with $\epsilon = 0.1$, trigger size 10\%, polarity 3, static trigger in the middle for 20 epochs, freezing the last convolutional layer and achieving 79\% ASR. After fine-pruning, clean accuracy is 90\%, and ASR is 89\%. We also observe this behavior for CIFAR10-DVS with the same settings, achieving 68\% clean accuracy and 100\% ASR before fine-pruning and 67\% clean accuracy and 100\% ASR after fine-pruning.

\section{Stealthiness Evaluation}
\label{sec:stealthiness}

Quantifying image quality is often used for applications where the end user is a human. Subjective evaluation is commonly unsuitable for specific applications due to time constraints or expensive costs. 

\subsection{User Study}

We conducted a user study to validate the stealthiness of backdoor triggers identified using \ac{SSIM}. The study aimed to assess the effectiveness of four triggers---static, moving, smart, and dynamic---on the DVS128-Gesture dataset.
25 participants---with different backgrounds in \ac{DL}---including researchers, practitioners, and students from various geographical locations, were recruited to participate in the user study.

During the study, participants were presented with a series of images, each containing one of the four backdoor triggers. Three images were clean, while one was compromised. To ensure a comprehensive evaluation, we explored various trigger positions and polarities for static and moving attacks, with the trigger occupying 10\% of the image size, the largest in our experiments. We also assessed our smart attack at the least and most active locations and the least and most common polarities. Furthermore, our dynamic attack is evaluated for different $\gamma$ values: $\gamma = {0.001,0.01,0.1}$. The task assigned to the participants was to identify the compromised image.

The study recorded the selection frequency for each image as the compromised image by the participants. The analysis focused on calculating the percentage of times each image was chosen. The results revealed that the stealthiness of backdoor triggers varied based on location and color. The static, moving, and smart triggers exhibited varying stealthiness, heavily influenced by their parameter settings. An average of 50.6\% of the participants correctly noticed the trigger in static settings. In moving backdoors, an average of 73.3\% participants noticed the trigger correctly. An average of 84\% of the participants found the trigger in smart triggers. As discussed in previous sections, we also found that the participants failed to select the correct poisoned sample when the trigger was placed in the image region with the most activity. The smart trigger's stealthiness depends on the location and polarity. A smart trigger in the least common area with the least common polarity is more visible than a smart trigger in the most active location with the most common polarity. In contrast, the dynamic triggers displayed exceptional stealthiness, particularly at $\gamma = 0.01$, where only 4\% of the participants found the trigger. At the same time, larger values of $\gamma$, like 0.1, render the triggers more visible, where 96\% of the participants found the trigger. These findings were consistent with the stealthiness evaluation using the \ac{SSIM} metric.

The observed variations in stealthiness highlight the importance of considering the location, color, and parameter settings when assessing the effectiveness of backdoor triggers. The superior stealthiness exhibited by the dynamic triggers, which are not dependent on specific locations or colors, makes them particularly desirable in backdoor attack scenarios. Given that inspecting all data samples in a dataset is impractical, especially in real-life scenarios where models can be trained on billions of data samples~\cite{brown2020language}, we evaluate the usability of metrics for quantifying the stealthiness of triggers in the subsequent sections.

\begin{finding}
    A dynamic trigger cannot be detected by humans when generated using $\gamma = 0.01$.
\end{finding}

\subsection{Evaluated Metrics}
\Ac{MSE}~\cite{wang2009mean} compares two signals, e.g., image and audio, and measures the error or distortion between them. In our case, our signal is a frame sequence of images, where we compare a clean sample $\textbf{x}$ with a distorted (backdoored) sample $\hat{\textbf{x}}$. In \ac{MSE}, the error signal is given by $e = \textbf{x} - \hat{\textbf{x}}$, which is indeed the difference between pixels for two samples. However, \ac{MSE} has no context neighbor pixels, which could lead to misleading results~\cite{girod1993s, wang2002image}. For instance, a blurry image with an \ac{MSE} score of 0.2, i.e., 20\% of the pixels are modified, and a square of 20\% of the sample size on top of the image would give the same \ac{MSE} value. However, the blurry image is recognizable while the other is not. That is, two differently distorted images could have the same \ac{MSE} for some perturbations more visible than others. Therefore, \ac{MSE} cannot be the best measurement for backdoor attacks. Still, it could provide sufficient insights for quantifying stealthiness.

To overcome the locality of \ac{MSE}, Wang et al.~\cite{wang2004image} proposed a measure called \ac{SSIM} that compares local patterns of pixel intensities rather than single pixels, as in \ac{MSE}. Images are highly structured, whereas pixels exhibit strong dependencies carrying meaningful information. \Ac{SSIM} computes the changes between two windows instead of the pixel-by-pixel calculations given by:
\begin{equation*}
    SSIM(\mathbf{x}, \hat{\mathbf{x}}) = \frac{(2\mu_\mathbf{x}\mu_{\hat{\mathbf{x}}} + c_1) (2\sigma_{\mathbf{x}\hat{\mathbf{x}}} +  c_2)}{(\mu_{\mathbf{x}}^2 + \mu_{\hat{\mathbf{x}}}^2 +c_1) (\sigma_{\mathbf{x}}^2 + \sigma_{\hat{\mathbf{x}}}^2 +c_2)},
\end{equation*}
where $\mu_\mathbf{x}$ is the pixel sample mean of $\mathbf{x}$, $\mu_{\hat{\mathbf{x}}}$ is the pixel sample mean of $\hat{\mathbf{x}}$, and $c_1$ and $c_2$ are two variables to stabilize the division.

\subsection{Evaluating Stealthiness}

In this section, having analyzed metrics for comparing the variation between the clean and the backdoor images, we select \ac{SSIM} as the most useful for our case. We evaluate the stealthiness of our different attacks based on the \ac{SSIM} between the clean and backdoored images. The \ac{SSIM} values are averaged over 16 (as the batch size) randomly selected images from the test set. Precisely, we compare each clean frame with its backdoor frame counterpart. Then, the \ac{SSIM} per frame is averaged. To the best of our knowledge, this is the first application of \ac{SSIM} for comparing similarities in neuromorphic data, used for backdoor attacks in \acp{SNN}, or used in the \ac{SNN} domain overall.
This addresses Challenge~\ref{itm:stealth}.

\subsubsection{Static and Moving Triggers}

\begin{figure}
    \centering
    \begin{subfigure}[b]{0.5\linewidth}
        \includegraphics[width=\linewidth]{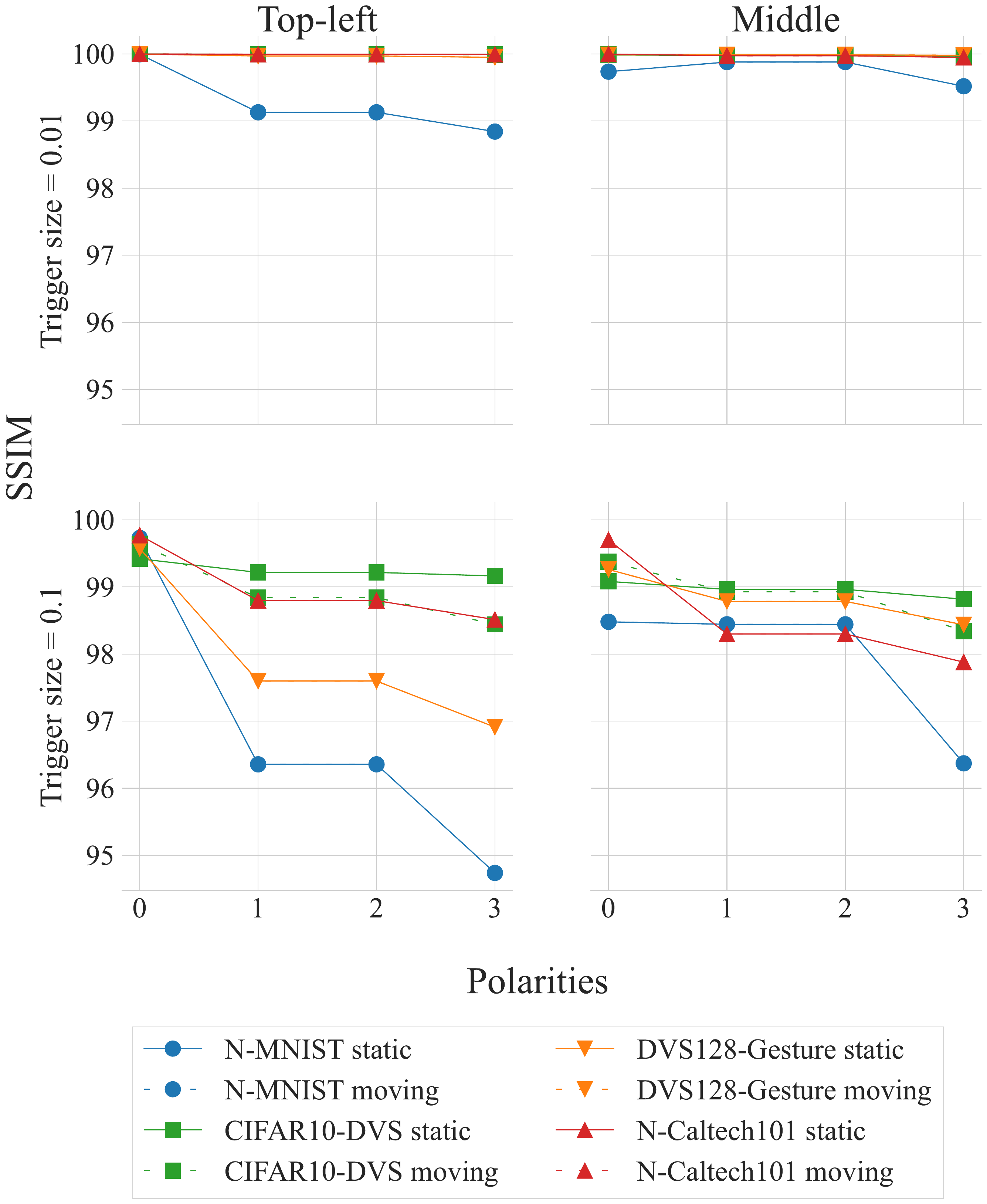}
        \caption{Static and moving triggers.}
        \label{fig:SSIM_static_moving}
    \end{subfigure}
    \begin{subfigure}[b]{0.48\linewidth}
        \includegraphics[width=\linewidth]{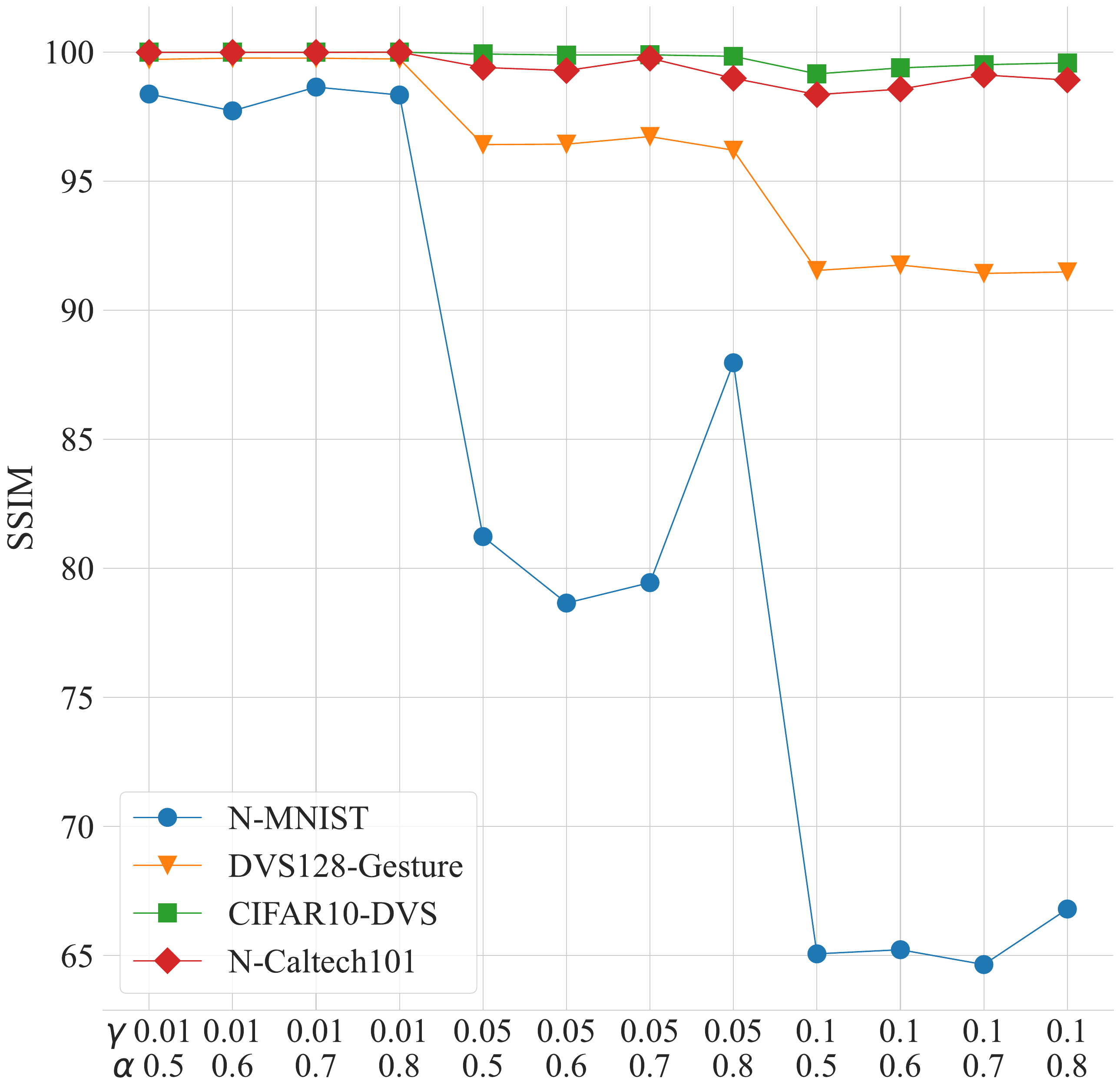}
        \caption{Dynamic triggers.}
        \label{fig:SSIM_dynamic}
    \end{subfigure}
    \vfill
    \caption{SSIM of different triggers.}
\end{figure}

We first analyze the static and moving triggers in two positions: corner (top-left) and middle, see~\autoref{fig:SSIM_static_moving}. We observe that the simpler the dataset, the more the stealthiness reduction, i.e., \ac{SSIM} is lower. Additionally, the trigger size and the polarity affect the stealthiness. Indeed, the larger the trigger size, the less \ac{SSIM}, which is expected. However, polarity also plays a crucial role in stealthiness. We observe a noticeable similarity downgrade related to the trigger polarity. The largest similarity degradation is observed for the N-MNIST dataset, with a trigger size of 0.01, and placing the trigger in the top-left corner. The background polarity, i.e., $p=0$, shows high \ac{SSIM}; however, with $p=3$, the \ac{SSIM} lowers to 94.5\%.
This could be directly linked to the number of pixels of a given polarity in an area. The less polarity in an area, the more ``contrast'' it would create, being less stealthy.

Comparing the static and moving triggers, we observe a more significant degradation when the trigger is moving, although it is negligible in some datasets or settings. Triggers in noisy datasets like CIFAR10-DVS and N-Caltech101 are more tolerable to input perturbations. Modifications in simpler datasets, such as N-MNIST, significantly change the image's overall structure, achieving a lower \ac{SSIM}.

\subsubsection{Smart Triggers}

Smart triggers select the trigger polarity and location by themselves, based on the image's least or most active area and the least prominent polarity in an area. We observe a larger degradation based on the trigger size and when placed in the least active area (see~\autoref{fig:ssim_appendix} in Appendix~\ref{sec:app_ssim}). This is expected as the trigger in the most active area gets hidden by the high activity, i.e., motion. Thus, the performance of triggers in the most active area gets lowered, but gains trigger stealthiness, which must be considered a trade-off between stealthiness and performance.


\subsubsection{Dynamic Triggers}

Lastly, dynamic triggers (see~\autoref{fig:SSIM_dynamic}) show impressive stealthiness as $\gamma$ gets smaller, even to the point of being indistinguishable from the clean image. We also observe that the more complex the dataset, the less the reduction in the stealthiness, contrary to N-MNIST, where the degradation is notable with $\gamma = 0.1$. This effect is related to the number of pixel changes and the noise in the data. A dataset with large noise has much activity, thus making it easier for the trigger to be hidden, as in CIFAR10-DVS and N-Caltech101. However, with ``clean'' datasets that contain little noise as N-MNIST, even the subtlest perturbation makes a noticeable change. Still, with $\gamma = 0.01$, the perturbation in every tested dataset is invisible.

\begin{figure}
    \centering
    \begin{subfigure}[b]{0.29\linewidth}
        \includegraphics[width=\linewidth]{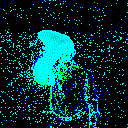}
        \caption{Clean.}
    \end{subfigure}
    \hfill
    \begin{subfigure}[b]{0.29\linewidth}
        \includegraphics[width=\linewidth]{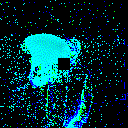}
        \caption{Static trigger.}
    \end{subfigure}
    \hfill
    \begin{subfigure}[b]{0.29\linewidth}
        \includegraphics[width=\linewidth]{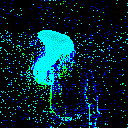}
        \caption{Dynamic trigger.}
    \end{subfigure}
    \caption{Comparison of triggers.}
    \label{fig:comparison}
\end{figure}

Overall, note that even if the \ac{SSIM} of the static trigger (\autoref{fig:SSIM_static_moving}) and the dynamic trigger (\autoref{fig:SSIM_dynamic}) are similar, as seen in~\autoref{fig:comparison} the visibility of the triggers are rather different. The static trigger is highly noticeable in the middle of the figure, while the dynamic trigger is indistinguishable from the clean sample. Although \ac{SSIM} has been previously used for comparing images, we require more robust metrics to compare neuromorphic data.

In general, employing metrics to identify outliers in data similarity holds potential for practical applications and large-scale scenarios. However, it should be noted that while certain evaluations using \ac{SSIM} yield comparable outcomes to those observed in our user study, \ac{SSIM} alone cannot serve as an ad-hoc defense against backdoor attacks. This approach tends to produce many false negatives and can be easily circumvented by sophisticated adversaries. Addressing this limitation is an essential direction for our future research.

\section{Related Work}
\label{sec:related}


\textbf{SNNs.}
In recent years, numerous efforts have been made to develop supervised learning algorithms for \acp{SNN} to make them more practical and widely applicable. One of the first such algorithms was Spike Prop~\cite{bohte2002error}, which was based on backpropagation and could be used to train single-layer \acp{SNN}. However, it was not until more recent developments that \acp{SNN} could be applied to multi-layer setups.
Despite such advances, most existing \ac{SNN} training methods still require manual tuning of the spiking neuron membrane, which can be time-consuming and may limit the performance of \ac{SNN}. To overcome this limitation, Fang et al.~\cite{fang2021incorporating} proposed a method that can learn the weights and hyperparameters of the membranes in an automated way, eliminating the need for manual tuning. This advancement may make \acp{SNN} more practical and easier to use for a broader range of applications.

Several other notable developments in the field of \acp{SNN} are worth mentioning. One such development is event-driven update rules, allowing \acp{SNN} to operate more efficiently by only updating the network when necessary~\cite{zhang2021event}. This contrasts traditional neural networks that require continuous updates and can be computationally expensive.
Another area of research in \acp{SNN} is structural plasticity, which refers to the network's ability to change its structure during training~\cite{zhang2021event, weerasinghe2021incorporating}. This can be accomplished through the addition or removal of connections between neurons or through the creation of new neurons altogether. Structural plasticity can improve the learning efficiency and generalization capabilities of \acp{SNN} and is effective in various tasks.
There are also ongoing efforts to develop unsupervised learning algorithms for \acp{SNN}, allowing them to learn from data without needing labeled examples~\cite{hazan2018unsupervised}. Unsupervised learning is a critical component of the brain's learning process and can significantly expand the range of tasks that \acp{SNN} can perform.

\textbf{Backdoor Attacks.}
Backdoor attacks were first introduced by Gu et al., where the authors presented BadNets~\cite{gu2019badnets}. BadNets uses a square-shaped trigger on a fixed location to inject the backdoor task; it was the first to show backdoor vulnerabilities in machine learning. BadNets requires access to the training data for injecting the backdoor, contrary to the work by Liu et al.~\cite{liu2018trojaning}, which alleviated this assumption. The authors presented a novel work where access to the training data was not needed. The authors systematically reconstructed training samples to inject the backdoor by adding the trigger on top of the samples and retraining the model.
The discussed approaches use static triggers, i.e., the trigger is in the exact location for all the samples. Nguyen and Tran~\cite{nguyen2020input} developed a dynamic backdoor attack in which the trigger varies with the input. Specifically, a generator creates a pixel scatter that is then overlapped with the clean input. A similar approach was investigated by Salem et al.~\cite{salem2022dynamic}, who also constructed a dynamic backdoor attack. Instead of a pixel-scattered trigger, the trigger is a square, which has the advantage of applying it to physical objects.
Aiming to increase the stealthiness of the backdoor, Lie et al. created ReFool~\cite{liu2020reflection}, which includes benign-looking triggers as reflections in the clean sample. A similar approach was followed by Zhang et al.~\cite{zhang2022poison}, who proposed Poison Ink, where the structure of the image is extracted. The structure is then injected with a crafted pixel pattern and included in the clean image. The resulting poisoned image is indistinguishable from the clean sample. 

In \acp{SNN} and neuromorphic datasets, only~\cite{abad2022poster} explored backdoor attacks. However, their experimentation is limited to exploring static and moving triggers using simple datasets and models. Moreover, based on the results, the authors do not provide insights into why backdoors occur in \acp{SNN}. They also do not consider any more advanced triggers, stealthiness evaluation, or defense mechanisms. 

\textbf{Defenses Against Backdoor Attacks.} 
Backdoor attacks can be mitigated using either model-based or data-based defenses. Model-based defenses involve examining potentially infected models to identify and reconstruct the backdoor trigger. An example is \ac{NC}~\cite{wang2019neural}, which aims to reconstruct the smallest trigger capable of causing the model to misclassify a specific input. This approach is based on the premise that an infected model is more likely to misclassify an input with a trigger than one that does not have a trigger. Another model-based approach is \ac{ABS}~\cite{liu2019abs}, which involves activating every model neuron and analyzing anomalies in the resulting output.
Model-based defenses are designed to specifically target infected models, searching for signs of a trigger and attempting to reconstruct it. This approach is effective because the presence of a trigger is often a reliable indicator that the model has been compromised. However, it is also possible for a model to be infected without a trigger, where they are not applicable. Thus, model-based defenses are limited to cases with triggers.
Data-based defenses aim to detect the presence of a backdoor by analyzing the dataset without inspecting the model itself. One approach in this category is clustering techniques to differentiate between clean and poisoned data~\cite{chen2018detecting}. Another approach, \ac{STRIP}~\cite{gao2019strip}, combines the provided dataset with known infected data and queries the model on the resulting combined dataset. By measuring the entropy of the model's output on this combined dataset, \ac{STRIP} can detect backdoored inputs, which tend to have lower entropy than clean inputs.
Data-based defenses focus on analyzing the dataset to detect the presence of a backdoor. These approaches are helpful because they do not require knowledge of the specific trigger used to infect the model, making them more robust against variations in the method of infection. However, data-based defenses may not be as effective at detecting more subtle forms of backdoor attacks, which may not leave as clear a signature in the dataset.
Currently, no defenses are \ac{SNN} or neuromorphic data-specific. 
As shown, some well-performing defenses adapted from the image domain do not work well in \acp{SNN}. Thus, developing SNNs or neuromorphic data-specific defenses is necessary for future research.

\section{Conclusions \& Future Work}
\label{sec:conclusions}

This study explored the security of \acp{SNN} in the context of backdoor attacks. Despite the growing importance of \acp{SNN} as an emerging technology, this area has received limited attention. Our investigation utilizes neuromorphic data and triggers to launch backdoor attacks in \acp{SNN}.
We proposed several attack methods, including a novel dynamic trigger that evolves over time and remains undetectable to human inspection. We have also evaluated the attacks against different state-of-the-art defenses, which we have adapted from the image domain. Our results demonstrate that our attacks can achieve an \ac{ASR} of up to 100\% without causing noticeable degradation in clean accuracy, even when the defenses are employed.
Our findings show that \acp{SNN} are highly vulnerable to backdoor attacks, indicating a need for further improvements in existing defense mechanisms. 

Future research should focus on developing \acp{SNN} specifically designed to counter backdoor attacks and address the unique challenges posed by neuromorphic data. It is worth noting that \acp{SNN} are also gaining prominence in other domains, such as the graph domain, where backdoor attacks are becoming increasingly significant. In these domains, specific backdoor designs may be required to tackle the challenges that the spiking graph neural networks pose.
Finally, while our study has primarily focused on backdoor attacks, it is important to consider other common threats that can be adapted to \acp{SNN}, such as inference attacks and adversarial examples. 

\section*{Acknowledgment}

We thank the shepherd for helpful feedback and support in improving this work. The Horizon Europe, Spanish CDTI, and ELKARTEK programs fund this research. Grant agreements 101021911 (IDUNN), CER-20191012 (EGIDA), and KK-2021/00091 (REMEDY), respectively.

\newpage

\bibliographystyle{IEEEtranS}
\bibliography{references}

\appendices

\section{Additional Experiments}
\label{sec:appendix}

\subsection{Datasets}
\label{app:datasets}

We use four datasets: N-MNIST~\cite{orchard2015converting}, CIFAR10-DVS~\cite{li2017cifar10}, DVS128-Gesture~\cite{amir2017low}, and N-Caltech101~\cite{orchard2015converting}. We use N-MNIST, CIFAR10-DVS, and N-Caltech101 because their non-neuromorphic versions are common benchmarking datasets in computer vision for security/privacy in ML.
N-MNIST is a spiking version of MNIST~\cite{lecun1998mnist}, which contains $34\times34$ 60\,000 training, and 10\,000 test samples. An \ac{ATIS}~\cite{posch2010high} captured the dataset across the 10 MNIST digits on an LCD monitor. The CIFAR10-DVS dataset is the spiking version of the CIFAR10~\cite{krizhevsky2009learning} dataset, which contains 9\,000 training, and 1\,000 test $128\times128$ samples, corresponding to 10 classes. The N-Caltech101 dataset is the spiking version of the original Caltech101~\cite{fei2004learning} dataset. The dataset contains $180\times240$ 8\,709 train, and 823 test samples. To be consistent with the rest of the dataset, we square the data by cropping them to $180\time180$.
Lastly, the DVS128-Gesture dataset is a ``fully neuromorphic'' dataset created for \acp{SNN} tasks. 
The DVS128-Gesture dataset collects real-time motion captures from 29 subjects making 11 different hand gestures under three illumination conditions, creating 1\,176 $128\times128$ training samples and 288 test samples.
For all datasets, the samples' shape is $T\times P\times H\times W$, where $T$ is the time steps (we set it to $T=16$), $P$ is the polarity, $H$ is the height, and $W$ is the width.

\subsection{Network Architectures}
\label{app:network}

We consider three network architectures for the victim classifiers used in related works~\cite{fang2021incorporating}. The N-MNIST dataset's network comprises a single convolutional layer and a fully connected layer. For the CIFAR10-DVS dataset, the network contains two convolutional layers followed by batch normalization and max pooling layers. Then, two fully connected layers with dropout are added, and lastly, a voting layer---for improving the classification robustness~\cite{fang2021incorporating}---of size ten is incorporated. Finally, for the DVS128-Gesture and N-Caltech101 datasets, five convolutional layers with batch normalization and max pooling, two fully connected layers with dropout, and a voting layer compose the networks. For more details, see our code repository.

Based on previous work~\cite{doan2021lira}, the spiking \ac{AE} for the dynamic attack has four convolutional layers with batch normalization, four deconvolutional layers with batch normalization, and \textit{tanh} as the activation function for the DVS128-Gesture, CIFAR10-DVS, and N-Caltech101 datasets. For N-MNIST, we use two convolutional and two deconvolutional layers with batch normalization and \textit{tanh} as the activation function, which is the common \ac{AE}~\cite{Goodfellow-et-al-2016} structure.

\subsection{Default Training Settings}
\label{sec:train}

For training, we set a default \ac{LR} of 0.001, \ac{MSE} as the loss function, Adam as the optimizer, and we split the neuromorphic datasets in $T=16$ frames using the SpikingJelly framework~\cite{SpikingJelly}. For the N-MNIST dataset, we achieve a (clean) accuracy of 99\% on a holdout test set in 10 epochs. For the CIFAR10-DVS case, we achieved 68\% accuracy after 28 epochs. For N-Caltech101, we achieve 76\% accuracy after 30 epochs\footnote{Results for N-Caltech101 can be found in Appendix~\autoref{sec:caltech} due to space limitation.}, and a 93\% accuracy with 64 epochs for the DVS128-Gesture dataset, see~\autoref{tab:accuracy}. The results are aligned with the state-of-the-art~\cite{samadzadeh2020convolutional}.

\begin{table}[htb]
    \centering
    \caption{Baseline training results for different datasets.}
    \label{tab:accuracy}
	   \resizebox{0.6\columnwidth}{!}{%
            \begin{tabular}{@{}ccc@{}}
                \toprule
                Dataset        & \# Epochs & Accuracy (\%) \\ \midrule
                N-MNIST        &    10     & 99.4 $\pm$ 0.06    \\
                CIFAR10-DVS    &    28     & 68.3 $\pm$ 0.28 \\
                DVS128-Gesture &    64     & 92.5 $\pm$ 0.91    \\
                N-Caltech101   &    30     & 75.5 $\pm$ 0.04    \\
                \bottomrule
            \end{tabular}
        }
\end{table}

\subsection{Results for Static and Moving Backdoors}
\label{sec:static_moving_results}

\begin{figure*}
    \centering
    \begin{subfigure}[b]{0.31\linewidth}
        \includegraphics[width=\linewidth]{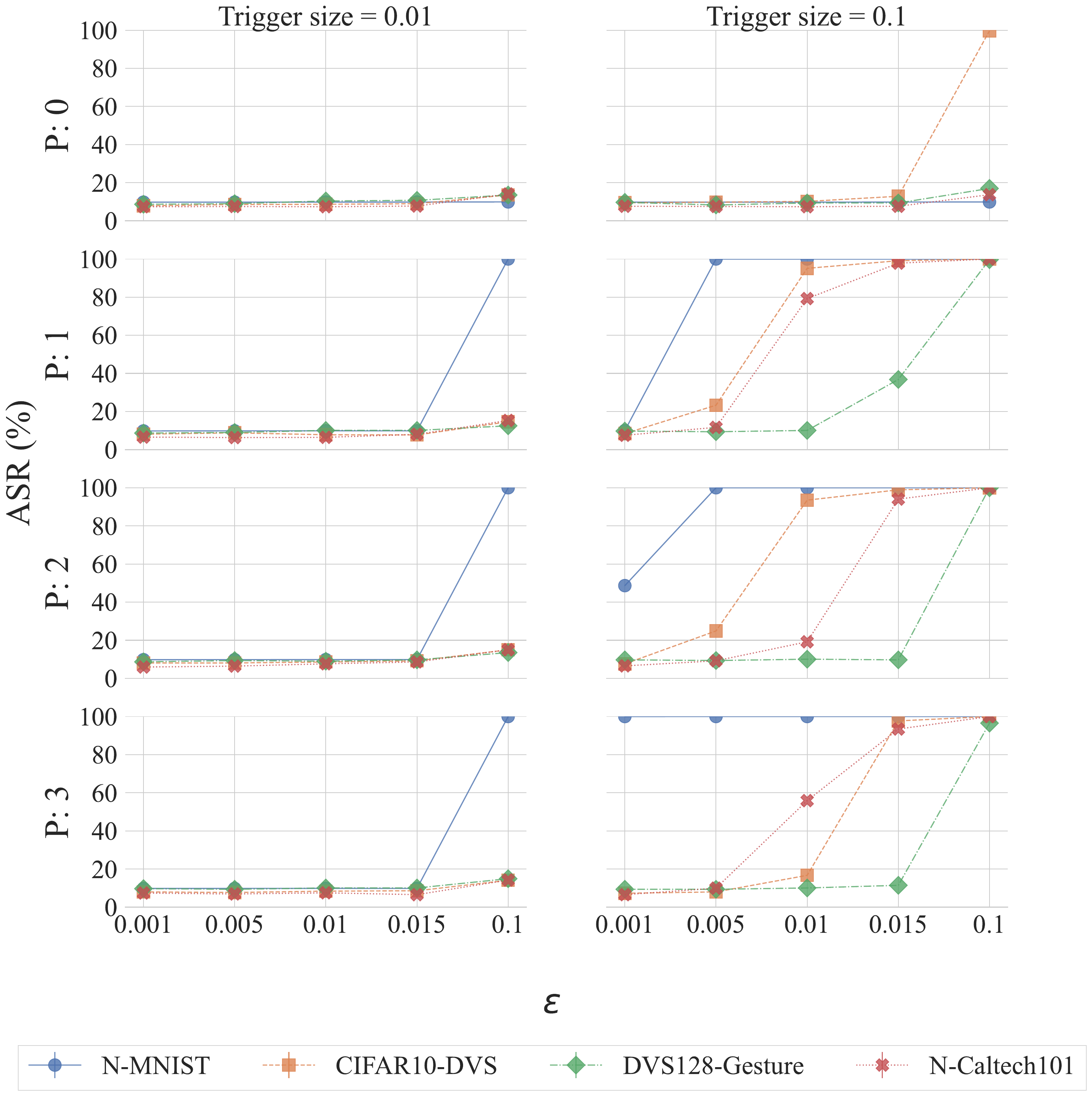}
        \caption{Bottom-right static backdoor.}
        \label{fig:static_bottom_right}
    \end{subfigure}
    \begin{subfigure}[b]{0.31\linewidth}
        \includegraphics[width=\linewidth]{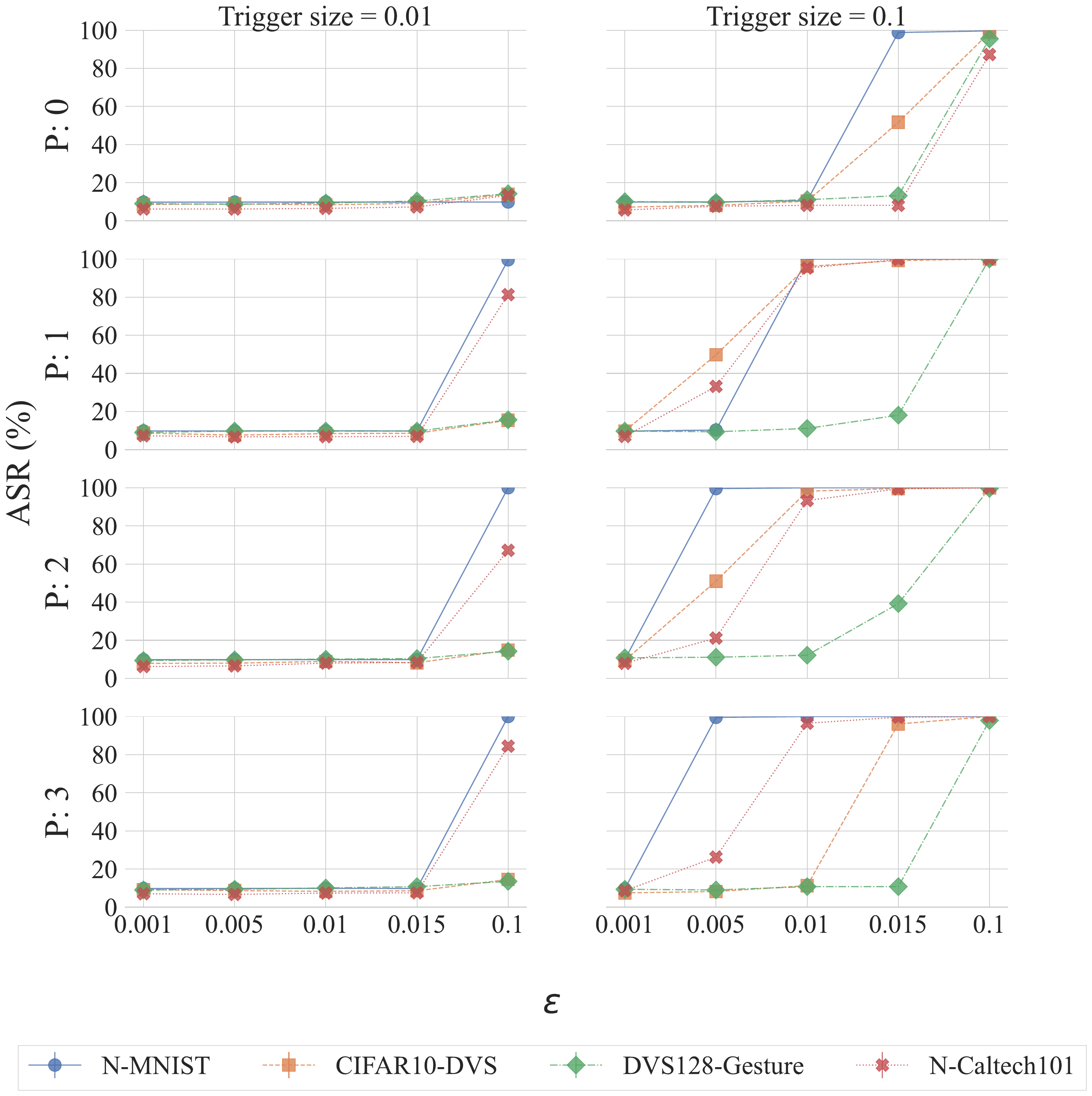}
        \caption{Middle static backdoor.}
        \label{fig:static_middle}
    \end{subfigure}
    \begin{subfigure}[b]{0.31\linewidth}
        \includegraphics[width=\linewidth]{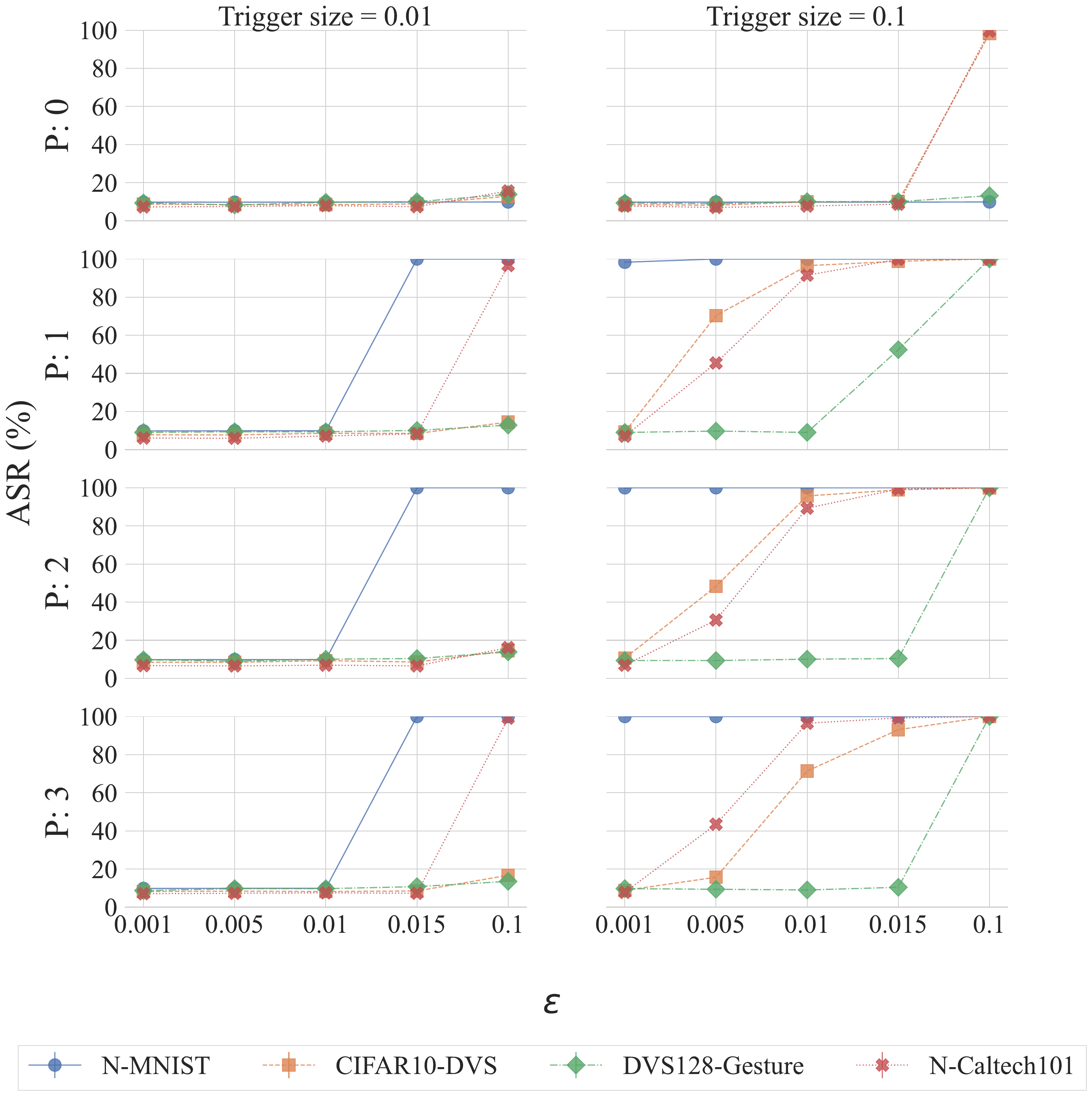}
        \caption{Top-left static backdoor.}
        \label{fig:static_top_left}
    \end{subfigure}
    \caption{ASR of static triggers.}
    \label{fig:static_backdoors_res}
\end{figure*}

\begin{figure*}
    \centering
    \begin{subfigure}[b]{0.31\linewidth}
        \includegraphics[width=\linewidth]{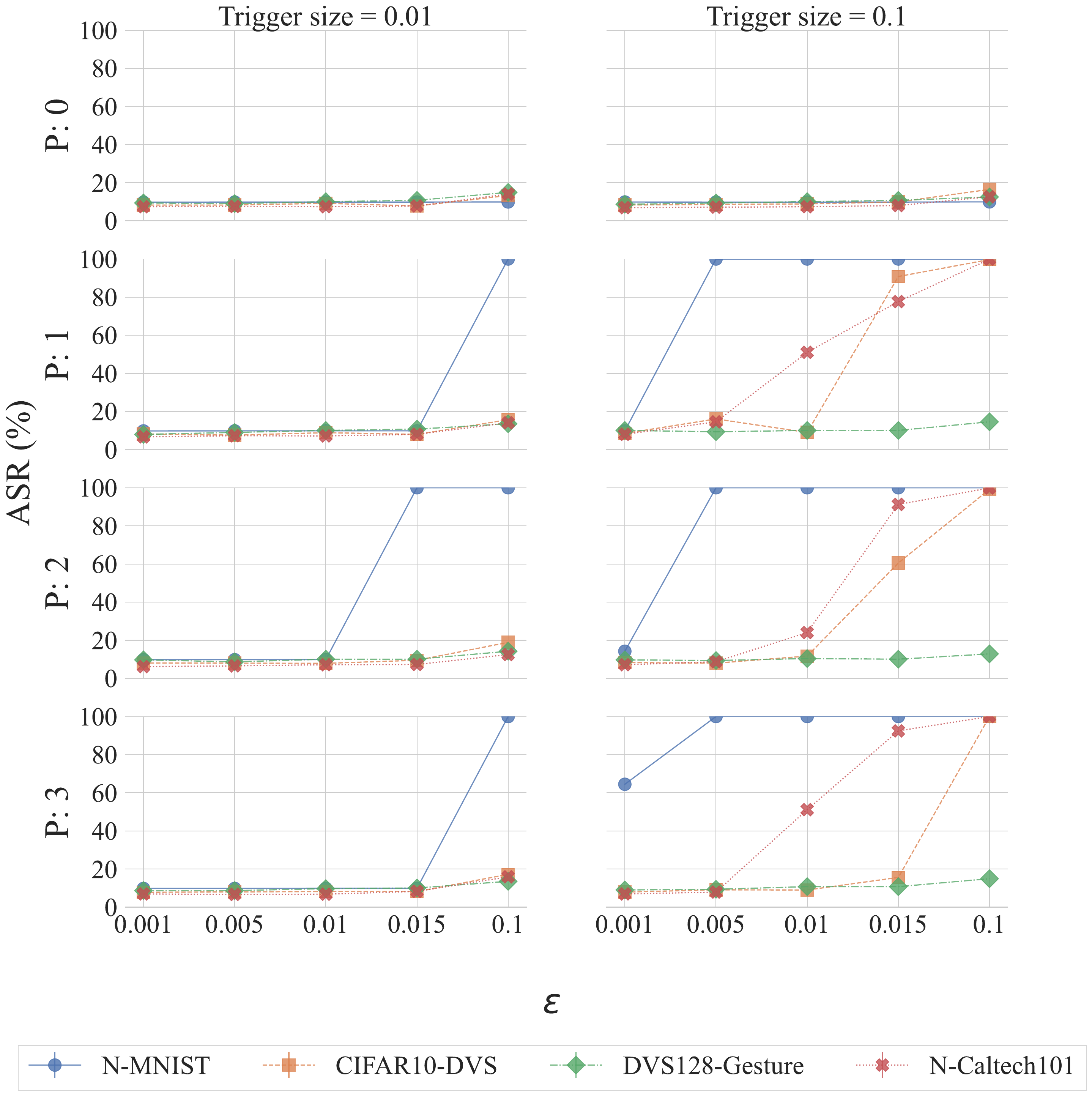}
        \caption{Bottom-right moving backdoor}
        \label{fig:moving_bottom_right}
    \end{subfigure}
    \begin{subfigure}[b]{0.31\linewidth}
        \includegraphics[width=\linewidth]{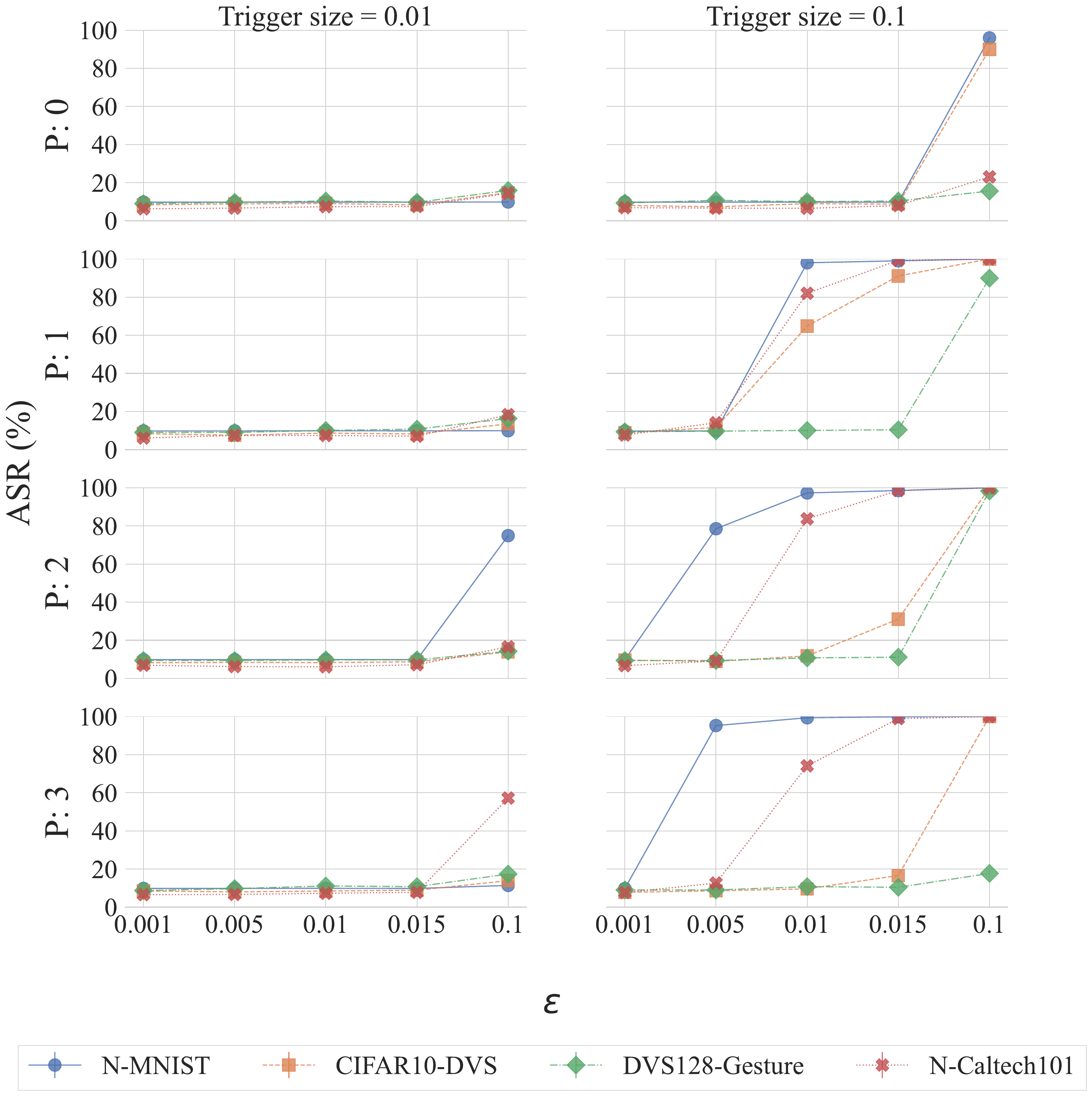}
        \caption{Middle moving backdoor}
        \label{fig:moving_middle}
    \end{subfigure}
    \begin{subfigure}[b]{0.31\linewidth}
        \includegraphics[width=\linewidth]{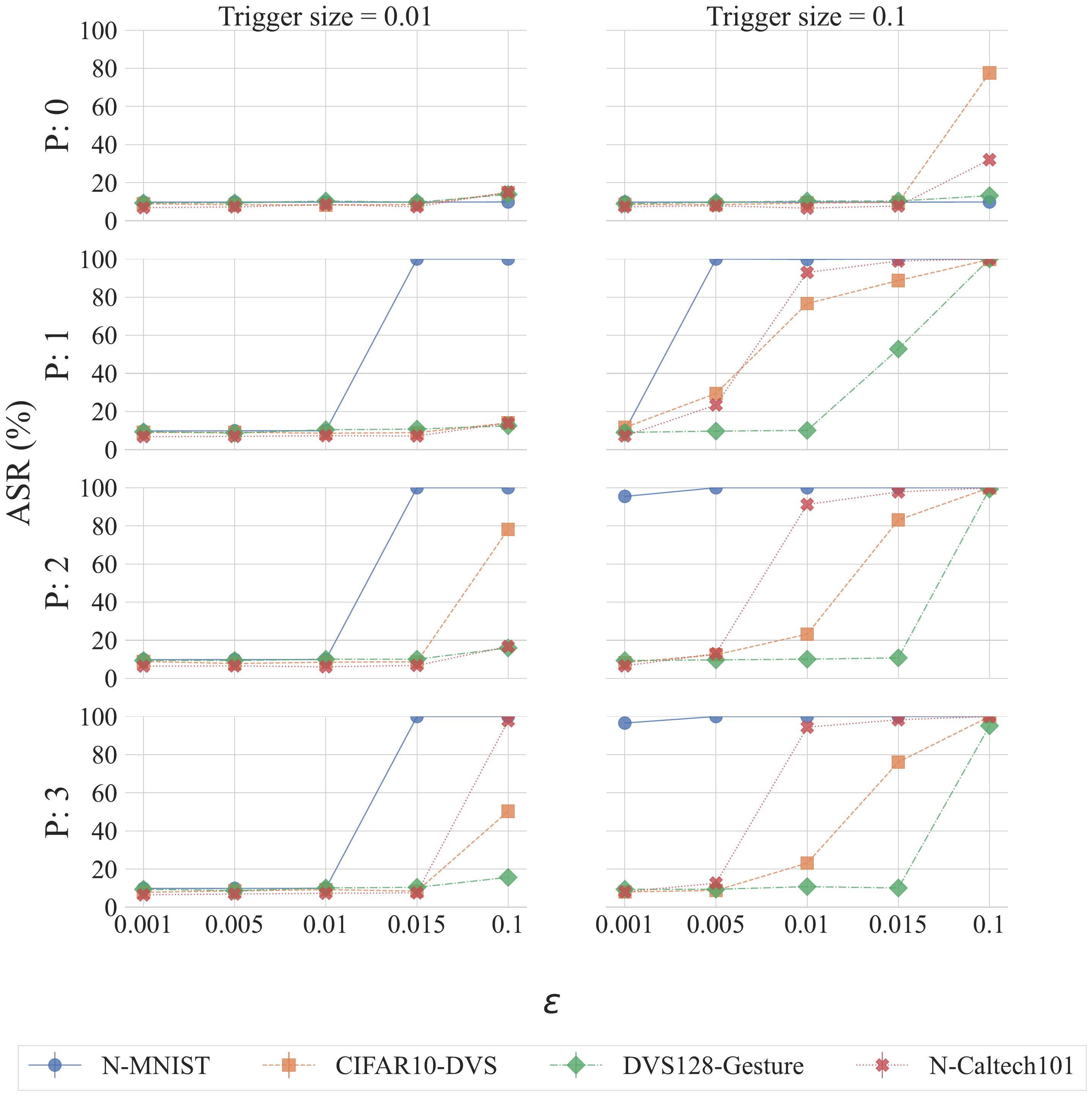}
        \caption{Top-left moving backdoor}
        \label{fig:moving_top_left}
    \end{subfigure}
    \caption{ASR of moving triggers.}
    \label{fig:moving_backdoors_res}
\end{figure*}

In~\autoref{fig:static_backdoors_res} and~\autoref{fig:moving_backdoors_res}, we provide results for static backdoors and moving backdoors, respectively.

\subsection{Clean Accuracy Degradation of Static Triggers}

\begin{figure*}
    \centering
    \begin{subfigure}[b]{0.31\linewidth}
        \includegraphics[width=\linewidth]{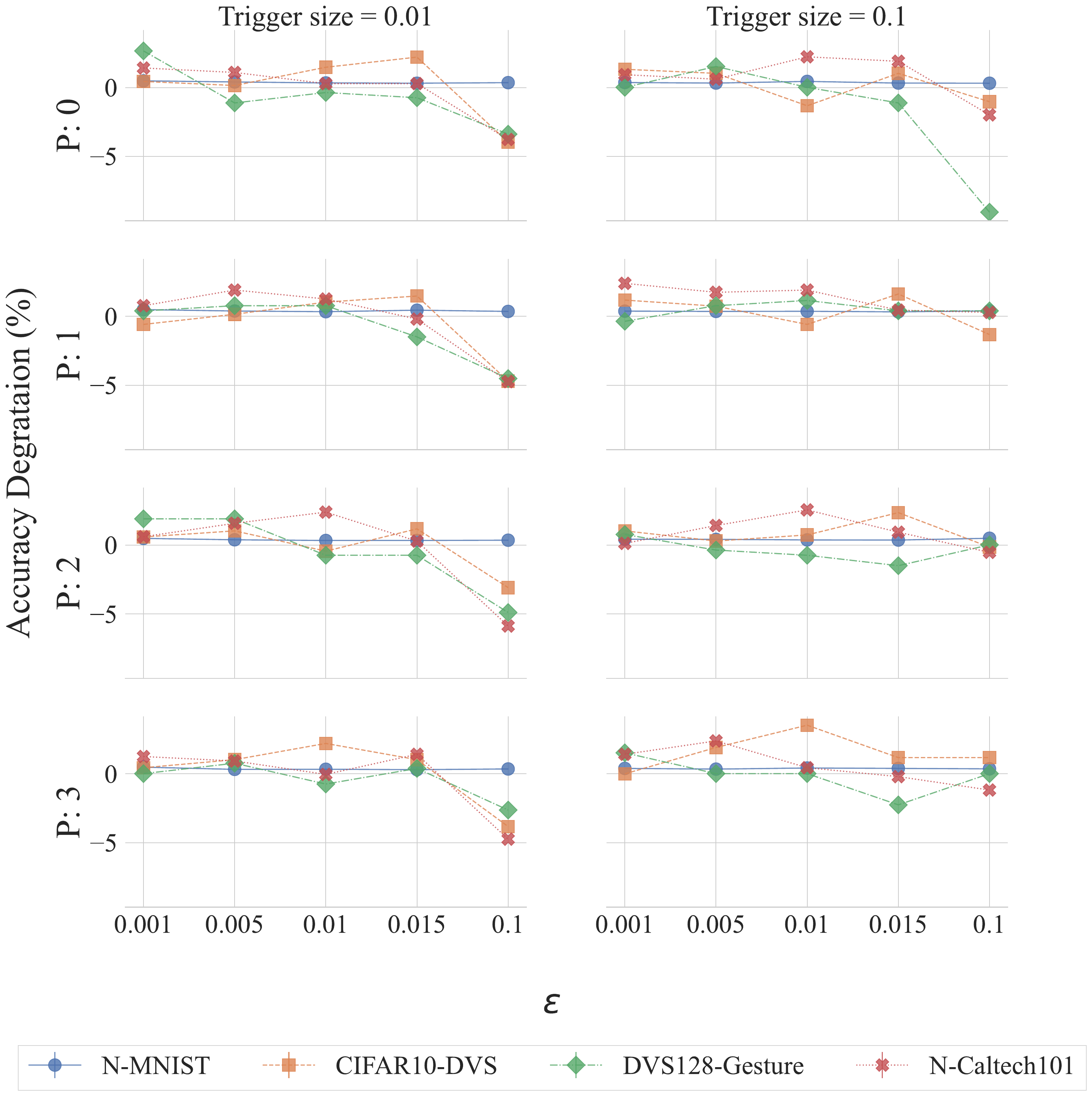}
        \caption{Bottom-right static backdoor}
        \label{fig:static_bottom_right_degradation}
    \end{subfigure}
    \begin{subfigure}[b]{0.31\linewidth}
        \includegraphics[width=\linewidth]{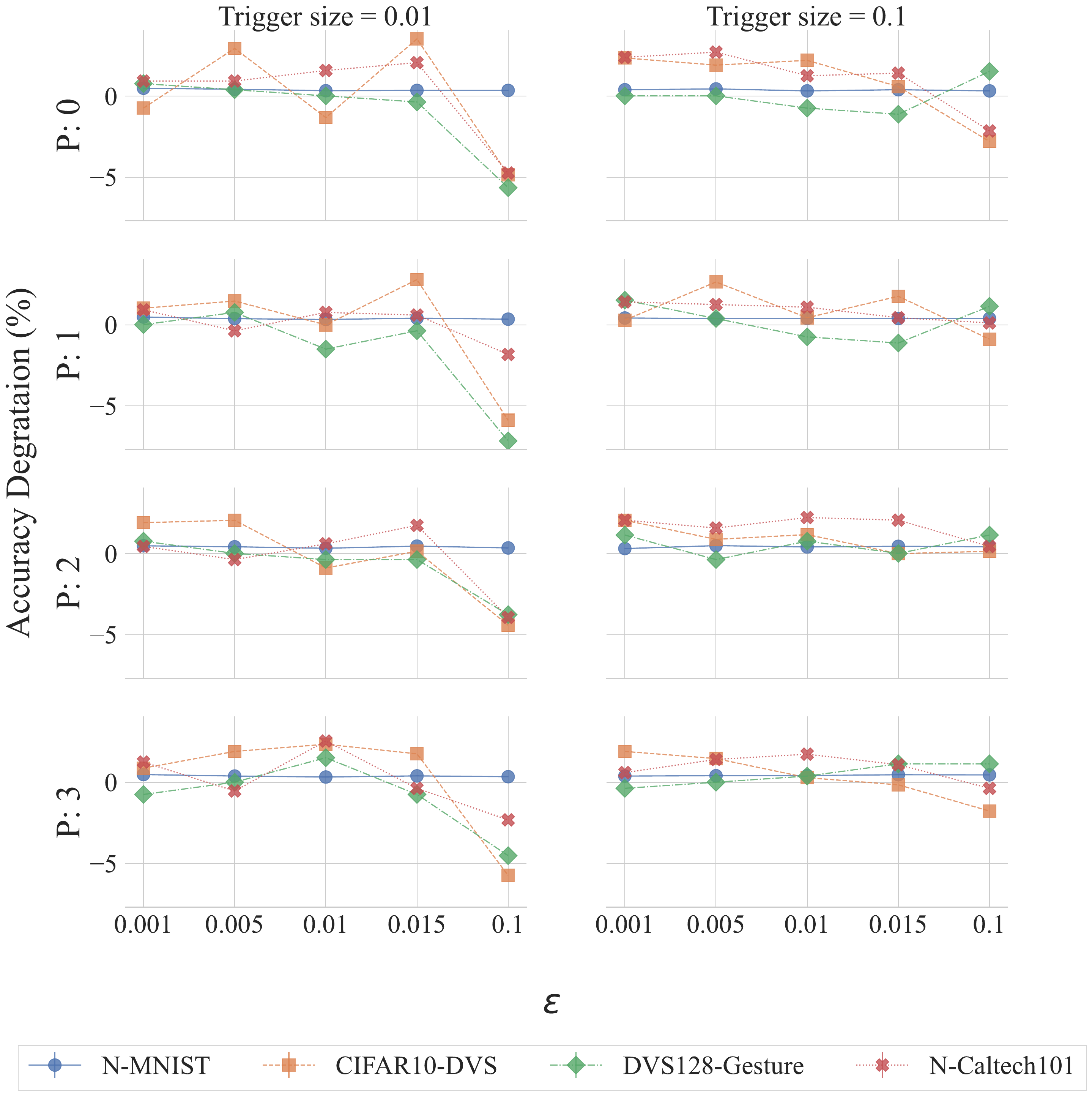}
        \caption{Middle static backdoor}
        \label{fig:static_middle_degradation}
    \end{subfigure}
    \begin{subfigure}[b]{0.31\linewidth}
        \includegraphics[width=\linewidth]{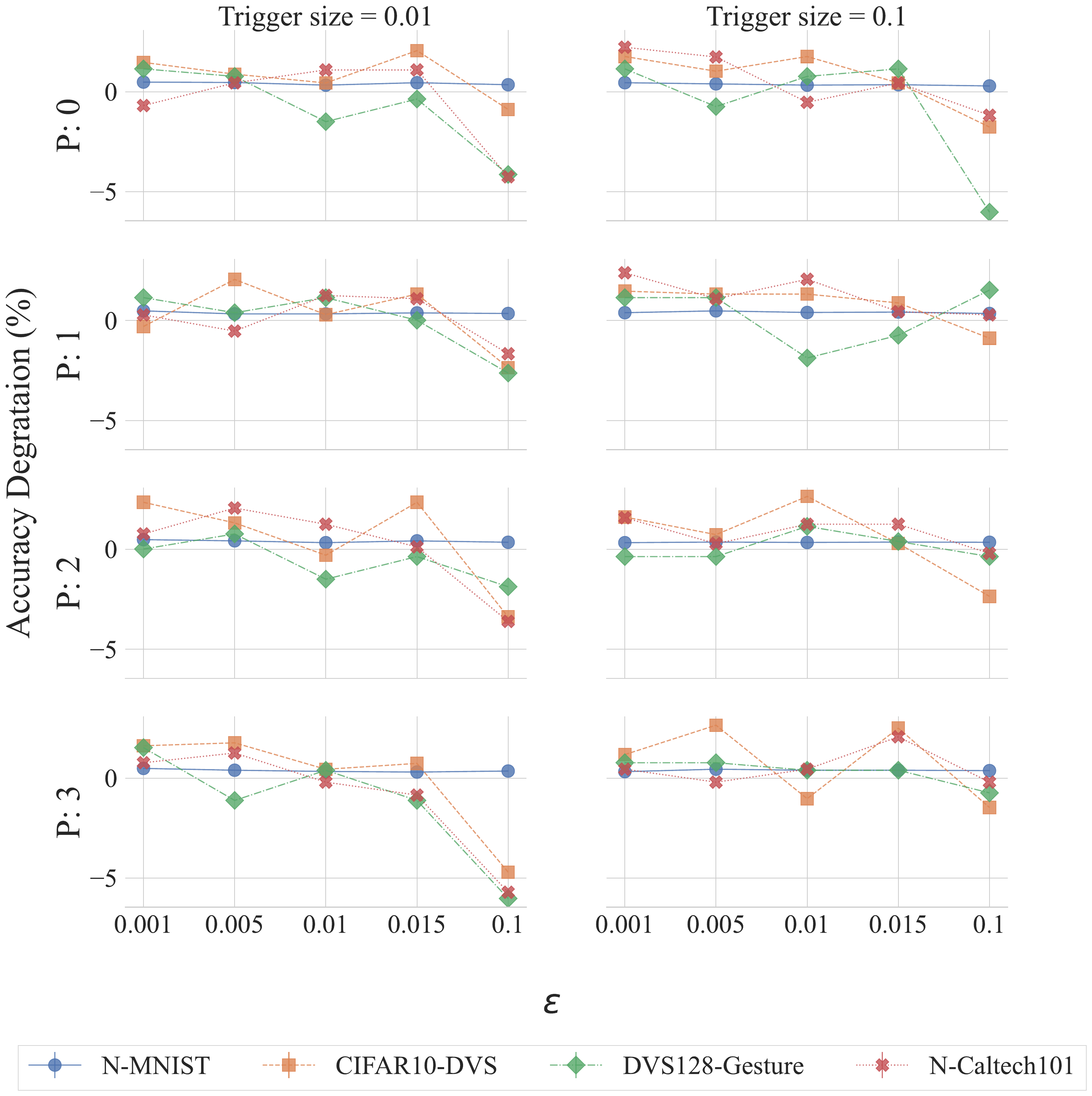}
        \caption{Top-left static backdoor}
        \label{fig:static_top_left_degradation}
    \end{subfigure}
    \caption{Clean accuracy degradation of static triggers.}
\end{figure*}

The clean accuracy degradation after the static attack in the bottom-right corner is shown in~\autoref{fig:static_bottom_right_degradation} and for the middle trigger in~\autoref{fig:static_middle_degradation}. Lastly, the degradation in the top-left corner is shown in~\autoref{fig:static_top_left_degradation}.

\subsection{Clean Accuracy Degradation of Moving Triggers}

\begin{figure*}
    \centering
    \begin{subfigure}[b]{0.31\linewidth}
        \includegraphics[width=\linewidth]{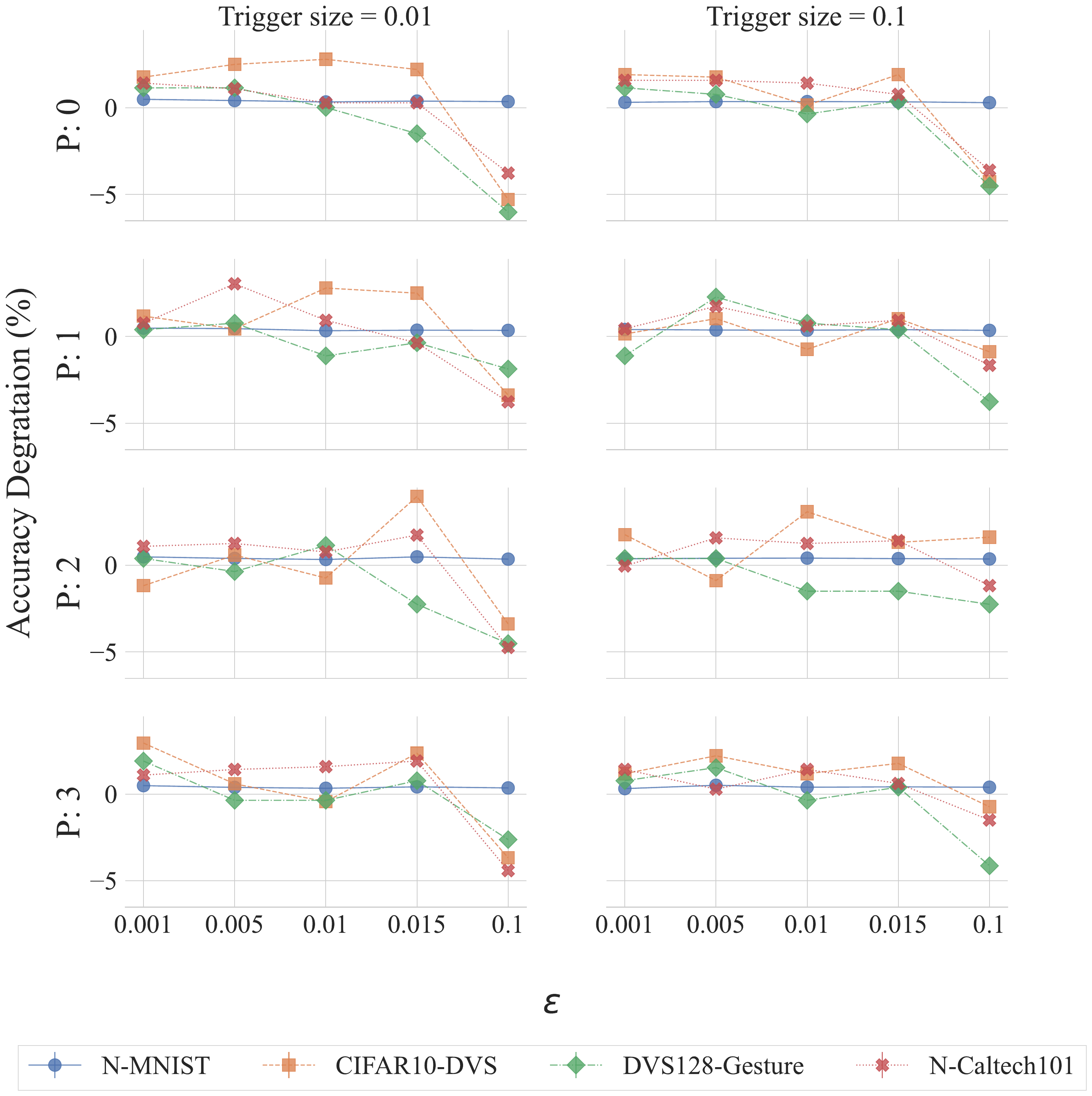}
        \caption{Bottom-right moving backdoor}
        \label{fig:moving_bottom_right_degradation}
    \end{subfigure}
    \begin{subfigure}[b]{0.31\linewidth}
        \includegraphics[width=\linewidth]{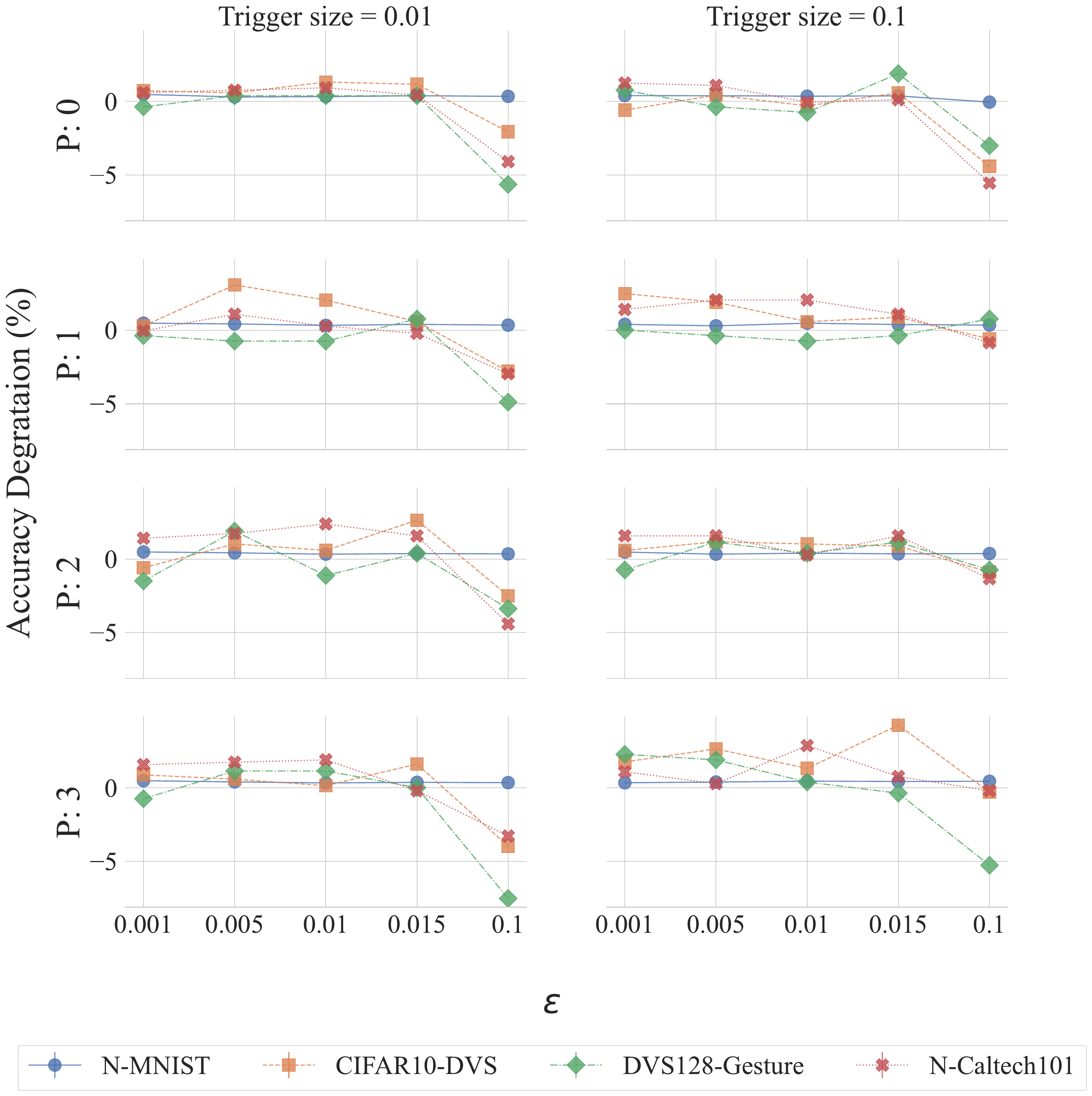}
        \caption{Middle moving backdoor}
        \label{fig:moving_middle_degradation}
    \end{subfigure}
    \begin{subfigure}[b]{0.31\linewidth}
        \includegraphics[width=\linewidth]{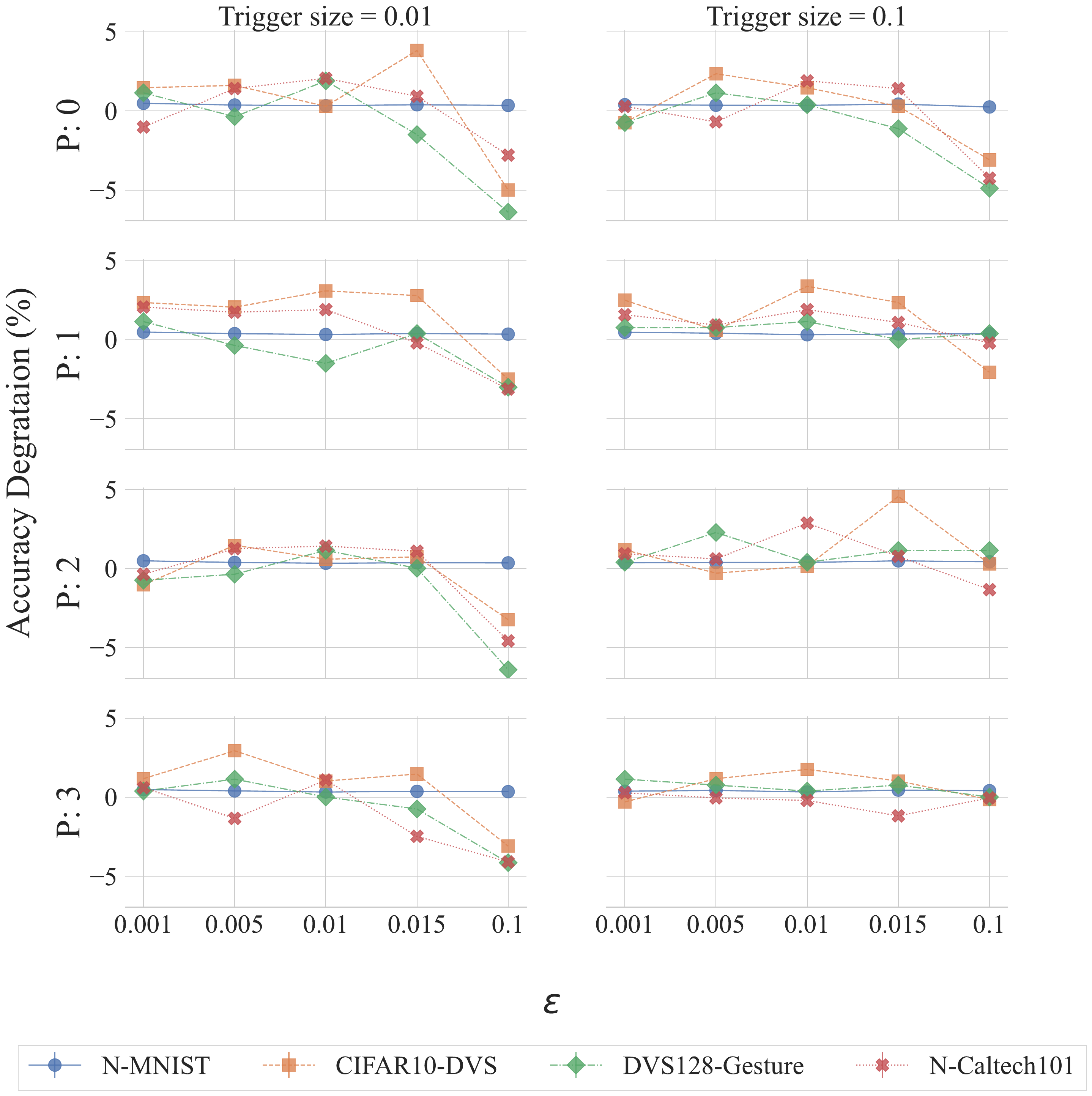}
        \caption{Top-left moving backdoor}
        \label{fig:moving_top_left_degradation}
    \end{subfigure}
    \caption{Clean accuracy degradation of moving triggers.}
\end{figure*}

The clean accuracy degradation after the moving attack in the bottom-right corner is shown in~\autoref{fig:moving_bottom_right_degradation} and for the middle trigger in~\autoref{fig:moving_middle_degradation}. Lastly, the degradation in the top-left corner is shown in~\autoref{fig:moving_top_left_degradation}.

\subsection{Clean Accuracy Degradation of Smart Triggers}

\begin{figure}
    \centering
    \begin{subfigure}[b]{0.49\linewidth}
        \includegraphics[width=\linewidth]{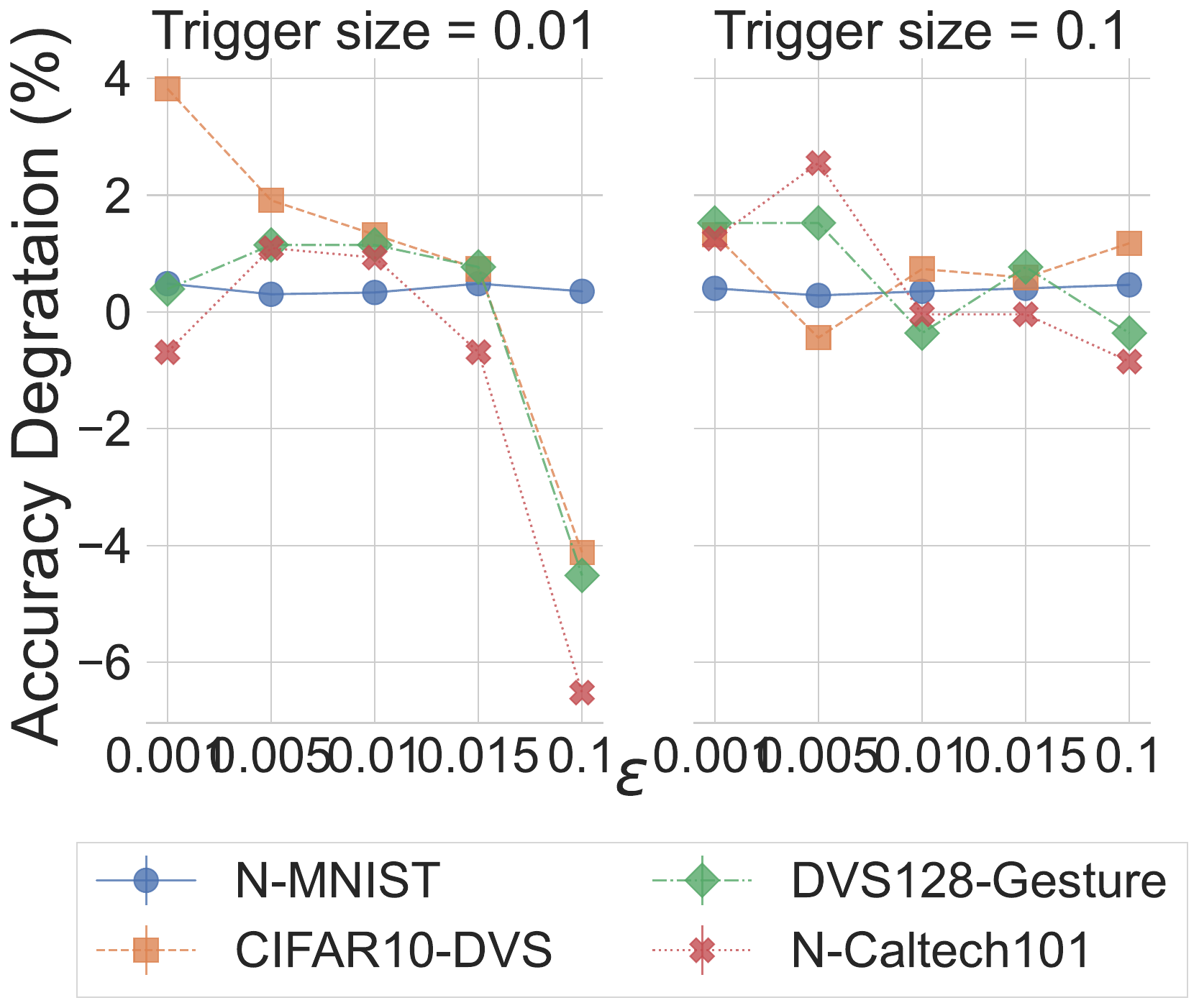}
        \caption{Most active area and least active trigger.}
        \label{fig:smart_False_False_degradation}
    \end{subfigure}
    \hfill
    \begin{subfigure}[b]{0.49\linewidth}
        \includegraphics[width=\linewidth]{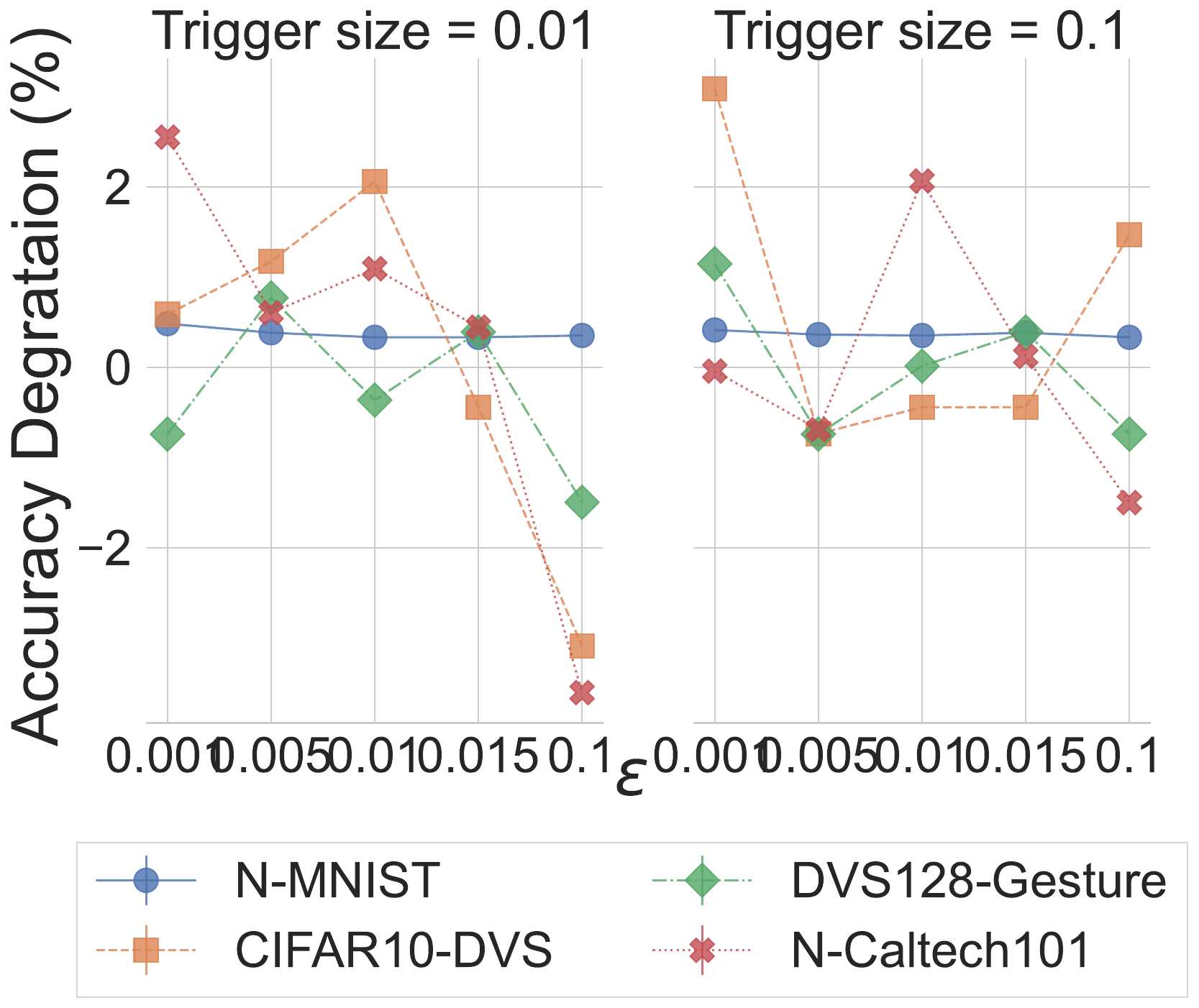}
        \caption{Least active area and least active trigger.}
        \label{fig:smart_True_False_degradation}    
    \end{subfigure}
    \caption{Clean accuracy degradation of the smart trigger in different settings.}
\end{figure}

\begin{figure}
    \centering
    \begin{subfigure}[b]{0.49\linewidth}
        \includegraphics[width=\linewidth]{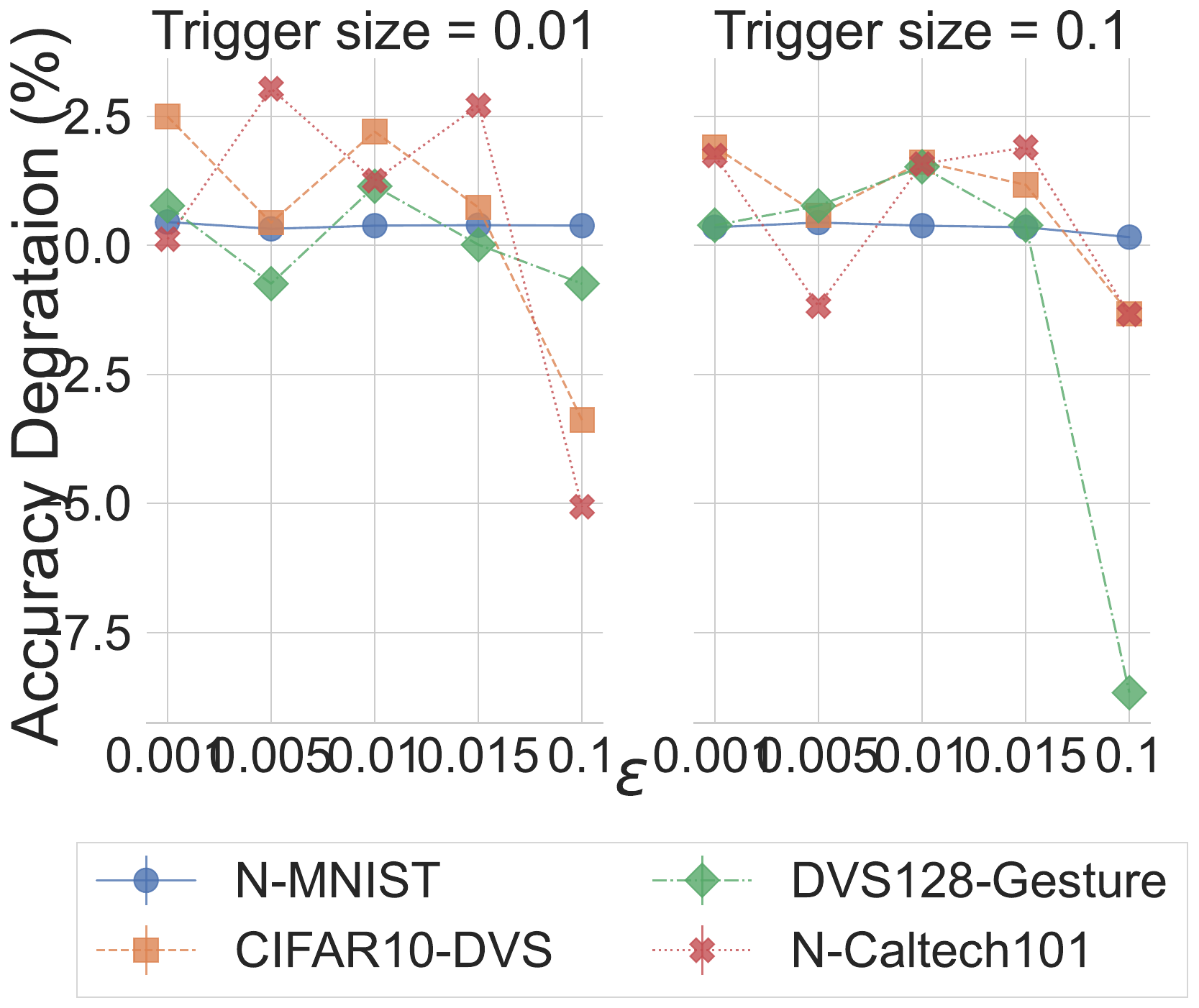}
        \caption{Most active area and most active trigger.}
        \label{fig:smart_False_True_degradation}
    \end{subfigure}
    \hfill
    \begin{subfigure}[b]{0.49\linewidth}
        \includegraphics[width=\linewidth]{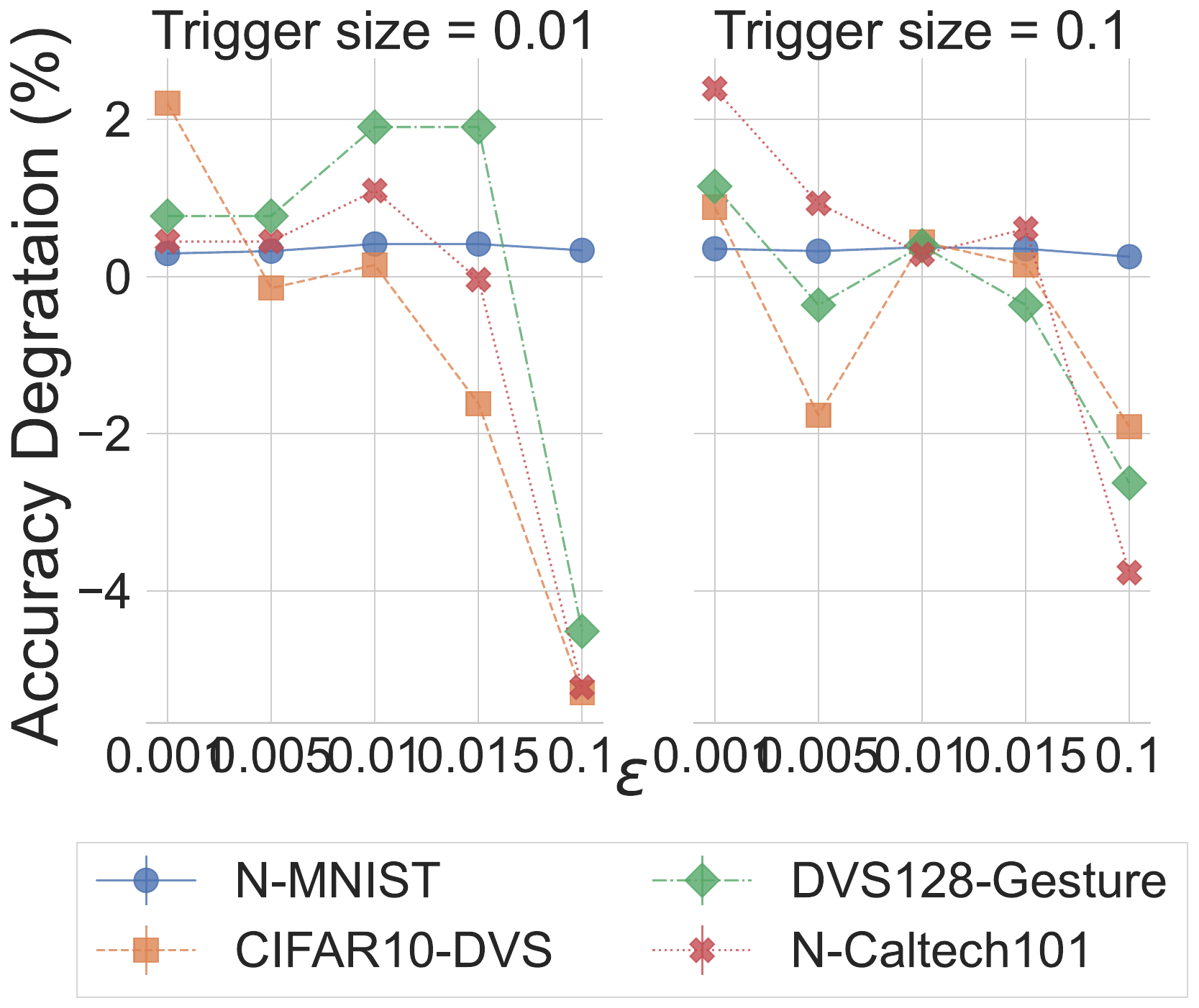}
        \caption{Most active area and most active trigger.}
        \label{fig:smart_True_True_degradation}
    \end{subfigure}
    \caption{Clean accuracy degradation of the smart trigger in different settings.}
\end{figure}

The clean accuracy degradation of the smart attack in the most active area and the least active trigger is shown in~\autoref{fig:smart_False_False_degradation}. For the least active area and the least active trigger, see~\autoref{fig:smart_True_False_degradation}. For the most active triggers in the most and least active areas, see~\autoref{fig:smart_False_True_degradation} and~\autoref{fig:smart_True_True_degradation}, respectively.

\subsection{Additional Experimentation on STRIP}
\label{sec:appdx_strip}

\begin{figure*}
    \centering
    \begin{subfigure}[b]{0.23\linewidth}
        \includegraphics[width=\linewidth]{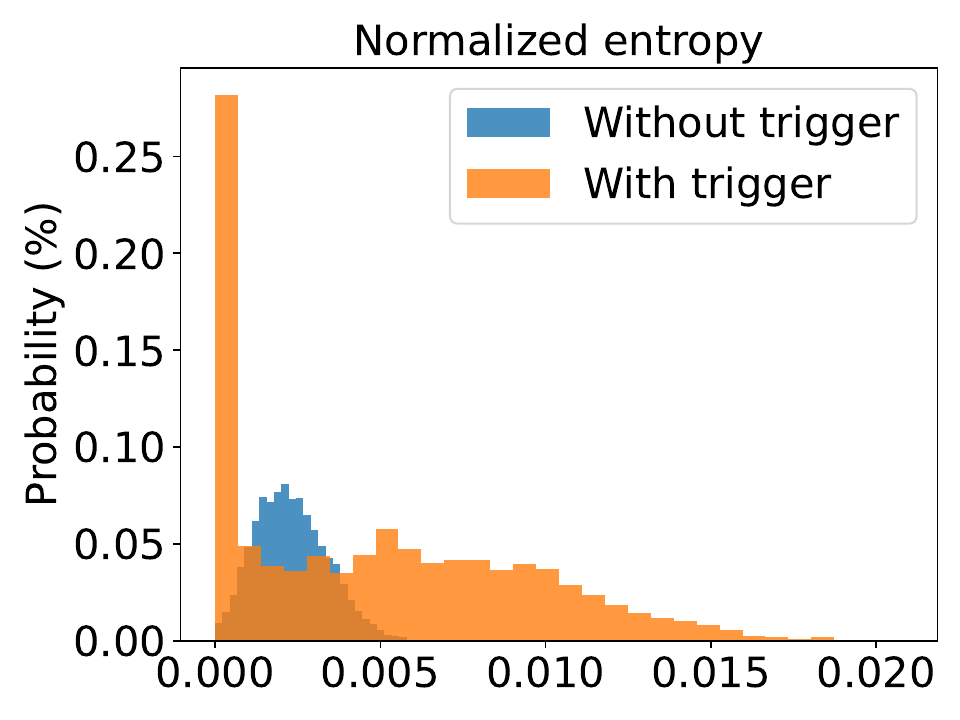}
        \caption{Moving N-MNIST}
    \end{subfigure}
    \begin{subfigure}[b]{0.23\linewidth}
        \includegraphics[width=\linewidth]{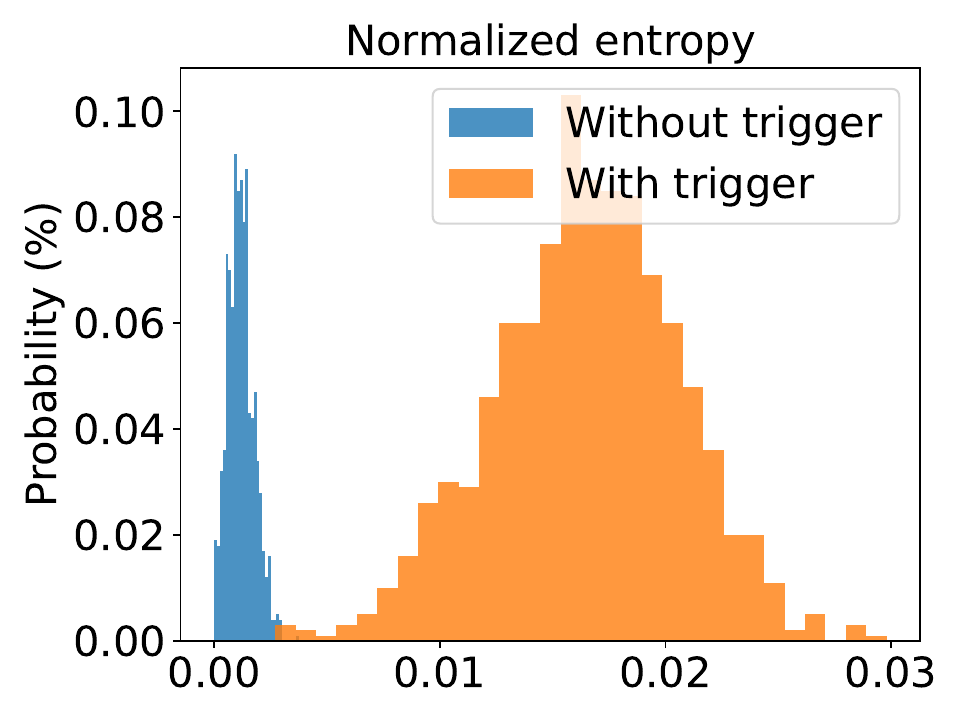}
        \caption{Moving CIFAR10-DVS}
    \end{subfigure}
    \begin{subfigure}[b]{0.23\linewidth}
        \includegraphics[width=\linewidth]{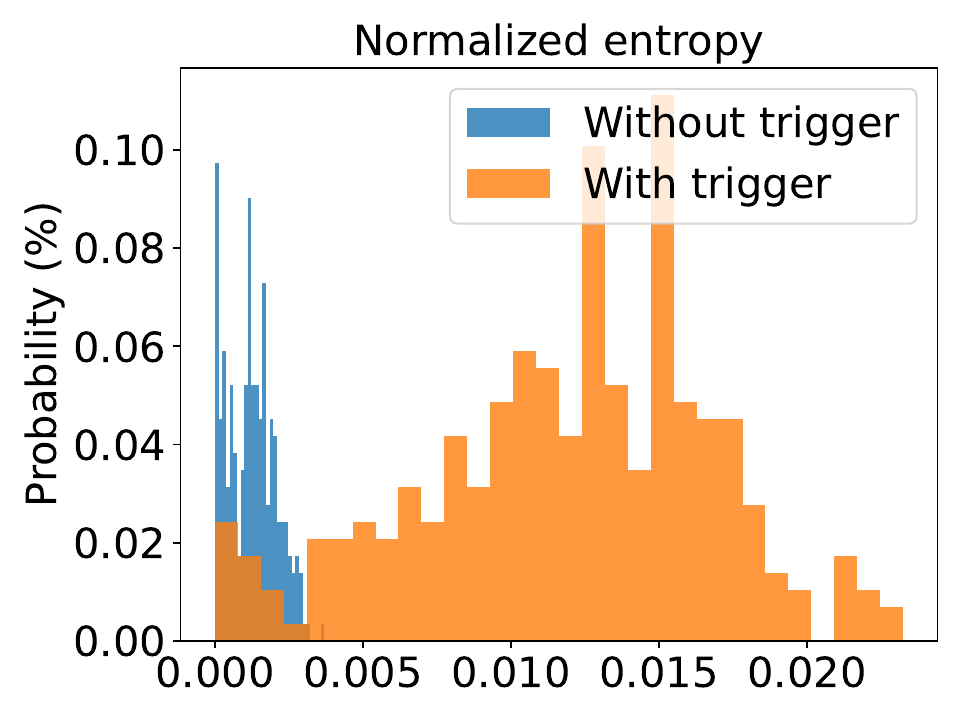}
        \caption{Moving Gesture}
    \end{subfigure}
    \begin{subfigure}[b]{0.23\linewidth}
        \includegraphics[width=\linewidth]{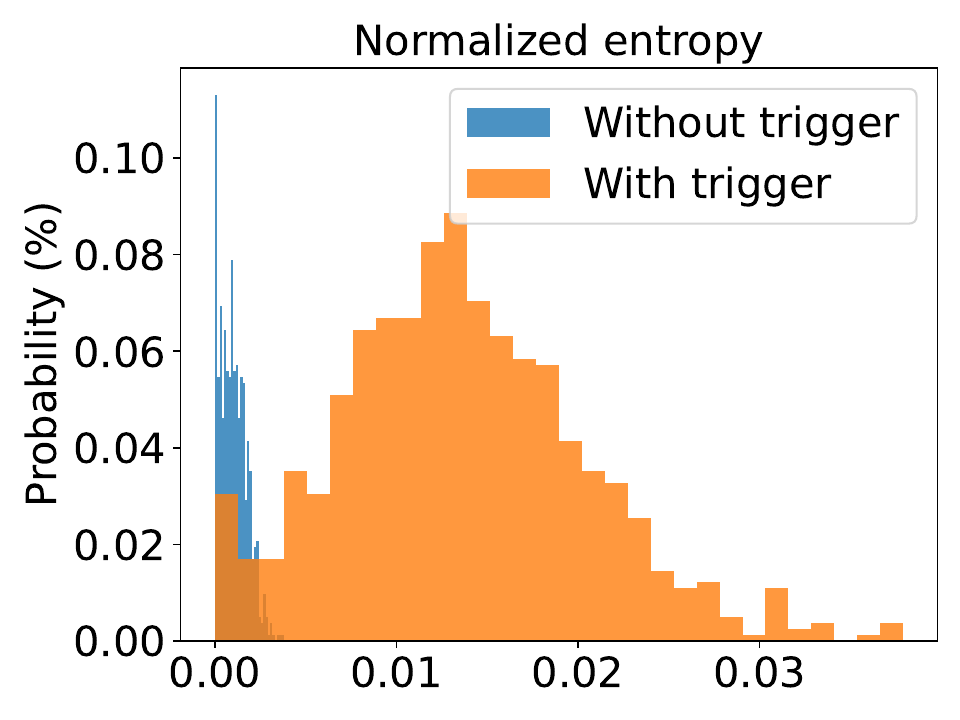}
        \caption{Moving Caltech}
    \end{subfigure}
    \vfill
    \begin{subfigure}[b]{0.23\linewidth}
        \includegraphics[width=\linewidth]{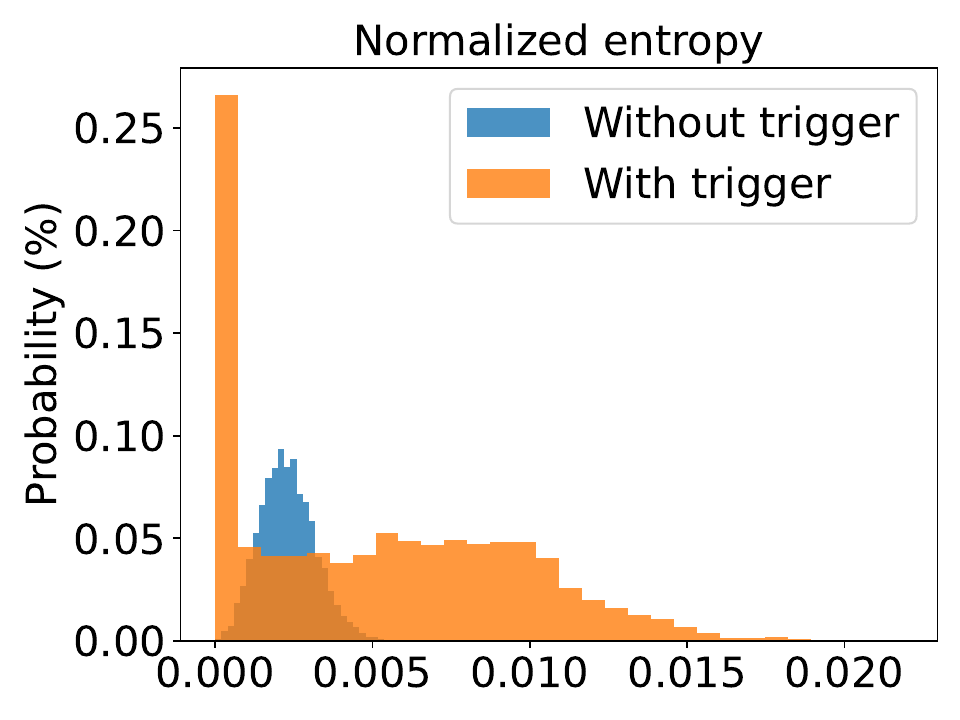}
        \caption{Smart N-MNIST}
    \end{subfigure}
     \begin{subfigure}[b]{0.23\linewidth}
        \includegraphics[width=\linewidth]{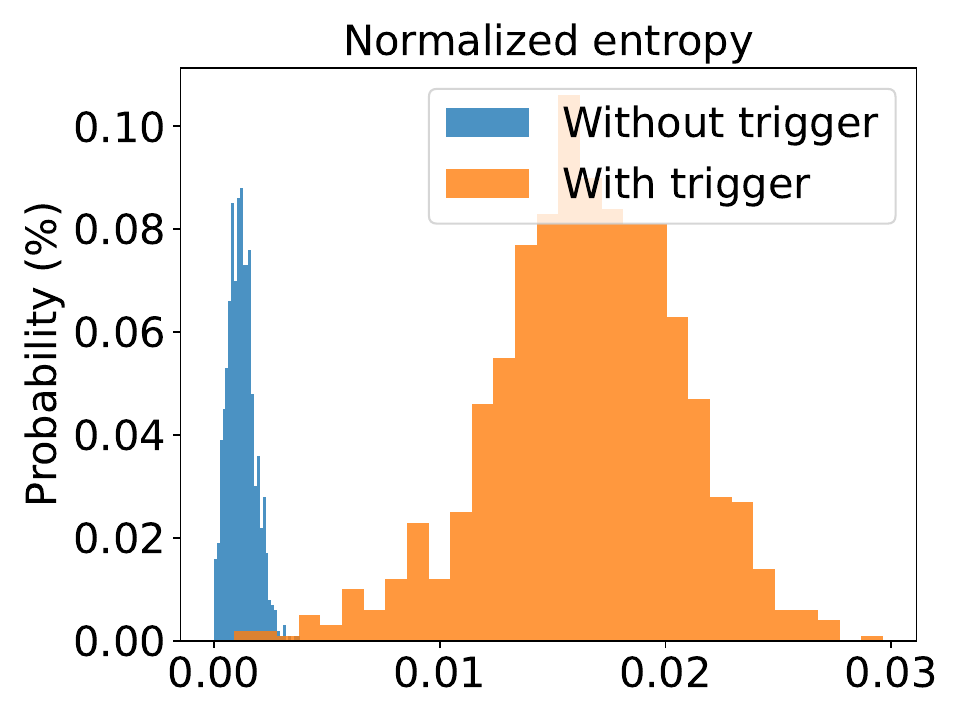}
        \caption{Smart CIFAR10-DVS}
    \end{subfigure}
    \begin{subfigure}[b]{0.23\linewidth}
        \includegraphics[width=\linewidth]{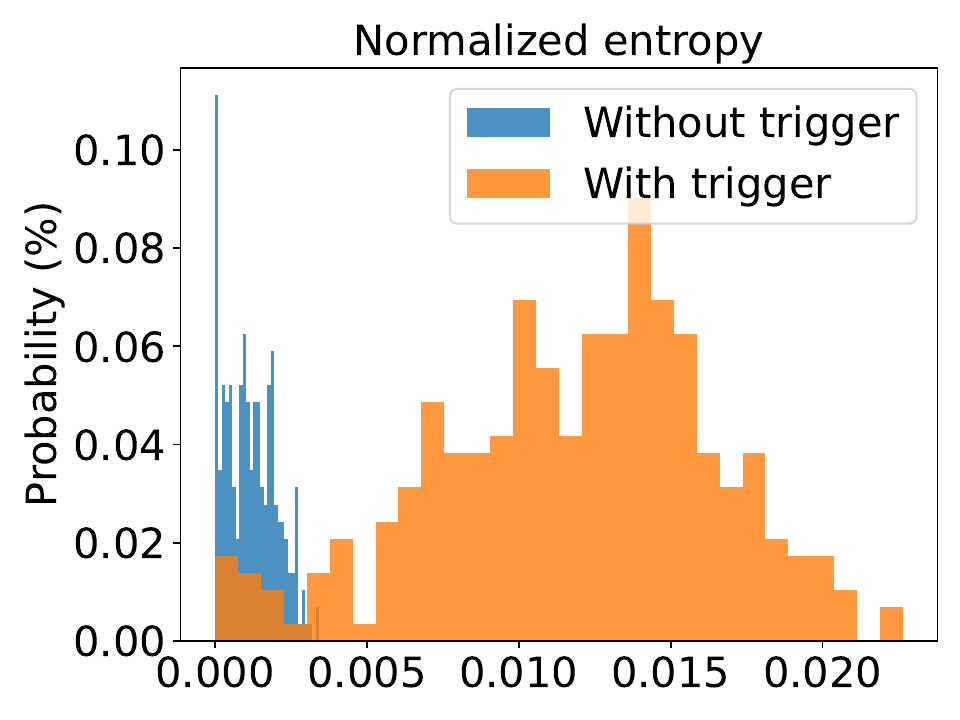}
        \caption{Smart Gesture}
    \end{subfigure}
    \begin{subfigure}[b]{0.23\linewidth}
        \includegraphics[width=\linewidth]{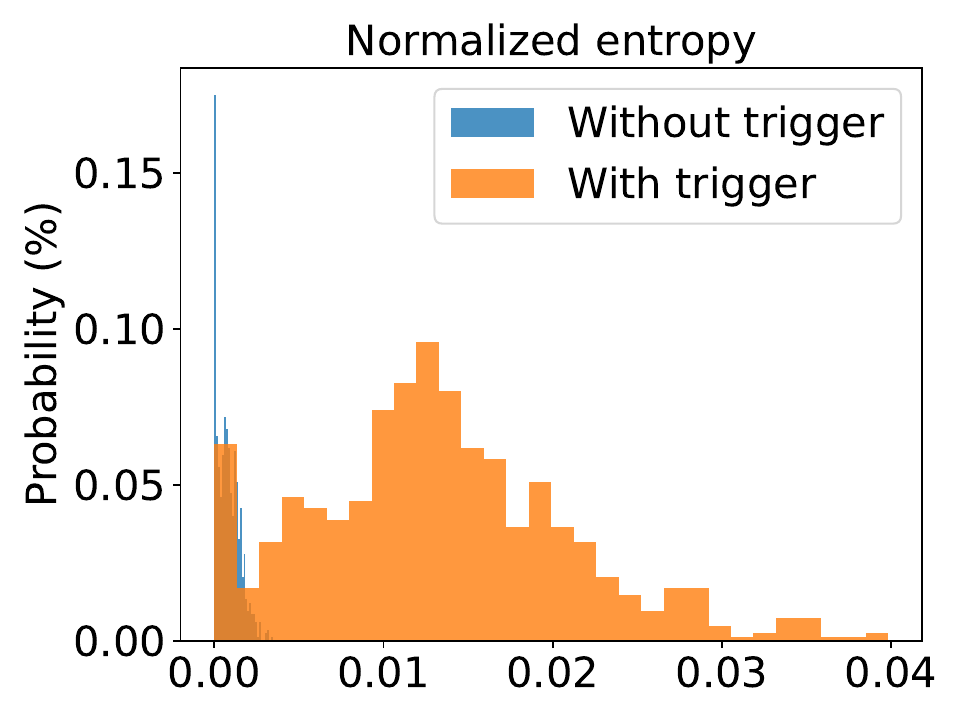}
        \caption{Smart Caltech}
    \end{subfigure}
    \caption{Normalized entropy of different triggers and datasets.}
    \label{fig:strip_appendix}
\end{figure*}

\autoref{fig:strip_appendix} shows the normalized entropy for moving and smart triggers on different datasets.

\subsection{Additional Experimentation on Pruning}

\begin{figure*}
    \centering
    \begin{subfigure}[b]{0.2\linewidth}
        \includegraphics[width=\linewidth]{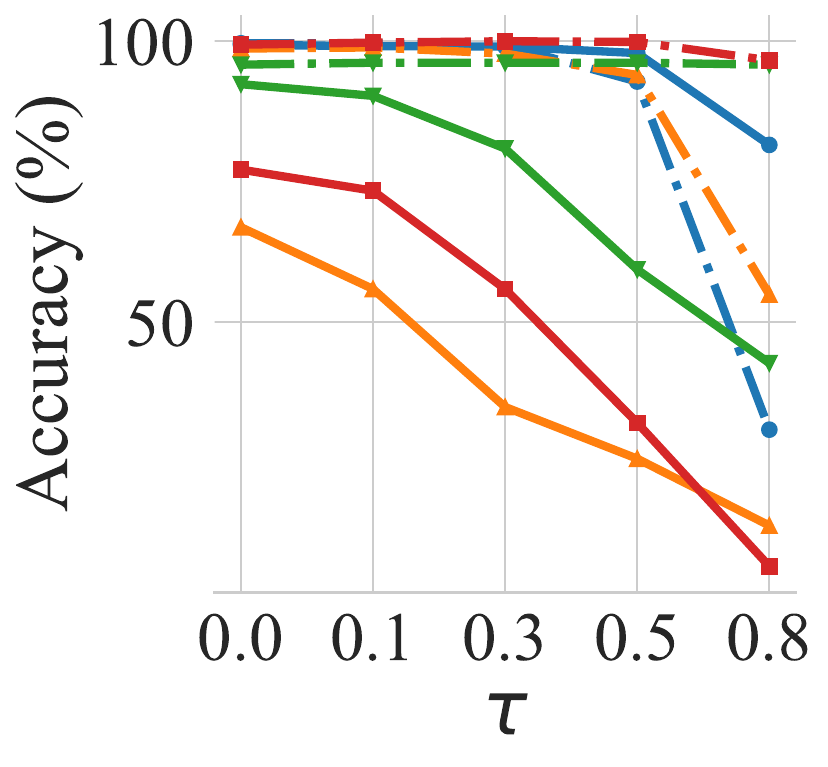}
        \caption{Pruning static}
        \label{fig:pruning_static}
    \end{subfigure}
    \begin{subfigure}[b]{0.18\linewidth}
        \includegraphics[width=\linewidth]{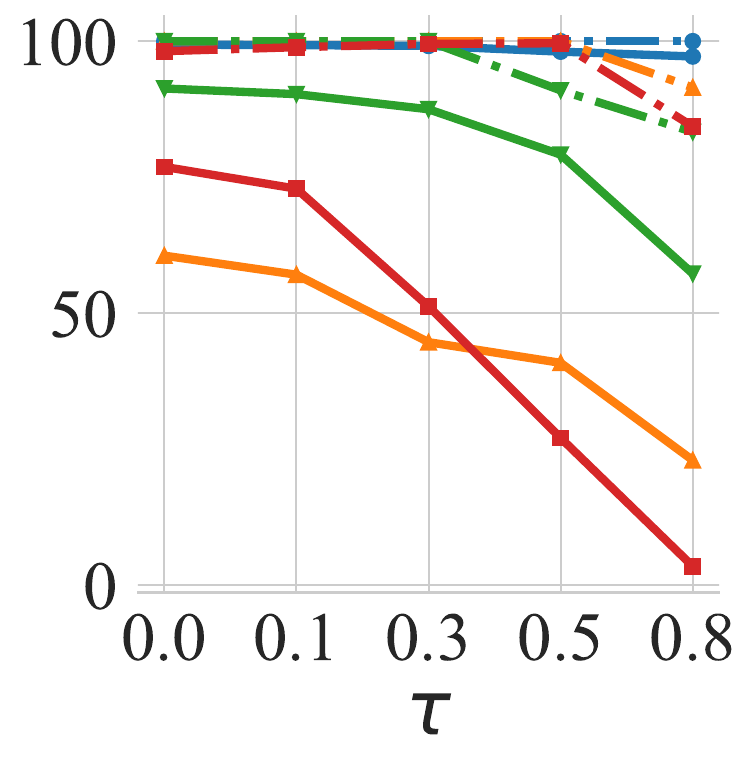}
        \caption{Pruning moving}
        \label{fig:pruning_moving}
    \end{subfigure}
    \begin{subfigure}[b]{0.18\linewidth}
        \includegraphics[width=\linewidth]{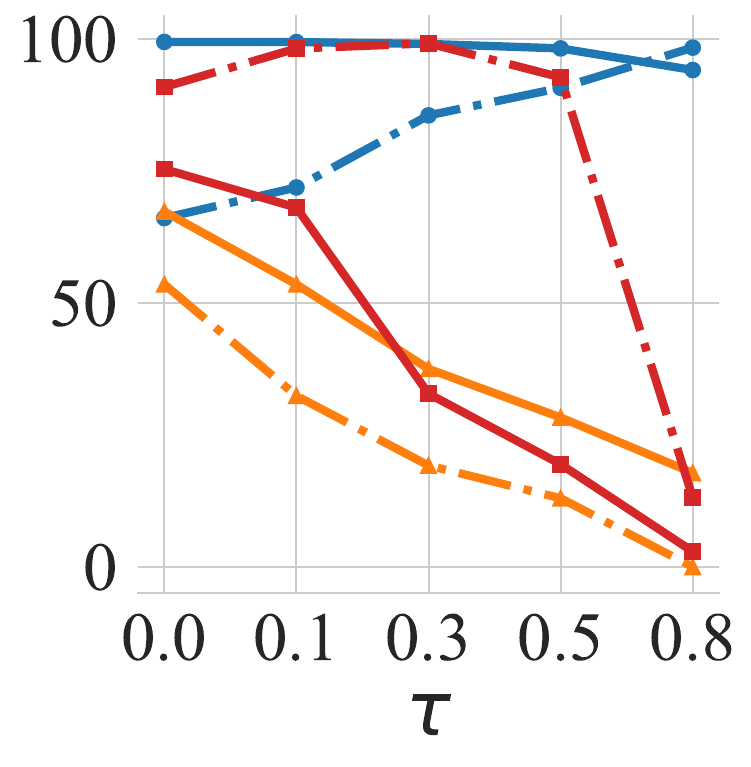}
        \caption{Pruning smart}
        \label{fig:pruning_smart}
    \end{subfigure}
    \begin{subfigure}[b]{0.18\linewidth}
        \includegraphics[width=\linewidth]{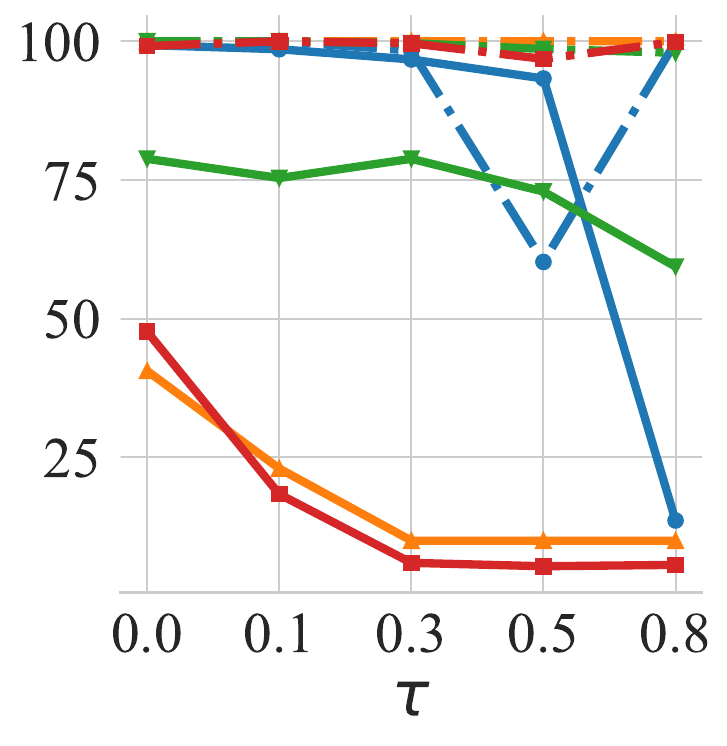}
        \caption{Pruning dynamic}
        \label{fig:pruning_dynamic}
    \end{subfigure}

    \caption{Effect of pruning on the \ac{ASR} (dashed lines) and clean accuracy (full line) for different types of attacks, i.e., static, moving, smart, and dynamic. Blue corresponds to N-MNIST, orange to CIFAR10-DVS, green to the DVS128-Gesture, and red to N-Caltech101.}
    \label{fig:pruning_appendix}
\end{figure*}

\autoref{fig:pruning_appendix} shows additional experimentation solely on pruning---without retraining for a few epochs on different trigger types and datasets.

\subsection{Additional Experimentation on SSIM}
\label{sec:app_ssim}

\begin{figure}
    \centering
    \includegraphics[width=0.9\linewidth]{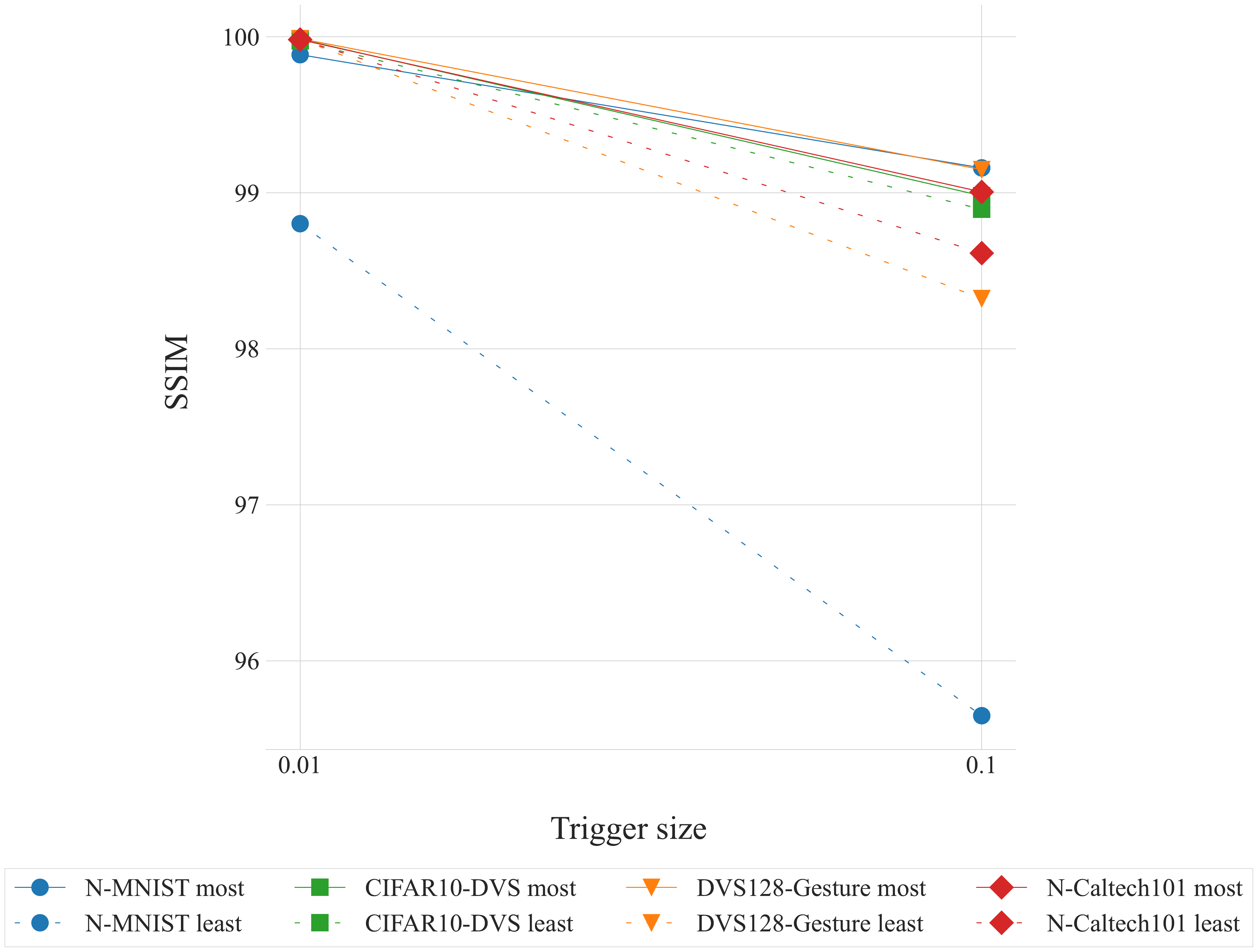}
    \caption{SSIM of smart triggers.}
    \label{fig:ssim_appendix}
\end{figure}

\autoref{fig:ssim_appendix} shows SSIM for smart triggers.

\subsection{Experimentation on N-Caltech101}
\label{sec:caltech}

This section discusses the results of different attacks on the N-Caltech101 dataset. 

\subsubsection{Static Backdoor}

For the N-Caltech101 dataset, we observe a similar behavior to the before-mentioned datasets. However, it is important to notice that the number of classes is ten times larger, making both main and backdoor tasks more challenging. Static triggers, see~\autoref{fig:static_backdoors_res}, show that a large trigger, e.g., 0.1, is needed to inject the backdoor. However, the backdoor can also be effective with smaller triggers in some settings where the trigger is placed in the middle or the top-left corner.

\subsubsection{Moving Backdoor}

Moving trigger for the N-Caltech101 dataset, see~\autoref{fig:moving_backdoors_res}, shows similar performance and properties as in previously evaluated datasets. Similarly, large triggers are required to achieve high backdoor performance. The importance of the polarity selection is also evident in this case, where $p=0$, i.e., background polarity, only works when the trigger is placed in the middle. This is because the image samples are centered, i.e., the image's action area is in the middle.

\subsubsection{Smart Backdoor}

In smart backdoors, the trigger location and polarity are selected automatically. A large dataset like N-Caltech101 allows a better selection of these parameters because the location better represents the whole population, i.e., the mean is more accurate. Thus, despite being a ``complex'' dataset with many classes and noise, it achieves excellent results even when a simpler dataset cannot achieve a successful backdoor, see~\autoref{fig:asr_smart_true_true}. Overall, as shown in~\autoref{fig:smart}, the results are aligned with the other datasets.

\subsubsection{Dynamic Backdoor}

Sample-specific dynamic triggers are also suitable for large datasets with many classes; see~\autoref{fig:dynamic}. Note that having many classes makes the backdoor injection and the main task more challenging even though the attack can still achieve 100\% \ac{ASR} in most settings. Similar to the datasets analyzed before, when the trigger is less stealthy, i.e., $\gamma = {0.05, 0.1}$,  \ac{ASR} is higher than in the most stealthy setup, i.e., $\gamma = 0.01$. Thus, there is a clear trade-off between stealthiness and backdoor performance.
We observe a behavior similar to the use case with CIFAR-10, where $\alpha \geq 0.9$ exhibits the best performance.

\section{Artifact Appendix}

In this section, we include information about the reproducibility of the results and general usage of the artifact.

\subsection{Description \& Requirements}

The repository\footnote{\url{https://github.com/GorkaAbad/Sneaky-Spikes}} contains a README showcasing the different attacks and general information. Additionally, a \emph{how to} guide is available. It contains detailed steps on how to prepare and download the datasets, as well as examples of how to run the code.

\subsubsection{How to access}
We provide access to the repository, including a guide on how to run the code. A permanent storage is also available\footnote{\url{https://doi.org/10.5281/zenodo.10156889}}.

\subsubsection{Hardware dependencies}
To run the code, a GPU is strongly recommended. The code is tested on a machine with 1 NVIDIA A100 GPU with 40GB.

\subsubsection{Software dependencies} 
The experiments were run on a Ubuntu 20.04 machine, using Python 3.8 and CUDA 11.7. We use the SpikingJelly\footnote{\url{https://github.com/fangwei123456/spikingjelly}} framework to ease the training of SNNs and the processing of neuromorphic data. 

\subsubsection{Benchmarks} 
Some datasets are automatically downloaded. However, some others have to be downloaded manually. This is a restriction of SpikingJelly. To simplify the process of preparing the data, we provided a detailed \textit{how to} guide in the repository.

\subsection{Artifact Installation \& Configuration}
We require the standard libraries used in DL, e.g., \verb|torch, numpy, pandas|. These can be easily installed with \verb|pip|. We provide a guide on how to install the requirements in the repository.

\subsection{Major Claims}

This paper provides four types of attacks, i.e., static, moving, smart, and dynamic. Each is composed of a different type of backdoor trigger. We provide examples for reproducing the claims on the DVS128-Gesture and N-MNIST datasets (as they are the fastest to run).
However, examples with additional datasets are available in the repository (in the \verb|scripts/| folder).

\begin{itemize}
    \item (C1): The static backdoor performance using a trigger of 10\% the size of the input image, on the top-left corner, with a poisoning rate of 10\%  of the dataset, achieves around 100\% ASR.
    This is proven by the experiment (E1) whose results are illustrated/reported in [Figure 10].
    \item (C2): The moving backdoor performance using a trigger of 10\% the size of the input image, on the top-left corner, with a poisoning rate of 10\%  of the dataset, achieves around 100\% ASR.
    This is proven by the experiment (E2) whose results are illustrated/reported in [Figure 11].
    \item (C3): The smart backdoor performance using a trigger of 10\% using the most common polarity with a poisoning rate of 10\%  of the dataset, achieves around 100\% ASR.
    This is proven by the experiment (E3) whose results are illustrated/reported in [Figure 3.B].
    \item (C4): The dynamic backdoor achieves around 100\% ASR.
    This is proven by the experiment (E4) whose results are illustrated/reported in [Figure 4].
\end{itemize}

\subsection{Evaluation}

\subsubsection{Experiment (E1)}
[Static] [0 human-minutes + 1 compute-hour]: The static attack achieves around 100\% ASR. 

\textit{[How to]}
See [Execution] for executing the experiment.

\textit{[Preparation]}
Prepare the DVS128-Gesture dataset.

\textit{[Execution]}
\begin{verbatim}
python main.py --dataset gesture --polarity 1 
--pos top-left --trigger_size 0.1 --cupy
--type static --epsilon 0.1 --epochs 64
\end{verbatim}

\textit{[Results]}
The results are stored in \verb|results/|.

\subsubsection{Experiment (E2)}
[Moving] [0 human-minutes + 1 compute-hour]: The moving attack achieves around 100\% ASR.

\textit{[How to]}
See [Execution] for executing the experiment.

\textit{[Preparation]}
Prepare the DVS128-Gesture dataset.

\textit{[Execution]}
\begin{verbatim}
python main.py --dataset gesture 
--polarity 1  --pos top-left 
--trigger_size 0.1 --epsilon 0.1 
--type moving --cupy --epochs 64
\end{verbatim}

\textit{[Results]}
The results are stored in \verb|results/|.

\subsubsection{Experiment (E3)}
[Smart] [0 human-minutes + 1 compute-hour]: The smart attack achieves around 100\% ASR.

\textit{[How to]}
See [Execution] for executing the experiment.

\textit{[Preparation]}
Prepare the N-MNIST dataset.

\textit{[Execution]}
\begin{verbatim}
python main.py --dataset mnist 
--trigger_size 0.1 --epsilon 0.1 
--type smart --most_polarity
--cupy --epochs 10
\end{verbatim}

\textit{[Results]}
The results are stored in \verb|results/|.

\subsubsection{Experiment (E4)}
[Dynamic] [0 human-minutes + 2 compute-hour]: The dynamic attack achieves around 100\% ASR.

\textit{[How to]}
See [Execution] for executing the experiment.

\textit{[Preparation]}
Prepare the N-MNIST dataset.

\textit{[Execution]}
\begin{verbatim}
python dynamic.py --dataset mnist
--cupy --epochs 10
--alpha 0.5 --beta 0.01
\end{verbatim}

\textit{[Results]}
The results are stored in \verb|results/|.

\subsection{Notes}
To execute the experiments for different types of attacks and datasets, we provided some scripts to ease the process. These can be found in the \verb|scripts/| folder in the repository. Note that selecting the parameters during the execution is important and can affect the performance.

\end{document}